\documentclass[reprint, superscriptaddress]{revtex4-1}
\usepackage{amsmath, amssymb, amsfonts}
\usepackage{bm}
\usepackage{color}
\usepackage{float}
\usepackage{graphicx}
\usepackage{hyperref}
\usepackage{dcolumn}

\hypersetup{
	colorlinks = true,
	linkcolor = blue,
	citecolor = blue,
	urlcolor = blue
}

\graphicspath{{paper_figures_final/}}

\begin{document}

\title{Development of structural correlations and synchronization from adaptive rewiring in networks of Kuramoto oscillators}

\author{Lia Papadopoulos} 
\affiliation{Department of Physics and Astronomy, University of Pennsylvania, Pennsylvania, PA 19104, USA}
\author{Jason Kim}
\affiliation{Department of Bioengineering, University of Pennsylvania, Pennsylvania, PA 19104, USA}
\author{J{\"u}rgen Kurths}
\affiliation{Potsdam Institute for Climate Impact Research - Telegraphenberg A 31, 14473 Potsdam, Germany}
\affiliation{Department of Physics, Humboldt University Berlin - 12489 Berlin, Germany}
\affiliation{Institute for Complex Systems and Mathematical Biology, University of Aberdeen - Aberdeen AB24 3UE, UK}
\author{Danielle S. Bassett}
\affiliation{Department of Bioengineering, University of Pennsylvania, Pennsylvania, PA 19104, USA}
\affiliation{Department of Electrical \& Systems Engineering, University of Pennsylvania, Pennsylvania, PA 19104, USA}
\email{dsb@seas.upenn.edu}

\date{\today}

\begin{abstract}

Synchronization of non-identical oscillators coupled through complex networks is an important example of collective behavior, and it is interesting to ask how the structural organization of network interactions influences this process. Several studies have explored and uncovered optimal topologies for synchronization by making purposeful alterations to a network. On the other hand, the connectivity patterns of many natural systems are often not static, but are rather modulated over time according to their dynamics. However, this co-evolution - and the extent to which the dynamics of the individual units can shape the organization of the network itself - are less well understood. Here, we study initially randomly connected but locally adaptive networks of Kuramoto oscillators. In particular, the system employs a co-evolutionary rewiring strategy that depends only on the instantaneous, pairwise phase differences of neighboring oscillators, and that conserves the total number of edges, allowing the effects of local reorganization to be isolated. We find that a simple rule - which preserves connections between more out-of-phase oscillators while rewiring connections between more in-phase oscillators - can cause initially disordered networks to organize into more structured topologies that support enhanced synchronization dynamics. We examine how this process unfolds over time, finding a dependence on the intrinsic frequencies of the oscillators, the global coupling, and the network density, in terms of how the adaptive mechanism reorganizes the network and influences the dynamics. Importantly, for large enough coupling and after sufficient adaptation, the resulting networks exhibit interesting characteristics, including degree - frequency and frequency - neighbor frequency correlations. These properties have previously been associated with optimal synchronization or explosive transitions in which the networks were constructed using global information. On the contrary, by considering a time-dependent interplay between structure and dynamics, this work offers a mechanism through which emergent phenomena and organization can arise in complex systems utilizing local rules.

\end{abstract}

\maketitle

\textbf{Electrical activity in populations of neurons, or the spread of information or disease across social networks are common examples of dynamical processes unfolding on networks. In many real-world systems, synchronization of dynamics, and its dependence on network architecture, is of critical interest. Importantly, network interactions can also be adaptive, such that connections rearrange over time according to the dynamical states of the nodes. A prevalent example of adaptation in connectivity takes place in biological neuronal networks, which reconfigure via local plasticity mechanisms that are informed by the dynamical relationships between directly coupled elements. Motivated by these types of systems, here we address the question of how both structured network topology and global synchronization can develop through a co-evolution of the underlying network connectivity and the dynamics. Using the canonical Kuramoto model, we suggest a simple, adaptive rewiring scheme based on local phase information that can evolve initially unstructured random graphs towards topologies with specific organizational principles. In turn, we find that this process simultaneously enhances synchronization. The structure that arises in the co-evolving systems manifests as specific relationships between the natural frequencies of the oscillators and the network topology, properties that have previously been associated with optimal synchronization in which networks were purposefully constructed using complete information of the network and node properties. Yet here, without any global information, we consider a time-dependent interplay between structure and dynamics, and offer a mechanism through which organization can emerge in complex systems utilizing local rules.}

\section{Introduction \label{s:intro}}

Exactly how dynamical processes unfold on networks with non-trivial coupling between individual units remains an important question in complex systems science \cite{Newman:2003aa, Boccaletti:2006a, Barthelemy:2008a, Vespignani:2012a}. Examples of such dynamical systems on networks include the time-dependent patterns of electrical activity in populations of neurons \cite{Ermentrout:2010a,Brunel:1999a,Brunel:2000a,Guardiola:2000a,Lago-Fernandez:2000a,Denker:2004a,Roxin:2004a}, the spread of information or disease across social networks \cite{bassett2012collective, Miller:2014a, Pastor-Satorras:2015a, Moreno:2002a}, or regulatory mechanisms in biological networks \cite{Barabasi:2004a, Karlebach:2008a, Elowitz:2000a,Luscombe:2004a}. In each case, the way the system evolves over time is dependent on the specific form of the dynamics, intrinsic properties of each element (either of nodes or edges), and the architecture of connectivity. Intriguing questions are if and how collective behavior can emerge in these systems. A significant and widespread manifestation of this is synchronization \cite{Winfree:1980a, Strogatz:2003a, Pikovsky:2003a, Manrubia:2004a, Arenas:2008a}, in which a group of interacting elements converge to the same state or evolve in unison over time. Real-world illustrations of this phenomenon range from circadian clock cycles \cite{Glass:2001a, Strogatz:2003a}, to the rhythmic patterns of functional activity in the human brain \cite{Fries:2005a,bassett2006adaptive,khambhati2015dynamic,Deco2011:EmergingConcepts,bassett2009cognitive,Sporns:2000a,Stam:2004a,Kopell:2000a,Engel:2001a,Fox:2005a,Buzsaki:2004a,Varela:2001a}, and synchronization in power-grid networks \cite{Lozano:2012a,Dorfler:2012a,Rohden:2012a,Motter:2013a}.

One of the most common and useful models for studying synchronization is the canonical Kuramoto model \cite{Kuramoto:1975a, Kuramoto:1984a}, which originally described the evolution of a population of $N$ all-to-all coupled phase oscillators that were in general non-identical (see \cite{Strogatz:2000a, Acebron:2005a} for reviews). In recent years, this model has been extended to study systems with heterogeneous network topologies, in order to investigate how the architecture of complex connectivity affects the onset of synchronization in diverse oscillator populations \cite{Arenas:2008a, Rodrigues:2016a}. Such efforts have provided important insights into the nature of the synchronization transition in different graph models including those that display a scale-free degree distribution \cite{Moreno:2004a, Lee:2005a, Gomez-Gardenes:2007a, Gomez-Gardenes:2007b}, those with small-world architecture \cite{Hong:2002a}, and those with community structure \cite{Oh:2005a, Arenas:2006a, Arenas:2007a, Arenas:2006b, Guan:2008a, Skardal:2012a}. 

Within this body of work, particular attention has been paid to understanding what features of a network inhibit or enhance the ability to support collective dynamics. For the case of identical oscillators, this is often studied from the perspective of minimizing a ratio of eigenvalues that depend only on the structure of the network \cite{Pecora:1998a, Barahona:2002a}. However, the question of ``optimal" networks for synchronization can be more interesting and complex when the oscillators' natural frequencies are heterogeneous, a characteristic of many real-world systems. In particular, for non-identical oscillators, a crucial consideration becomes how the structure of the network is intertwined with dynamical properties. For example, in the Kuramoto model, synchronization can be enhanced when there are specific types of correlations between node degrees and oscillator frequencies or between the natural frequencies of adjacent oscillators \cite{Brede:2008b, Brede:2009a}. \textcite{Gomez-Gardenes:2011aa} demonstrated that in scale-free networks, positive frequency-degree correlations can lead to a first-order, or explosive, transition to synchronization. More recently, discontinuous transitions have been found by imposing constraints on the minimal difference between connected nodes' natural frequencies \cite{Leyva:2013a,Leyva:2013b}. There has also been progress in analytical work determining network topologies that enhance synchronization. For example, it has been shown that optimal networks for synchronizing collections of non-identical oscillators exhibit particular relationships between Laplacian eigenvectors and oscillator frequencies \cite{Skardal:2014a,Skardal:2015a,Skardal:2016a}. In addition, dimensionality reduction approaches \cite{Gottwald:2015a} have recovered many previous numerical results, and have been used to derive analytical conditions for optimizing synchronization of Kuramoto oscillators in networks with attractive and repulsive interactions \cite{Pinto:2015a}.

Notably, while the coupling structure between oscillators can drive their dynamics, network dynamics can also modulate structure. Specifically, in adaptive systems, the pattern of connectivity itself is continuously updated and modified in response to the dynamics that occur on top of it \cite{Gross:2009a,Gross:2008a,Sayama:2013a}. Systems that display these processes can be observed across biological, ecological, social, and distribution networks \cite{Gross:2008a,Gross:2009a}, and collectively they can be characterized by topology and dynamical states that co-evolve with one another. Kuramoto-like models in particular provide a useful framework in which to explore the effects of co-evolution and adaptation \cite{Yuan:2011a, Yuan:2013a,Zhou:2006a, Zhu:2010a,Aoki:2009a,Aoki:2011a,Rubinov:2009a,Gleiser:2006a,Seliger:2002a,Nadal:2009a,Nadal:2008a,Assenza:2011a,Gutierrez:2011a}, and allow one to address questions such as \emph{(i)} can and how might these systems organize themselves towards network configurations that enhance local or global synchronization?, and \emph{(ii)} from an initially unstructured topology, can and how do different adaptive mechanisms lead to the emergence of certain architectural patterns or correlations between dynamical properties and network structure?

In addition to being adaptive, it is important to note that the evolution of many real-world networks is often governed by \textit{local} rules, in which node dynamics update as a function of only neighboring node states, and in turn, the placement or weights of edges update primarily as a function of the states of the nodes they directly couple \cite{Gross:2009a,Gross:2008a,Sayama:2013a}. This type of behavior is especially pertinent in biological systems, which typically co-evolve in the absence of global or top-down controllers of node states and/or network structure. A particularly salient example of this occurs in biological neuronal networks, where, under Hebbian plasticity rules, increases in synaptic weights occur when connected neurons exhibit correlated dynamics \cite{Markram:1996a,Song:2000a,Caporale:2008a}, while under anti-Hebbian plasticity rules, the opposite occurs \cite{Lamsa:2007a,Roberts:2010a}. Other systems that obey such local adaptation are prevalent, and studied examples include models of reconfiguration of social networks under disease propagation \cite{Gross:2006a} and opinion formation \cite{Holme:2006a}, or reorganization under feedback mechanisms in boolean models with applications to gene regulatory or neural networks \cite{Bornholdt:2000a,Bornholdt:2003a}. 

In this study, we use the Kuramoto model to investigate how a simple, adaptive rewiring scheme can evolve initially unstructured random graphs towards ordered topologies, and also simultaneously lead to enhanced synchronization in the system. Importantly, the rule is informed by only local information of neighboring nodes' states at a given instant of time, and works by regularly breaking and randomly rewiring connections between more instantaneously phase-synchronized oscillators, while maintaining connections between more de-synchronized pairs of oscillators. This process repeats continually over time, and can be thought of as a repulsive mechanism, or one that tends to represses assortativity (in terms of nodes connecting to other nodes with similar instantaneous states). We find that co-evolution of the network and dynamics can promote the degree of synchronization in the system, which occurs in tandem with the development of specific correlations between the topology and the natural frequencies of the oscillators. In previous work, these features have been imposed by purposeful selection, or have been shown to arise in work on optimizing synchronization. Here, however, the properties emerge from the interplay between network structure and dynamics. Focusing on the simplest situation of binary, undirected networks, we isolate the effects of adaptive reconfiguration alone, and thereby uncover a process through which heightened collective dynamics and organized network structure simultaneously arise in a local, unsupervised way.

The remainder of this paper is organized as follows. Sec.~\ref{s:kuramoto_dynamics} states the formulation of Kuramoto dynamics on complex networks. In Sec.~\ref{s:motivation_and_methods} we first briefly outline past work to motivate the specific mechanism studied here, and then detail the proposed co-evolutionary strategy. Sec.~\ref{s:results} describes several interesting results of the adaptive process, and in Sec.~\ref{s:discussion} we discuss the implications of our findings and conclude.

\section{The Kuramoto model on complex networks}
\label{s:kuramoto_dynamics}

Of the many models that exist for studying synchronization phenomena on complex networks, one of the most useful has been the paradigmatic Kuramoto model \cite{Kuramoto:1975a,Kuramoto:1984a}. It describes the dynamical evolution of a population of $N$ phase oscillators coupled on a network according to the following equation:

\begin{equation}
\label{eq:kuramoto}
\dot{\theta}_{i} = \omega_{i} + \alpha \sum_{j = 1}^{N} A_{ij} \sin(\theta_{j} - \theta_{i}).
\end{equation}

\noindent In this formulation, $\theta_{i}$ is the instantaneous phase of the $i^{th}$ oscillator, $\omega_{i}$ is its natural frequency, $\alpha$ is the overall coupling strength, and $\mathbf{A}$ is the $N \times N$ adjacency matrix describing the connectivity of the network. In this report, we consider binary, undirected networks, such that

\begin{equation}
\label{eq:adjacency}
A_{ij} = \left\{\begin{array}{ll}
         1 \text{ if there is an edge between nodes \it{i} and \it{j},} \\
       	0 \text{ otherwise.}
        \end{array} \right.\
\end{equation}

\noindent The natural frequencies $\{\omega_{i}\}$ are distributed according to a probability density $g(\omega)$, which we will take to be symmetric and centered around a mean frequency of zero.

The overall amount of synchrony in the population at a given time $t$ is typically quantified with the Kuramoto order parameter \cite{Kuramoto:1975a,Kuramoto:1984a}
\begin{equation}
R(t)e^{i\psi(t)} = \frac{1}{N}\sum_{j = 1}^{N} e^{i\theta_{j}(t)},
\end{equation}
which can be thought of as the centroid of the $N$ phases on a unit circle in the complex plane \cite{Rodrigues:2016a}. Here, $\psi$ is the average phase of the population, and the modulus $R$, given by,
\begin{equation}
\label{eq:R}
R(t) = \frac{1}{N} \left \lvert \sum_{j = 1}^{N} e^{i\theta_{j}(t)} \right \rvert,
\end{equation}
quantifies the amount of phase coherence. When the oscillators' phases are uniformly spread, $R \approx 0$ and the system exhibits low synchrony. On the other hand, when the phases become tightly clustered, $R \approx 1$ and the system exhibits high levels of synchrony. 

We can also use this order parameter, which ranges from $R = 0$ (complete incoherence) to $R = 1$ (complete phase synchronization), to monitor the global degree of synchrony in the system as a function of the coupling $\alpha$. In this case, one typically reports a time-averaged value
\begin{equation}
\label{eq:R_time_avg}
\langle R \rangle = \frac{1}{T} \int_{T_{R}}^{T_{R}+T} R(t) dt
\end{equation}
computed on an interval of length $T$ after several transient or relaxation time steps $T_{R}$ have been discarded. 

\section{Motivation and the co-evolutionary model}
\label{s:motivation_and_methods}

In this study, we consider a feedback process between dynamics and the restructuring of network topology that integrates different ideas and results from previous studies on adaptation or enhancing synchronization in networks of non-identical Kuramoto oscillators. To better frame and motivate our contributions, we briefly outline some past work on these topics below.

\subsection{Inspiration from prior investigations}

In general, it is expected that different adaptive strategies will lead to the emergence of different patterns in both the network topology and the dynamics, and a number of studies have explored these ideas using, for example, chaotic dynamics \cite{Gong:2004a,Zhou:2006a,Sorrentino:2008a,Zhu:2010a,Yuan:2013a}, models of neuronal dynamics \cite{Kwok:2006a,Rubinov:2009a,Rubinov:2011a}, and non-identical Kuramoto oscillators \cite{Seliger:2002a,Gleiser:2006a,Ren:2007a, Ren:2014a,Aoki:2009a,Aoki:2011a,Gutierrez:2011a, Assenza:2011a, Avalos-Gaytan:2012a,Yuan:2011a,Nadal:2008a,Nadal:2009a,Timms:2014a,Zhang:2015a,Singh:2015a}. Focusing on the latter of these classes, one early study found that dynamical rewiring to force links between nodes with more similar time-averaged frequencies creates strongly synchronized clusters of nodes, and the network reaches a small-world configuration \cite{Gleiser:2006a}. More recently local, competitive adaptation mechanisms, which tend to strengthen (weaken) connections between more dynamically coherent (incoherent) oscillators have been shown to lead to the emergence of modular organization in Kuramoto networks \cite{Gutierrez:2011a, Assenza:2011a, Avalos-Gaytan:2012a,Yuan:2011a}, and positive feedback can simultaneously enhance synchronization and percolation in initially fragmented networks \cite{Eom:2016a}. In complementary efforts, \textcite{Nadal:2008a,Nadal:2009a} studied a growth process in which heterogeneous oscillators make connections to external pace-maker nodes so as to become locked with the pace-maker dynamics. When the attachment process is preferential and determined from differences in the dynamical states of the heterogeneous oscillators and pace-maker nodes, entrainment can occur simultaneously with the emergence of a power law degree distribution. In addition, adaptive processes that favor the strengthening of edges between more out of phase oscillators, can significantly improve global synchronization in non-identical Kuramoto networks \cite{Ren:2007a, Ren:2014a}. However, the resulting networks were not necessarily evolved under fixed total weight and the topology was not analyzed in depth for relationships between structure and dynamical properties as a function of the global coupling, both of which are aims of the present work.

Another line of inquiry revolves around the problem of optimizing synchronization of non-identical systems of Kuramoto oscillators in order to understand what structural properties of the network are important. For example, using an optimization procedure to maximize the order parameter $R$, \textcite{Brede:2008b, Brede:2009a} uncovered key features that can enhance synchronization. These include the placement of more edges on oscillators with natural frequencies further from the mean (yielding positive correlations between degrees and frequency magnitudes), as well as the preferential attachment of oscillators with positive frequencies to other oscillators with negative frequencies (and vice-versa, yielding negative correlations between the intrinsic frequencies of adjacent oscillators. These findings have been corroborated in several other studies on optimizing networks of non-identical Kuramoto oscillators as well \cite{Fan:2009a, Buzna:2009a, Carareto:2009a,Kelly:2011a}. Furthermore, forcing positive \emph{versus} negative correlations between adjacent frequencies appears to change the critical coupling and exponents for the synchronization transition \cite{Brede:2010a}. Other work has considered the enhancement of both global and local phase synchronization \cite{Brede:2008a, Freitas:2015a}, finding that local synchronization leads to an onset of collective dynamics at lower couplings -- facilitated by clustering and the grouping together of nodes with similar frequencies --  but makes the state of full synchronization more difficult to achieve. More recently, spectral analyses have been employed to show that synchronization is optimized under specific couplings of the natural frequencies to eigenvectors of the Laplacian, i.e., when the frequencies are maximally aligned with the largest eigenvector \cite{Skardal:2014a,Skardal:2015a,Skardal:2016a}. Importantly, the previously reviewed works indeed uncover several crucial network features for optimizing collective dynamics. However, the question of if or via what type of mechanism such networks could be generated or evolved for using solely local adaptive strategies - and how such a process occurs over time and at different couplings - remains open.

In this study, we numerically investigate the interplay between network structure and oscillator dynamics. We report on a local, state-dependent rewiring mechanism that can \emph{(i)} evolve initially random and uncorrelated networks towards structured configurations with specific relationships between dynamical and topological properties and \emph{(ii)} through a reciprocal process, simultaneously improve synchronization. While the ideas of dynamical self-organization from local rules and global network optimization strategies have been studied on separate fronts, here, we attempt to specifically consider them in tandem. In what follows, we first state the initial setup and parameters of the system (Sec.~\ref{s:network_construction}) and then describe the co-evolutionary process in detail (Sec.~\ref{s:adaptive_mechanism}).

\subsection{Initial network construction}
\label{s:network_construction}

All simulations were carried out with a $4^{th}$ order Runge-Kutta method using a time step of $\Delta t = 0.02$. Initial phases $\{\theta_{i}(0)\}$ were distributed at random in the interval $[-\pi,\pi]$, and the natural frequencies $\{ \omega_{i} \}$ were drawn at random from a uniform distribution in the range $[-2,2]$, denoted as $\{ \omega_{U}\}$. In Appendix~\ref{a:normal_frequencies}), we also show results for the case of a Gaussian distribution with zero mean and unit standard deviation, denoted as $\{ \omega_{G}\}$. 

The initial network configurations are binary and undirected Erd\"{o}s-Renyi (ER) random graphs of type $G(N,L)$ (i.e., the network is drawn from the distribution of random graphs with $N$ nodes and $L$ edges), with corresponding average degree $\langle k \rangle = 2L/N$. Importantly, this initialization produces networks without special topological characteristics and without relationships between network properties and dynamical properties. We will denote the initial network configurations as $\mathcal{G}_{o}$ and the networks at the end of the adaptive process as $\mathcal{G}_{\star}$, and similarly we will use `o' and `${\star}$' to denote quantities computed on each network. For the remainder of the main text, we fix $N = 100$ and examine two different values of $L$ chosen such that $\langle k \rangle = 12.5$ or $\langle k \rangle = 25$. (In the \emph{S.M.} \cite{KuramotoSupplement}, we also explore the robustness of several results with respect to variations in system size, the network density, the initial network topology, and the presence of asymmetry in the frequency distribution). Reported measures correspond to ensemble averages over independent simulations using different instantiations of the initial network, initial phases, and node frequencies.

\subsection{Mechanism of adaptive rewiring}
\label{s:adaptive_mechanism}

\begin{figure}
\centering
\includegraphics[width=\columnwidth]{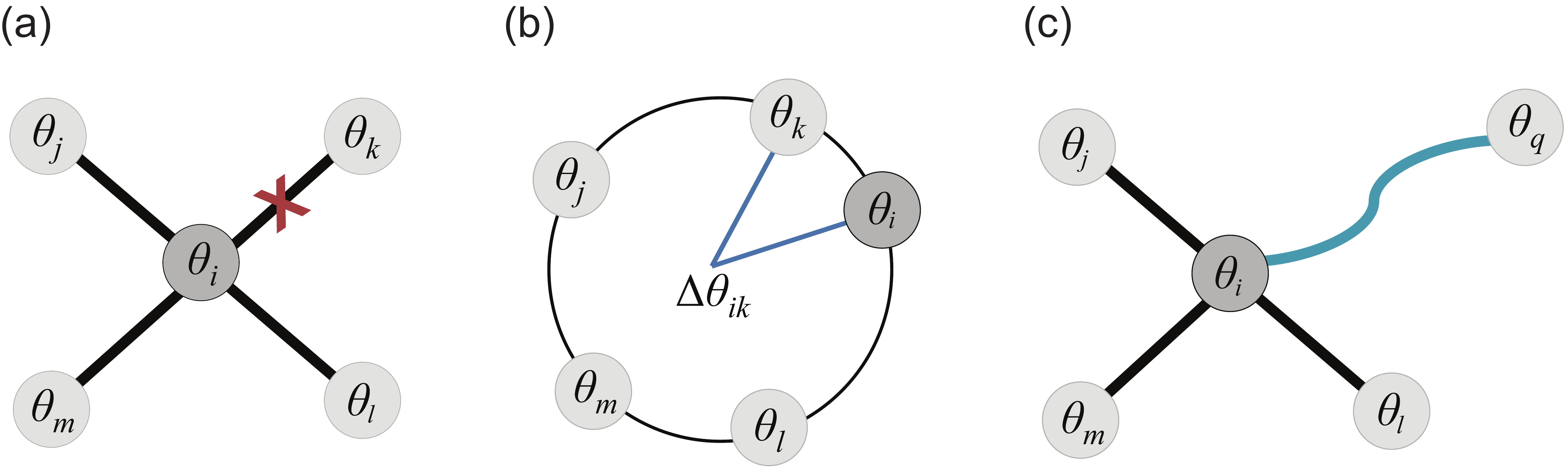}
\caption{A schematic of the network update process. \emph{(a)} At a time $t_{m}  = T_{R} + m T$, node $i$ is selected at random from all nodes in the network, and it breaks its connection to the node $k$, which \emph{(b)} is the node it is instantaneously most in phase with. \emph{(c)} A new edge $(i,q)$ is then created between $i$ and a randomly selected node $q$ with phase $\theta_{q}$}
\label{f:rewiring_schematic}
\end{figure}

We turn now to an explanation of the co-evolutionary scheme, whereby the network is restructured according to the dynamics. Motivated by previous literature \cite{Ren:2007a, Ren:2014a}, we take as our starting point a mechanism that aims to maintain connectivity between a given node and its neighbors with which it is more instantaneously out of phase, and rewire connections between a given node and its neighbor that it is most instantaneously synchronized with. After setting up the initial conditions and network $\mathcal{G}_{o}$ as described in Sec.~\ref{s:network_construction}, the dynamics are first run for a relaxation interval $T_{R}$. Then at regular times $t_{m}  = T_{R} + m T$, where $m$ is the number of attempted rewirings and $T$ is an associated interval characterizing the adaptation process (which will be some multiple of the time step), rewiring is invoked as follows. A node $i$ is chosen at random from the network and the quantity 
\begin{equation}
\label{eq:adapt_function}
f_{ij} = \frac{1}{2}[1 - \cos(\theta_{i}(t_{m}) - \theta_{j}(t_{m}))] 
\end{equation}
is computed for all $j \in \mathcal{N}_{i}$, where $\mathcal{N}_{i}$ denotes the set of nodes directly connected to $i$. Note that this function is \textit{local} in the sense that it depends only on the instantaneous phases of node $i$ and of the nodes $j \in \mathcal{N}_{i}$ that are neighbors of $i$ at the current time $t_{m}$. It takes on a value of $1$ when $\theta_{i} - \theta_{j} = \pm \pi$ (maximal phase separation) and a value of $0$ when $\theta_{i} - \theta_{j} \mod{2\pi} = 0$ (perfectly in-phase). The edge $(i,k)$, where $k$ is the node in $\mathcal{N}_{i}$ minimizing $f_{ik}$, is then broken and a new link $(i,q)$ is formed between node $i$ and a randomly selected node $q$ that was not one of $i's$ neighbors. This step corresponds to the breaking and rewiring of the link between node $i$ and the neighbor $k$ with which it is currently most in phase (Fig.~\ref{f:rewiring_schematic}). These dynamical update rules are repeated $m$ times, resulting in an evolved network $\mathcal{G}_{\star}$. We use $m = 2.5\times10^{4}$, which allows us to observe interesting dynamical and structural changes in the system, while being within enough computational reason to allow for the exploration of several different network and natural frequency parameter changes (however, we also point out places where the capability to run more adaptation steps may offer addition insight). Finally, once rewiring has ceased, the dynamics continue to run atop $\mathcal{G}_{\star}$ for another relaxation interval. For clarity, in the figures that follow, all quantities that change every \textit{time step} will be labeled as ``time, \emph{t}", and all quantities that change only when the network is \textit{rewired} will be labeled as ``rewiring step, \emph{m}".

Before continuing, we wish to point out some features of this mechanism. First, it applies to the case of binary, undirected connectivity, and maintains the density of the network throughout co-evolution. Setting these constraints disambiguates the role of rearrangements in network topology in shaping results from other factors such as increases in the total number of edges, heterogeneous weighting, or directionality. Furthermore, it allows for the cleanest comparison against much of the previous work on optimizing synchronization via rewiring. Second, we note that this adaptive process is a type of repulsive or suppressive strategy, but unlike \cite{Zhou:2006a,Ren:2007a, Aoki:2009a,Zhu:2010a, Aoki:2011a, Yuan:2013a,Ren:2014a}, which consider weight plasticity, or \cite{Nadal:2008a,Nadal:2009a} which consider a growth mechanism, we consider connectivity reorganization where edges in an initially random configuration are periodically pruned and rewired between the most instantaneously and locally in-phase oscillators, but remain in place between the more locally dissonant oscillators. The random rewiring after edge deletion introduces realistic stochasticity and allows for a sampling of the network, without breaking the locality condition for determining which edge is rewired \cite{Gross:2008a}. Though not a model of a specific system, this type of disassortative mechanism has real-world counterparts. For example, it is in the same vein as anti-Hebbian learning rules in neural systems, where synapses weaken between more dynamically correlated neurons and strengthen between more incoherent neurons \cite{Lamsa:2007a,Roberts:2010a}. It may also mimic some strategies of opinion formation and influence on social networks, where people preferentially link with those of more different opinions from themselves \cite{Kimura:2008a}. 

\section{Results}
\label{s:results}

\subsection{Co-evolved networks exhibit enhanced global synchronization}
\label{s:enhanced_synchronization_uniform}

\begin{figure}
\centering
\includegraphics[width=\columnwidth]{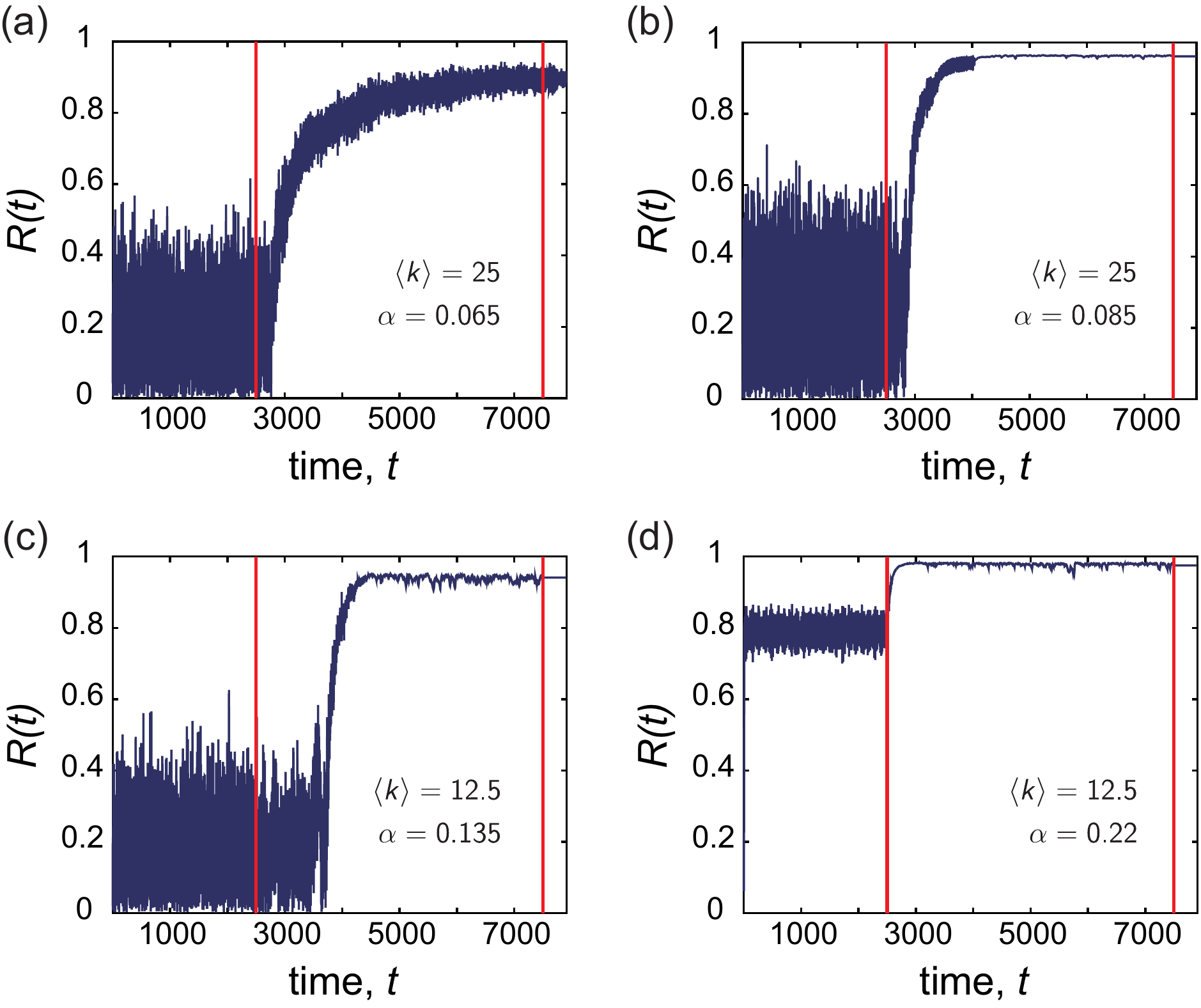}
\caption{Examples of the global order parameter $R(t)$ \emph{vs.} time $t$ for various representative couplings $\alpha$. In each case, the dynamics were first run atop an initially ER random graph with average degree $\langle k \rangle$, after which co-evolution of the network and dynamics took place between the two red lines. The natural frequencies were drawn from the uniform distribution $\{ \omega_{U} \}$, and the mean degree $\langle k \rangle$ and coupling $\alpha$ used for each panel were \emph{(a)} $\langle k \rangle = 25$, $\alpha = 0.065$, \emph{(b)} $\langle k \rangle = 25$, $\alpha = 0.085$, \emph{(c)} $\langle k \rangle  = 12.5$, $\alpha = 0.135$, \emph{(d)} $\langle k \rangle = 12.5$, $\alpha = 0.22$. During the adaptation period, the network was continually rewired once every $T = 0.2$ time units. The co-evolving networks exhibit enhanced collective dynamics, as observed by increases in the global order parameter.}
\label{f:R_vs_T_Uniform}
\end{figure}

Our first main result is that the adaptive rewiring mechanism is able to enhance global synchronization over a broad range of coupling values. This result holds for both frequency distributions (See Appendix~\ref{a:normal_frequencies} for the analysis with the normally distributed frequencies) and values of average degree. In addition, the result is quite robust to an order of magnitude difference in the structural reconfiguration interval, $T$. 

We first discuss the time evolution of the order parameter $R(t)$ for the co-evolving networks. Fig.~\ref{f:R_vs_T_Uniform} depicts examples of $R(t)$ \emph{vs.} time $t$ for the case of uniformly distributed frequencies $\{ \omega_{U} \}$ and $T = 0.2$. The top row corresponds to networks with $\langle k \rangle = 25$, at representative couplings \emph{(a)} $\alpha = 0.065$ and \emph{(b)} $\alpha = 0.085$, and the bottom row corresponds to networks with $\langle k \rangle = 12.5$, at representative couplings \emph{(c)} $\alpha = 0.135$ and \emph{(d)} $\alpha = 0.22$. In each case, the dynamics are first run atop the initially static ER network for several time steps. Adaptation begins at the time denoted by the first red line, and ends after several time steps at the second red line. We observe that at lower coupling values - where the dynamics on the initial network exhibit little coherence across time - the adaptive strategy is able to significantly increase $R$ to an intermediate value during the rewiring stage, though the order parameter may still exhibit fluctuations. As $\alpha$ is increased, the order parameter in the non-adaptive regime sits at an intermediate average value, but once co-evolution begins, the self-organizing network rearranges such that $R$ again increases and reaches a value near 1. When rewiring ceases after several cycles, the resulting networks are able to maintain these states of heightened collective dynamics, and the global order parameter remains at an increased value from its initial location. However, though the time-averaged value $\langle R \rangle$ remains high for the networks with $\langle k \rangle = 12.5$, we find that in some cases the order parameter still exhibits fluctuations, even after the adaptation period has ceased. This intuitively suggests that the co-evolved networks with higher average degree are more robust to the stochasticity in the rewiring and the exact placement of edges in the network, in terms of their ability to support a smooth, frequency-synchronized steady-state. Appendix~\ref{a:R_vs_time_alpha} contains additional figures of $R(t)$ \emph{vs.} time for the case of normally distributed frequencies. Also, in the \emph{S.M.}, we examine another measure of synchronization that quantifies how the number of \textit{locally-synchronized clusters} \cite{Arenas:2006a} of oscillators changes as a function of time due to the network rearranging \cite{KuramotoSupplement}.

In order to obtain a more complete picture of the effect of adaptive rewiring, we performed a sweep over a comprehensive coupling range. For $\langle k \rangle = 12.5$ we considered a range $\alpha \in [0, 0.4]$, and for $\langle k \rangle = 25$, we considered $\alpha \in [0, 0.2]$; in both cases, couplings were sampled at a resolution $\Delta \alpha = 0.005$. At each value of $\alpha$, networks and initial conditions were initialized as described in Sec.~\ref{s:network_construction}. We then ran a set of simulations on the original, uncorrelated networks $\mathcal{G}_{o}$, and obtained a time-averaged value of the global order parameter $\langle R \rangle$ (Eq.~\ref{eq:R_time_avg}) over the last $1\times10^{4}$ time steps. A second set of simulations were then run with the same initial conditions, but under the co-evolutionary scheme (i.e. the network topology was allowed to co-evolve with the dynamics for $m$ rewiring steps, after which a time-averaged order parameter was computed on the final adapted network $\mathcal{G}_{\star}$). 

\begin{figure}[h]
\centering
\includegraphics[width=\columnwidth]{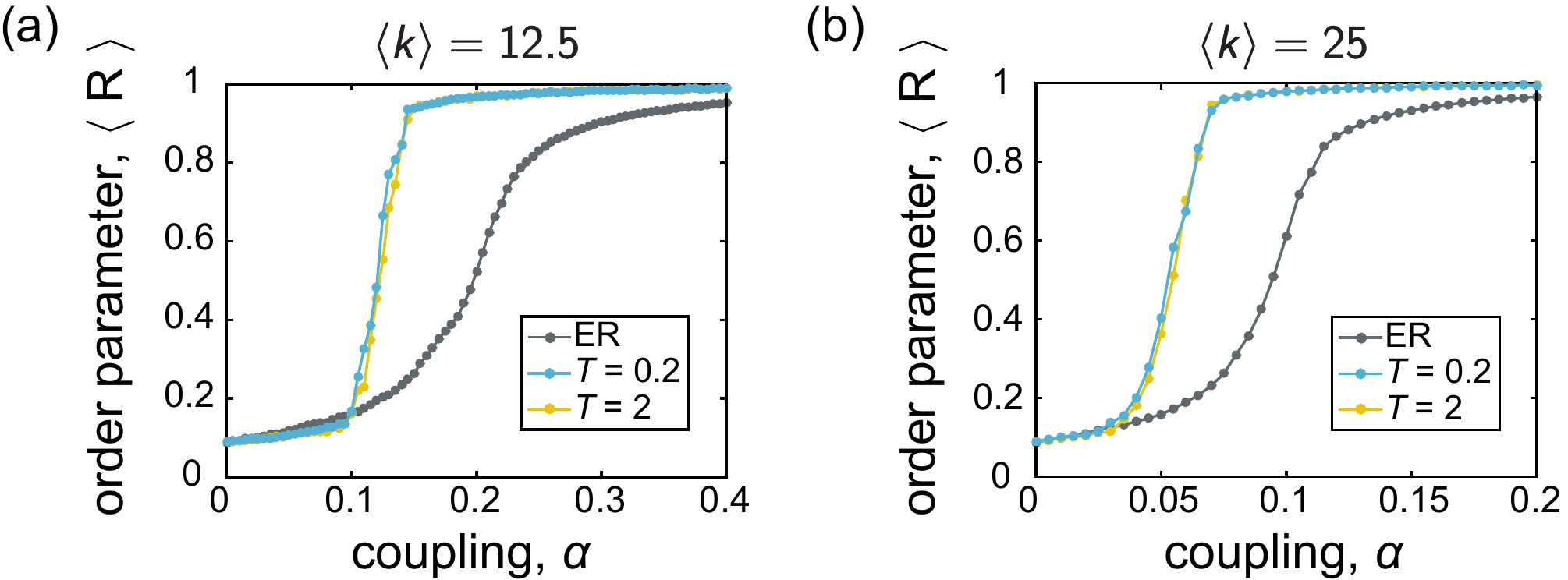}
\caption{The time-averaged order parameter $\langle R \rangle$ \emph{vs.} coupling $\alpha$. In each panel, the gray data points correspond to static ER random graphs $\mathcal{G}_{o}$, and the blue and yellow points correspond to the adapted networks $\mathcal{G}_{\star}$ evolved under rewiring time scales of $T = 0.2$ and $T = 2$, respectively.  The frequencies were drawn from the uniform distribution $\{\omega_{U}\}$, and the mean degree of the networks are \emph{(a)} $\langle k \rangle = 12.5$, and \emph{(b)} $\langle k \rangle = 25$. All curves depict averages over 25 instantiations, and the lines between data points serve as guides for the eye.}
\label{f:R_vs_alpha_Uniform}
\end{figure}

We show the outcome of this analysis in Fig.~\ref{f:R_vs_alpha_Uniform}, where each panel is a plot of $\langle R \rangle$ \emph{vs.} $\alpha$. Panels \emph{(a)} and \emph{(b)} correspond to networks with $\langle k \rangle = 12.5$ and $\langle k \rangle = 25$, respectively. Furthermore, we plot curves for two different values of the waiting time between structural changes to the network: $T = 0.2$ and $T = 2$. As can be seen from each of the curves, there is a broad range of $\alpha$ over which the rewiring mechanism for network restructuring leads to improvements (higher values) of the time-averaged global order parameter compared to the random networks. This enhancement does not begin immediately at $\alpha = 0$, but once the coupling is high enough (still at a relatively low value), the order parameter for the adaptive networks $\mathcal{G}_{\star}$ begins to increase at a much steeper rate. Once this begins, we find that $\langle R \rangle$ remains noticeably higher for the adaptive networks compared to the static networks $\mathcal{G}_{o}$, across all couplings beyond a certain point. These trends are robust for both rewiring intervals and average degree values, and as can been seen in Sec.~\ref{a:R_vs_time_alpha}, the results hold for the case of normally distributed natural frequencies as well. (The \emph{S.M.} \cite{KuramotoSupplement} shows additional and qualitatively similar findings for simulations on slightly larger networks or those with lower mean degree, and for the case of a non-symmetric frequency distribution). It is also worth noting that -- although the total number of times the network is allowed to rewire is limited by computational constraints -- since there is no built-in condition for adaptation to cease and since there is a stochastic component to the network reconfiguration, the ability to run more adaptation steps may yield even further improved results.

\subsection{Emerging structure and correlations between network topology and dynamical properties}
\label{s:emerging_correlations}

Given that dynamical reconfiguration of the network - informed by local information on the states of connected oscillators - can improve the overall amount of synchronization in the system, a second line of inquiry is understanding what properties of the evolved networks lend themselves to this capability. In particular, does the self-enacted rewiring mechanism generate the interesting topological features and correlations that arise with global optimization schemes, which in a cyclic process, then allow synchronization to occur? 

We begin by investigating the networks $\mathcal{G}_{\star}$ for certain relationships between their topology and the natural frequencies of the oscillators. Two properties in particular -- degree-frequency correlations and frequency-neighbor frequency correlations -- have been associated with networks optimized for synchronization \cite{Brede:2008b,Brede:2009a,Skardal:2014a} (see Sec.~\ref{s:intro} and Sec.~\ref{s:motivation_and_methods}). 

\begin{figure*}
\centering
\includegraphics[width=0.85\textwidth]{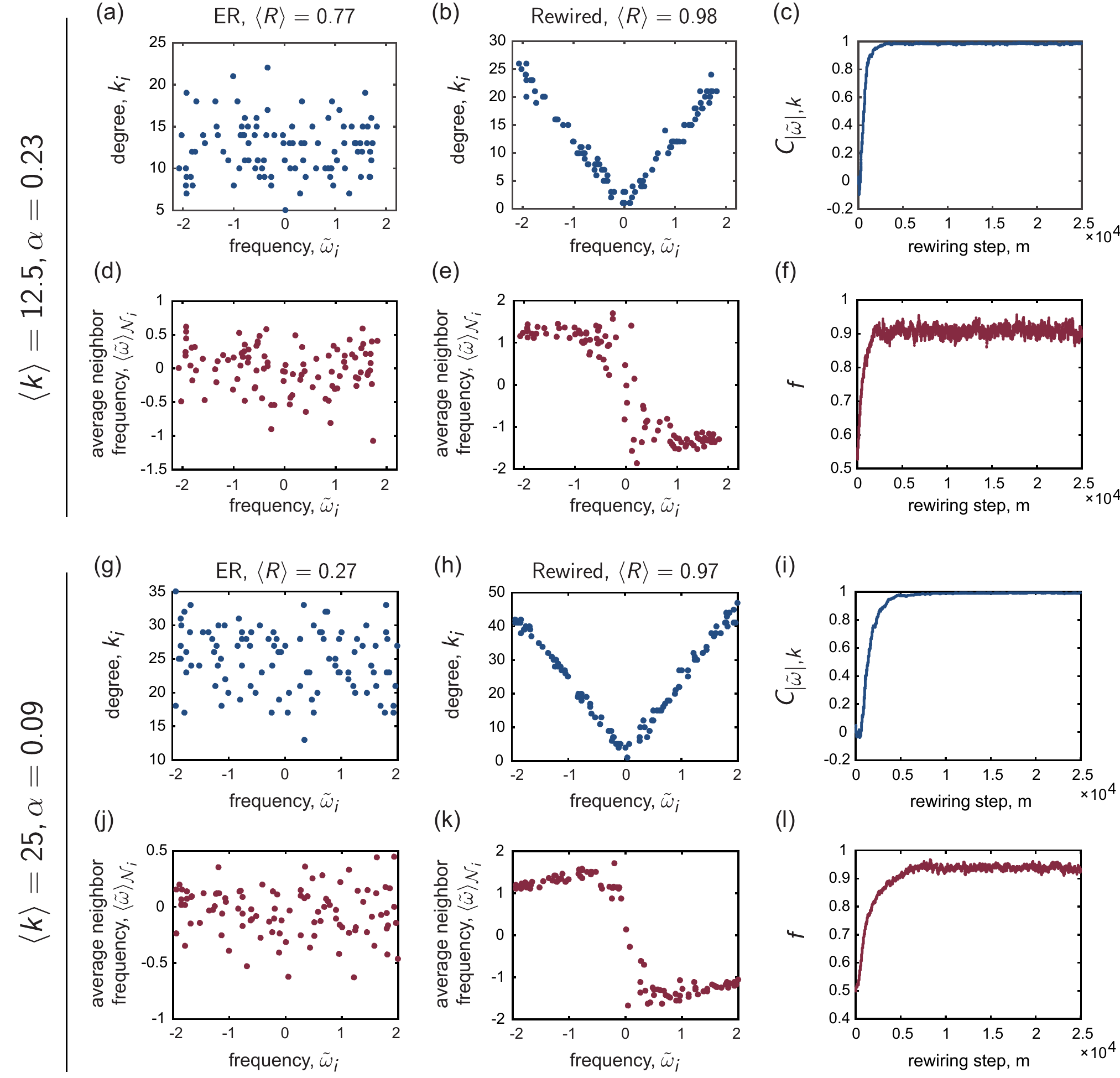}
\caption{Relationships between the network structure and the intrinsic frequencies of the oscillators. The top two rows show examples for a network with $\langle k \rangle = 12.5$, and the bottom two rows show examples for a network with $\langle k \rangle = 25$; in both cases, the frequencies were drawn from the uniform distribution $\{\omega_{U}\}$. For each network density, the first column corresponds to an ER random graph that exhibits only intermediate levels of synchrony at the displayed coupling $\alpha$ (as measured by $\langle R \rangle$), and the second column corresponds to the adapted network, which exhibits a higher level of synchrony. These plots highlight key relationships that emerge from the co-evolutionary network update rule. \emph{(a,b); (g,h)} Node degree $k_{i}$ \emph{vs.} frequency offset $\tilde{\omega}_{i}$. \emph{(d,e); (j,k)} Average neighbor frequency offset $\langle \tilde{\omega} \rangle_{\mathcal{N}_{i}}$ \emph{vs.} frequency offset $\tilde{\omega}_{i}$. \emph{(c); (i)} The correlation $C_{|\tilde{\omega}|, k}$ between node degree $k_{i}$ and the magnitude of the frequency offset $|\tilde{\omega}_{i}|$ \emph{vs.} the number of rewiring steps $m$, and \emph{(f); (l)} the mean fraction $f$ (i.e. averaged over all nodes in the network) of an oscillator's neighbors that have frequency offsets of opposite sign compared to that of the central oscillator \emph{vs.} the number of rewiring steps $m$.}
\label{f:correlations_uniform}
\end{figure*}

Though the natural frequencies of each node remain fixed here, the way oscillators of different frequencies are coupled to one another changes over time from the initial, random configuration to the end of adaptation. In Fig.~\ref{f:correlations_uniform}, we show examples of how two specific relationships manifest in an ER network $\mathcal{G}_{o}$ and the corresponding adaptively rewired network $\mathcal{G}_{\star}$. The top and bottom two rows of Fig.~\ref{f:correlations_uniform} correspond to networks with $\langle k \rangle = 12.5$ and $\langle k \rangle = 25$, respectively, and in both cases, the first column corresponds to the ER network and the second column corresponds to the rewired network. For each node $i$, we first computed the offset $\tilde{\omega}_{i}$ of $i$'s intrinsic frequency from the mean of the population, where $\tilde{\omega}_{i} = \omega_{i} - \langle \omega \rangle$ and $\langle \omega \rangle$ is the mean intrinsic frequency over all nodes in the system. Panels \emph{(a,b)} and \emph{(g,h)} then show node degree $k_{i}$ \emph{vs.} frequency offset $\tilde{\omega}_{i}$, and panels \emph{(d,e)} and \emph{(j,k)} show the average frequency offset of oscillator $i's$ neighbors, $\langle \tilde{\omega} \rangle_{\mathcal{N}_{i}} = \sum_{j\in\mathcal{N}_{i}}\tilde{\omega_{j}}/k_{i}$, \emph{vs.} frequency offset $\tilde{\omega}_{i}$. Note that the coupling values $\alpha$ are such that the original networks exhibited intermediate levels of synchrony, and the reorganized networks were able to entrain the population to a higher level of synchrony.

As expected, initially there is little correlation present between the topology of the network and the dynamical property of oscillator frequency, as illustrated by the lack of organization in the plots in the first column of Fig.~\ref{f:correlations_uniform}. However, from observation of the second column, it is evident that when the co-evolutionary mechanism can enhance synchronization, it simultaneously leads to the emergence of very specific correlations between the network connectivity and the intrinsic frequencies of the oscillators. We first note the appearance of the marked \textit{v}-shaped curves characterizing the plots of $k_{i}$ \emph{vs.} $\tilde{\omega}_{i}$ (panels \emph{(b)} and \emph{(h)}), which signify that the node degree becomes positively correlated with the absolute value of the oscillator frequency offset. In other words, nodes with natural frequencies further from the mean natural frequency of the population gather proportionately more edges. 

A second finding is that when increased global synchrony arises from a restructuring of the network, the final arrangement exhibits distinct relationships between the frequency offsets of a given oscillator and the frequency offsets of that central oscillator's direct neighbors on the network. The patterns in panel \emph{(e)} and \emph{(k)} showing $\langle \tilde{\omega} \rangle_{\mathcal{N}_{i}}$ \emph{vs.} $\tilde{\omega}_{i}$ point to the fact that oscillators with positive natural frequency offsets tend to become connected to other oscillators with, on average, negative natural frequency offsets. Since the mean frequency $\langle \omega \rangle \approx 0$ for the distributions we consider, this implies that oscillators with positive frequencies tend to become neighbored by oscillators with negative intrinsic frequencies, and vice-versa. In Sec.~\ref{a:freq_totalFreq} we also define an additional measure of frequency-neighbor frequency organization as the correlation $C_{\tilde{\omega},\sum\tilde{\omega}}$ between natural frequency offsets and the \textit{sum} of neighbor frequency offsets. Fig.~\ref{f:freq_sumFreq_corr_uniform} shows examples of this quantity for the same networks as those in Fig.~\ref{f:correlations_uniform}.

To quantify these relationships and study how they evolve with the number of rewiring steps, we considered two summary statistics, following \cite{Brede:2008a,Brede:2008b,Brede:2009a,Brede:2010a}. The first of these measures is a simple (Pearson) correlation coefficient, $C_{|\tilde{\omega}|, k}$, to quantify the strength of the relationship between node degree and the magnitude of frequency offset. This measure increases steadily throughout co-evolution of the network and dynamics (Fig.~\ref{f:correlations_uniform}\emph{c},\emph{i}). In addition, for each node $i$, we calculated the fraction of its neighbors $f_{i}$ that had natural frequency offsets of the opposite sign as compared to oscillator $i's$ frequency offset, and then computed an average $f = \sum_{i}f_{i}/N$ over all nodes in the network. This metric also increases as the network is rewired, as observed in Fig.~\ref{f:correlations_uniform}\emph{f},\emph{l}. (See Fig.~\ref{f:freq_sumFreq_corr_uniform} for an example of the evolution of the additional measure,$C_{\tilde{\omega},\sum\tilde{\omega}}$).

\begin{figure}[h!]
\centering
\includegraphics[width=\columnwidth]{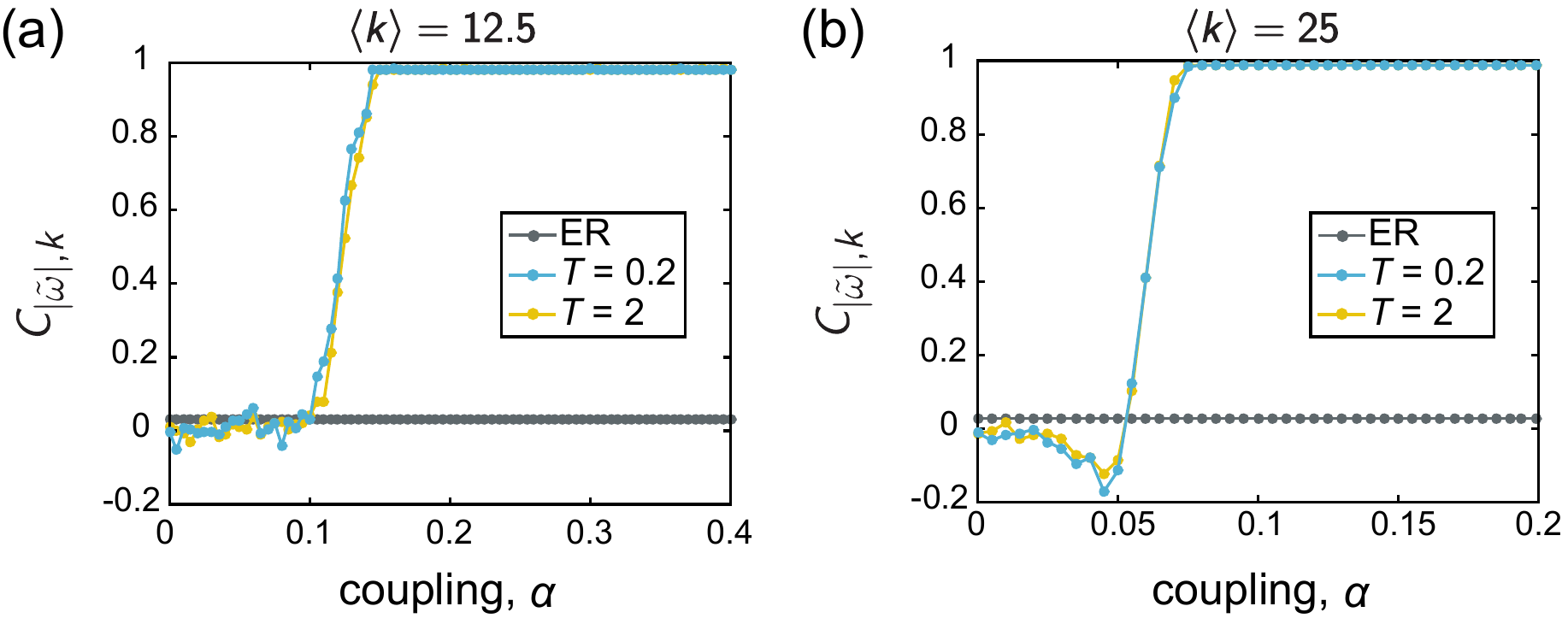}
\caption{These plots depict the correlation $C_{|\tilde{\omega}|, k}$ between node degree $k_{i}$ and the magnitude of the frequency offset $|\tilde{\omega}_{i}|$, as a function of the coupling. In each panel, the gray data points correspond to the initial, uncorrelated ER random graphs $\mathcal{G}_{o}$, and the blue and yellow points correspond to the adapted networks $G_{\star}$ evolved under rewiring time scales of $T = 0.2$ and $T = 2$, respectively. The frequencies were drawn from the uniform distribution $\{\omega_{U}\}$, and the mean degree of the networks are \emph{(a)} $\langle k \rangle = 12.5$, and \emph{(b)} $\langle k \rangle = 25$. The dip observed in \emph{(b)} at $\alpha \approx 0.5$ is due to the localization of edges on a cluster of oscillators with natural frequencies near the mean; this is examined further in Sec.~\ref{s:time_dependence_uniform}. All curves depict averages over 25 instantiations, and the lines between data points serve as guides for the eye.}
\label{f:degree_freq_corr_uniform}
\end{figure}

\begin{figure}[h!]
\centering
\includegraphics[width=\columnwidth]{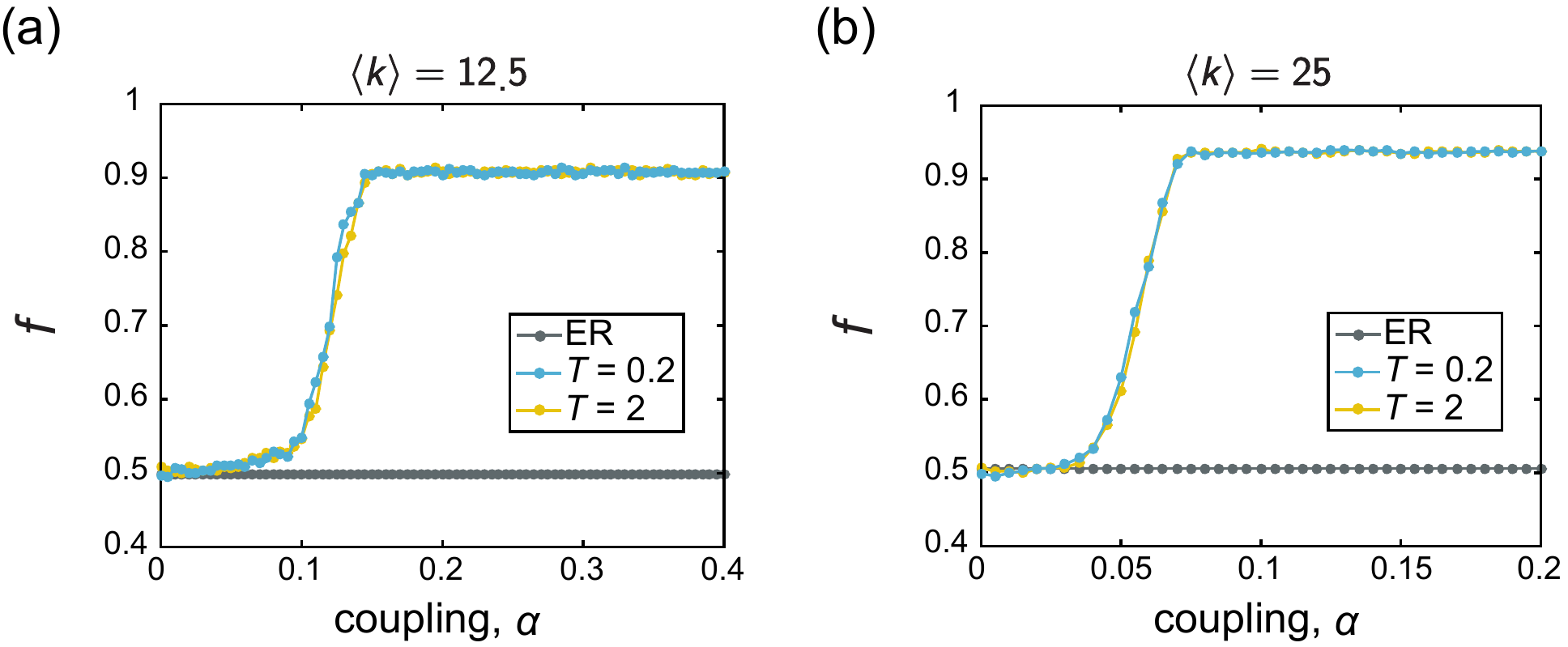}
\caption{These plots depict the mean fraction $f$ (averaged over all nodes in the network) of an oscillator's neighbors that have frequency offsets of opposite sign compared to that of the central oscillator, as a function of the coupling. In each panel, the gray data points correspond to the initial, uncorrelated ER random graphs $\mathcal{G}_{o}$, and the blue and yellow points correspond to the adapted networks $G_{\star}$ evolved under rewiring time scales of $T = 0.2$ and $T = 2$, respectively. The frequencies were drawn from the uniform distribution $\{\omega_{U}\}$, and the mean degree of the networks are \emph{(a)} $\langle k \rangle = 12.5$, and \emph{(b)} $\langle k \rangle = 25$. All curves depict averages over 25 instantiations, and the lines between data points serve as guides for the eye.}
\label{f:freq_neighbFreq_uniform}
\end{figure}

We have thus far shown examples of emerging structural patterns that arise when a co-evolved network is clearly able to entrain the oscillators to a state of higher synchrony. However, the ability of this behavior to occur is also dependent on the coupling $\alpha$. In order to better understand the appearance of the topological and dynamical correlations and their dependence on the overall coupling, we computed the same measures ($C_{|\tilde{\omega}|, k}$ and $f$) as a function of $\alpha$ (see Figs.~\ref{f:degree_freq_corr_uniform}, \ref{f:freq_neighbFreq_uniform}, respectively). In each case, the measures were computed on the final networks $\mathcal{G}_{\star}$ that exist at the end of the adaptation period. 

For each combination of natural frequency distribution, average degree, and rewiring interval, we observe robust trends. In particular, $C_{|\tilde{\omega}|, k}$ remains near zero at low coupling values, and then proceeds to quickly increase as a function of $\alpha$ until it saturates to a relatively steady value close to 1. In other words, as the overall coupling increases, the correlation between the node degrees and the magnitude of the frequency offsets becomes more apparent. The frequency-neighbor frequency relationship follows a similar evolution. Specifically, the mean fraction $f$ (averaged over all nodes in the network) of an oscillator's neighbors that have frequency offsets of opposite sign relative to that of the central oscillator also grows with $\alpha$, plateauing near a high value of $f \approx 0.9$. This points to a heightened mixing of oscillators with different intrinsic properties that arises with increased coupling. We also refer the reader to Fig.~\ref{f:freq_sumFreq_vs_alpha_uniform} for the analysis of the supplementary quantity, $C_{\tilde{\omega},\sum\tilde{\omega}}$, which supports and further aids in the understanding of these findings.

There are three main features in these trends worth pointing out explicitly. First, for the rewiring intervals considered here, the general form of the curves appears to be relatively independent of the time $T$ between a structural perturbation to the network. Second, the two quantities considered, $C_{|\tilde{\omega}|,k}$ and $f$, both exhibit a clear change in behavior (signified by the onset of a rapid increase in value) at approximately the same coupling. Thus, in the proposed co-evolutionary scheme, the emergence of these relationships between the network topology and intrinsic properties of the dynamics seem to arise in tandem to one another, in that the appearance of one feature implies the development of the other. A final important observation is that the coupling at which these patterns begin to take shape is near the coupling at which the global order parameter begins to rise (compare to Fig.~\ref{f:R_vs_alpha_Uniform}). (Fig.~\ref{f:freq_sumFreq_vs_alpha_uniform} shows that consistent behavior is found for $C_{\tilde{\omega},\sum\tilde{\omega}}$ as well). Thus, as found by \textcite{Brede:2008a,Brede:2008b,Brede:2009a,Brede:2010a} in his work on optimization of synchronization of non-identical oscillators, enhanced collective behavior co-occurs with the materialization of specific relationships between network connectivity and the intrinsic frequencies. Here, we have shown that organized structure in the form of these correlations can emerge in networks from a simple, adaptive mechanism based on a combination of local state information and stochastic rewiring. We also again note that allowing the co-evolution to run longer may give heightened results, especially at low to intermediate couplings. 

\subsection{Analysis of the time-dependence of the adaptive mechanism}
\label{s:time_dependence_uniform}

Our analysis thus far has considered the global order parameter and the correlations between the intrinsic frequencies and the network topology that arise in the adaptively rewired networks, and we have mainly focused on how these properties manifest after several rewirings of the network and how their strength depends on the overall coupling. While Figs.~\ref{f:R_vs_T_Uniform} and \ref{f:correlations_uniform} do briefly examine the time-dependence of these properties as well, in order to gain a further understanding of the adaptation mechanism, we carry out a more in-depth exploration of how the co-evolutionary process unfolds over time at different couplings, and also how individual oscillators with different intrinsic dynamical properties (i.e. intrinsic frequencies) are affected by the rewiring mechanism.

\subsubsection{Evolution of the instantaneous frequencies}

\begin{figure}
\centering
\includegraphics[width=\columnwidth]{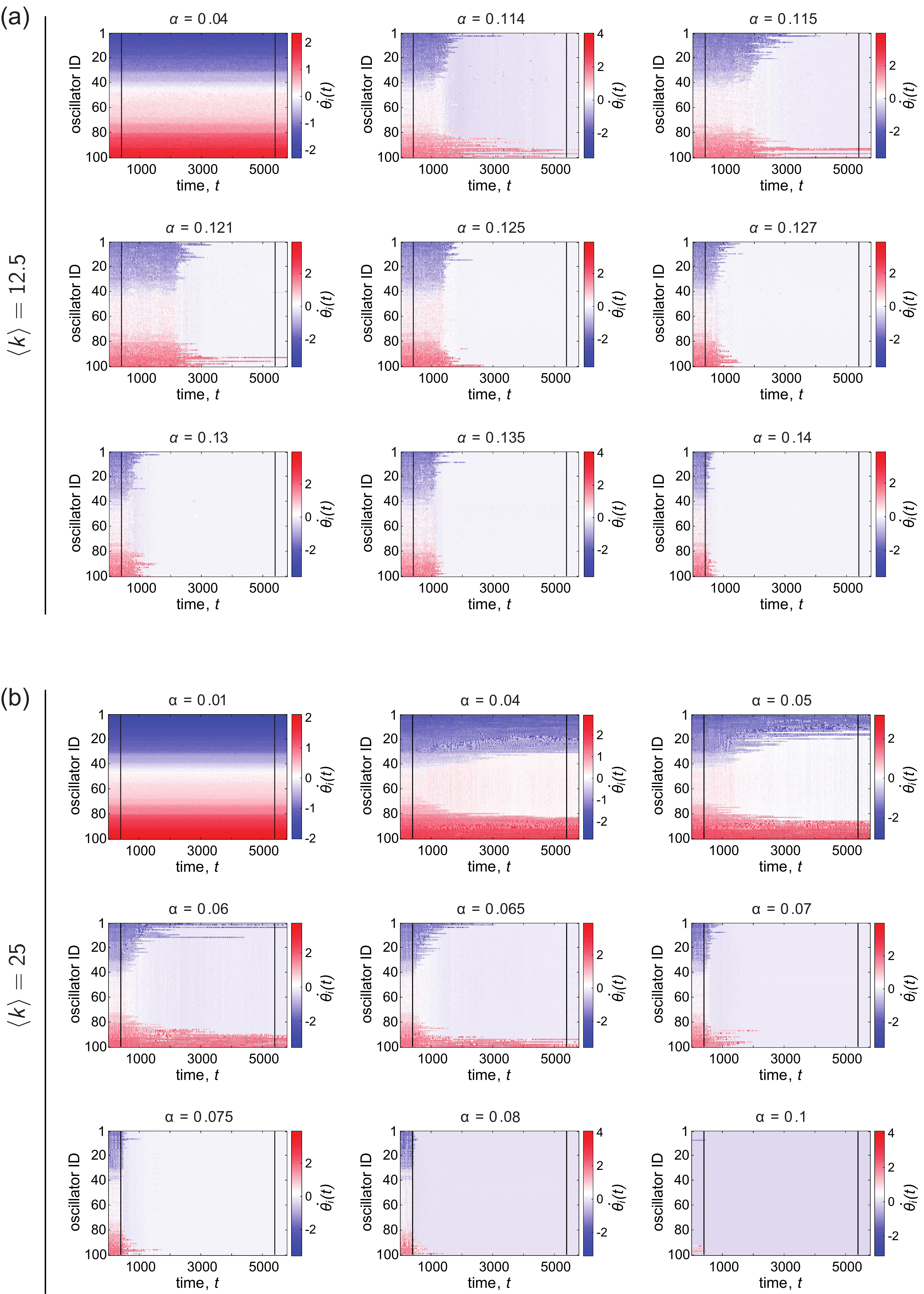}
\caption{Examples of the instantaneous frequencies $\dot{\theta}_{i}(t)$ \emph{vs.} time $t$, for various representative couplings $\alpha$. The mean degree of the networks are \emph{(a)} $\langle k \rangle = 12.5$ and \emph{(b)} $\langle k \rangle = 25$, and the natural frequencies $\{\omega_{U}\}$ were drawn from the uniform distribution. In all panels, each row corresponds to one oscillator, and the rows are ordered by the quantity $\tilde{\omega}_{i} = \omega_{i} - \langle \omega \rangle$ (i.e. the offset from the mean intrinsic frequency of the population). For each coupling, the dynamics were first run atop an initially ER random graph, after which co-evolution of the network and dynamics took place between the two black lines. During the adaptation period, the network was continually rewired once every $T = 0.2$ time units.}
\label{f:freq_vs_time_allCouplingSubPlot_N100_T10_typeOmegauniform_spreadOmega2_kavg25_trial1_kavg125_trial1}
\end{figure}

To investigate the temporal evolution of the system - and how the adaptive scheme works from a more local standpoint - we first examine the instantaneous frequencies $\dot{\theta_{i}}(t)$ as a function of time. Recall that the condition for a frequency-synchronized state corresponds to the instantaneous frequencies of all oscillators locking to the mean natural frequency of the population, $\langle \omega \rangle$. Thus, to better understand the mechanism of the co-evolution process, we specifically consider how oscillators of different \textit{intrinsic} frequencies (in terms of how close or far $\omega_{i}$ is to the average natural frequency $\langle \omega \rangle$) evolve as a function of time, and how they may be differentially affected by the time-dependent rewiring of the network. Fig.~\ref{f:freq_vs_time_allCouplingSubPlot_N100_T10_typeOmegauniform_spreadOmega2_kavg25_trial1_kavg125_trial1} shows examples of $\dot{\theta_{i}}(t)$ \emph{vs.} $t$ for several values of the coupling $\alpha$ around the point in which the dynamics transition to a synchronized state. The top set of panels \emph{(a)} correspond to a network with $\langle k \rangle = 12.5$, and the bottom set of panels \emph{(b)} correspond to a network with $\langle k \rangle = 25$; the frequencies are uniformly distributed and the same in both cases (Appendix~\ref{s:time_dependence_normal} contains additional figures for examples with normally distributed frequencies). Each row corresponds to one oscillator, and the rows are ordered by the quantity $\tilde{\omega}_{i} = \omega_{i} - \langle \omega \rangle$ (i.e. the offset from the mean natural frequency of the population). Adaptation of the network sets in at the time denoted by the first black line and ends at the second black line.  

At very low coupling, adaptively rewiring the network appears to have negligible effect on the instantaneous frequencies. But as the coupling increases to an intermediate value near the transition to synchronization (see Fig.~\ref{f:R_vs_alpha_Uniform}), we observe a change in the dynamics: oscillators with intrinsic frequencies near the center of the distribution start evolving with a frequency near the mean earliest, and then later in time, more outlying oscillators become incorporated into the coherent group. For the denser networks in particular, there are a few sampled coupling values for which the co-evolution results in partial synchronization of those oscillators around the center of the natural frequency distribution, but that coherent core does not extend to the most disparate oscillators. Thus, there does seem to be a dependence on the natural frequencies in terms of how the co-evolution develops and affects different oscillators over time. However, at even larger coupling, the transition in the dynamics becomes much faster, and the dependence on the intrinsic frequencies is less noticeable. The findings are similar for the case of normally distributed frequencies (Appendix~\ref{s:time_dependence_normal}, Fig.~\ref{f:freq_vs_time_allCouplingSubPlot_N100_T10_typeOmeganormal_spreadOmega1_kavg25_trial8_kavg125_trial8})

Before continuing, we note that the figures shown here (and throughout the rest of Sec.~\ref{s:time_dependence_uniform}) are for single realizations of the initial network topology and the intrinsic frequencies. For these particular instantiations, we have sampled coupling values that highlight different regimes -- in regards to the behavior over time and the overall outcome -- of the adaptive rewiring. However, it is important to state that over different realizations, we observe some variability in terms of how the co-evolution affects the dynamics and the network over time, and whether or not it is able to cause significant changes in the dynamics and the network structure at a given coupling. This is especially apparent for the sparser networks and low values of $\alpha$. Though it is beyond the scope of the current work, it may be interesting in future work to investigate the degree of this variability and its dependence on the frequency distribution and properties of the network such as density. 

\subsubsection{Evolution of the correlations between topology and natural frequencies}

We know from Sec.~\ref{s:emerging_correlations} that increases in the order parameter due to the adaptive mechanism are accompanied by the emergence of correlations between the network topology and the oscillator frequencies. Therefore, we now further examine how the network organization restructures over time as the system co-evolves. We focus in particular on the evolution of the degree of individual nodes across time, and also consider how the degree - natural frequency relationship proceeds as the network rearranges itself at different values of the global coupling.

\begin{figure}
\centering
\includegraphics[width=\columnwidth]{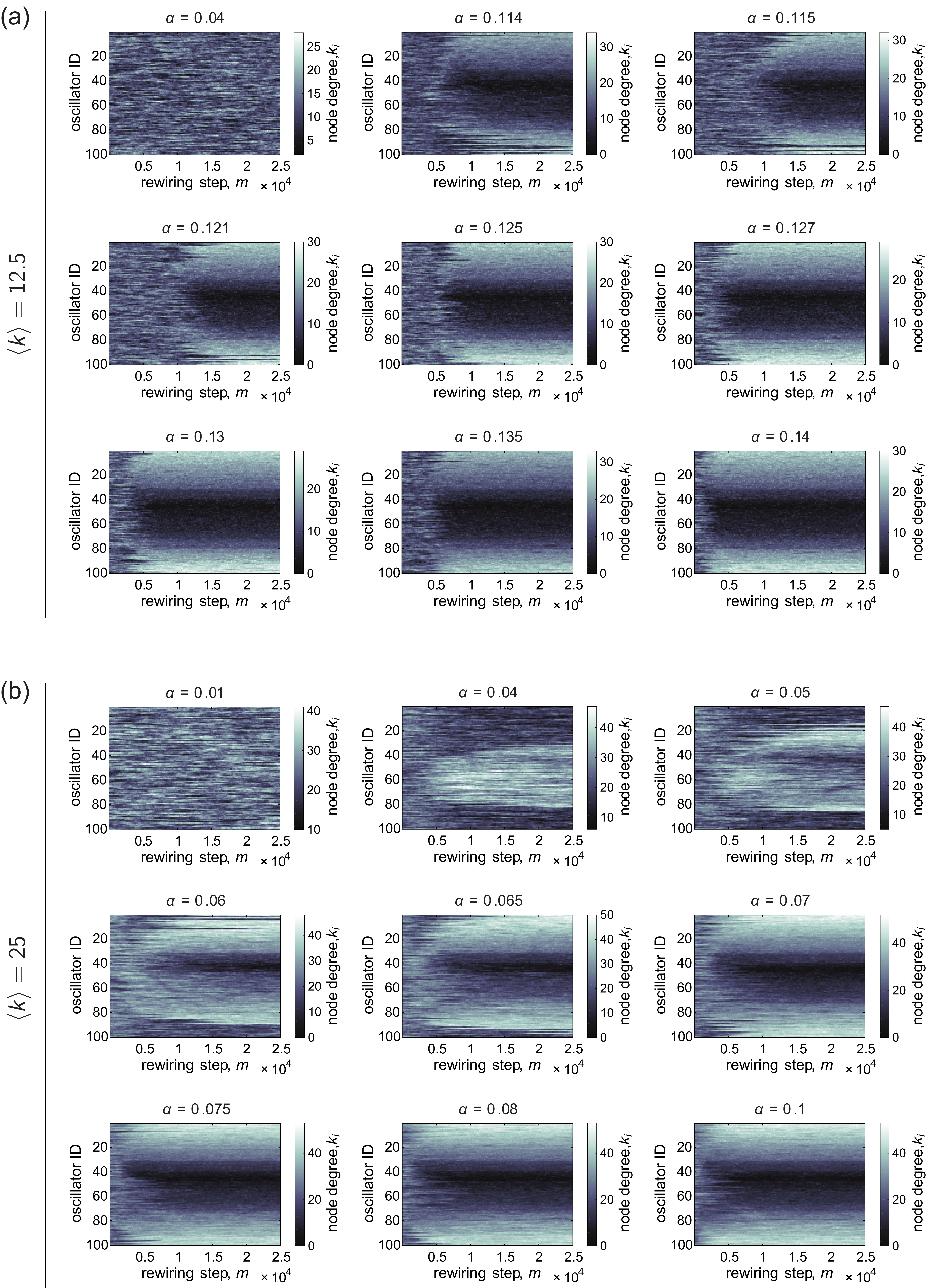}
\caption{Examples of the evolution of the node degree $k_{i}$ \emph{vs.} the rewiring step $m$, for various representative couplings $\alpha$. The mean degree of the networks are \emph{(a)} $\langle k \rangle = 12.5$ and \emph{(b)} $\langle k \rangle = 25$, and the natural frequencies $\{\omega_{U}\}$ were drawn from the uniform distribution. In all panels, each row corresponds to one oscillator, and the rows are ordered by the quantity $\tilde{\omega}_{i} = \omega_{i} - \langle \omega \rangle$ (i.e. the offset from the mean intrinsic frequency of the population). The network was continually rewired once every $T = 0.2$ time units.}
\label{f:degree_vs_time_allCouplingSubPlot_N100_T10_typeOmegauniform_spreadOmega2_kavg25_trial1_kavg125_trial1}
\end{figure}

\begin{figure}
\centering
\includegraphics[width=\columnwidth]{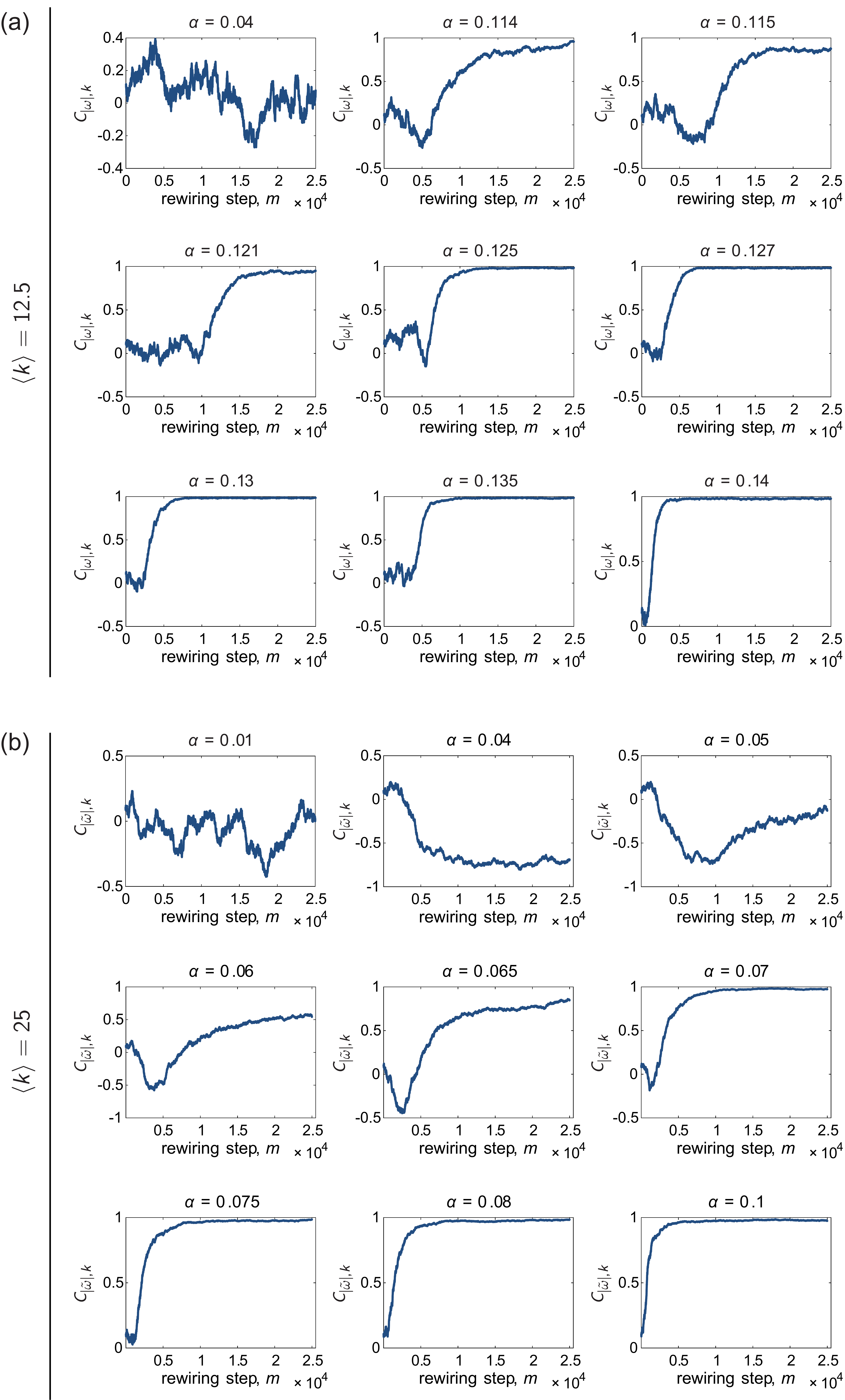}
\caption{Examples of the evolution of the correlation $C_{|\tilde{\omega}|, k}$ between oscillator frequency offset $\tilde{\omega}_{i}$ and the node degree $k_{i}$,\emph{vs.} the rewiring step $m$, for various representative couplings $\alpha$. The mean degree of the networks are \emph{(a)} $\langle k \rangle = 12.5$ and \emph{(b)} $\langle k \rangle = 25$, and the natural frequencies $\{\omega_{U}\}$ were drawn from the uniform distribution. In all panels, each row corresponds to one oscillator, and the rows are ordered by the quantity $\tilde{\omega}_{i} = \omega_{i} - \langle \omega \rangle$ (i.e. the offset from the mean intrinsic frequency of the population). The network was continually rewired once every $T = 0.2$ time units.}
\label{f:degree_freq_corr_vs_time_allCouplingSubPlot_N100_T10_typeOmegauniform_spreadOmega2_kavg25_trial1_kavg125_trial1}
\end{figure}

For the same initial networks, natural frequencies, and coupling values, Fig.~\ref{f:degree_vs_time_allCouplingSubPlot_N100_T10_typeOmegauniform_spreadOmega2_kavg25_trial1_kavg125_trial1} shows the node degree $k_{i}$ \emph{vs.} the rewiring step $m$ and Fig.~\ref{f:degree_freq_corr_vs_time_allCouplingSubPlot_N100_T10_typeOmegauniform_spreadOmega2_kavg25_trial1_kavg125_trial1} shows the correlation $C_{|\tilde{\omega}|,k}$ between the absolute value of the frequency offset $|\tilde{\omega}_{i}|$ and the node degree $k_{i}$ \emph{vs.} the rewiring step $m$. In each case, panel \emph{(a)} corresponds to a network with $\langle k \rangle = 12.5$, and panel \emph{(b)} corresponds to a network with $\langle k \rangle = 25$. The natural frequencies were uniformly distributed in both cases. Below, we discuss the observations for each of the densities in turn.

For $\langle k \rangle = 12.5$ and at low coupling (i.e. $\alpha = 0.04$), the rewiring does not noticeably affect how edges are distributed on particular oscillators, and $C_{|\tilde{\omega}|,k}$ fluctuates around zero. As the coupling increases, though, the co-evolution begins to cause significant changes in the network (see panels for $\alpha = 0.114$ and $\alpha = 0.0115$). In particular, there is a short period of time in which edges become concentrated on oscillators with natural frequencies near the mean, whereas oscillators with intrinsic frequencies on the ends of the distribution remain with relatively low degrees. This is followed by the gradual spreading out of edges onto the more outlying oscillators with subsequent rewiring. We can quantify this behavior by considering the value of the correlation $C_{|\tilde{\omega}|, k}$. For $\alpha = 0.114$ and $\alpha = 0.115$ in this example, we see that a slight negative degree-frequency correlation develops briefly in the initial stages of the rewiring, followed by an increase in this quantity to a high, positive value.

For the case of $\langle k \rangle = 25$, we again see that at very low coupling, the rewiring does not drive persistent changes in the system. Interestingly, though, for slightly larger coupling (i.e. $\alpha = 0.04$ and $\alpha = 0.05$ in this example) there is a regime during which edges consistently build up on oscillators with natural frequencies near the center of the distribution. This results in $C_{|\tilde{\omega}|,k}$ becoming significantly negative due to the co-evolution of the network and dynamics, and unlike the situation in the sparser networks, this behavior can persist across the entire rewiring period (see, for example, $\alpha = 0.04$). The concentration of edges on oscillators near the center of the distribution explains the dips in Fig.~\ref{f:degree_freq_corr_uniform}b, which shows $C_{|\tilde{\omega}|,k}$ \emph{vs.} $\alpha$ for the case of $\langle k \rangle = 25$. We also note that near the value of the coupling where this dip occurs, the order parameter for the rewired networks begins to deviate in a positive direction from that of the ER networks (Fig.~\ref{f:R_vs_alpha_Uniform}b), which is likely due to the formation of a locally synchronized cluster of oscillators with natural frequencies near the population average. We do not observe this type of organization for the sparser networks, suggesting an intricate dependence on the network density; this may be an interesting parameter to explore further in future work. As the coupling increases further to intermediate values (for example, $\alpha = 0.06$ and $\alpha = 0.065$), we find that there is first an increase in degree for oscillators with frequencies more closely surrounding the mean, which gives rise to a slightly negative degree-frequency correlation. But as the network continues to co-evolve, edges begin to localize on more outlying oscillators, and the degree-frequency correlation crosses zero and then starts to become positive. However, the most disparate oscillators may still not be able to gather enough edges, resulting in a positive correlation that is less than 1 (i.e., there is some scatter in the relationship).

As the coupling increases further, there is another shift in terms of how the adaptive mechanism affects the networks. For the example with $\langle k \rangle = 12.5$, we observe some fluctuation in $C_{|\tilde{\omega}|, k}$ when the rewiring begins, but eventually the period of marked negative degree-frequency correlation disappears and the edges rapidly become redistributed onto oscillators with the most disparate intrinsic frequencies. We also find that for the examples with $\langle k \rangle = 25$, $C_{|\tilde{\omega}|, k}$ almost immediately begins to rise -- rather than decreasing first -- at high coupling. In addition, the correlation in each case quickly saturates at a value close to 1, signifying a very strong degree-frequency relationship that extends to even the oscillators on the far edges of the natural frequency distribution. Thus, at large couplings the time evolution is similar for $\langle k \rangle = 12.5$ and $\langle k \rangle = 25$. If network co-evolution were allowed to increase for an even longer time, we may observe a plateau in $C_{|\tilde{\omega}|, k}$ for more intermediate values of $\alpha$, though not necessarily at a level corresponding to a near-perfect positive relationship. We repeat this analysis for the case of normally distributed frequencies $\{\omega_{G}\}$ in Sec.~\ref{s:time_dependence_normal} of the Appendix and find qualitatively similar results.
 
Although a rigorous mathematical treatment of the co-evolution is beyond the scope of this study, it is useful to postulate mechanisms that might be able to explain and complement at least some of the empirically-based descriptions discussed above. For example, one interesting observation is that, at low coupling and in the initial stages of rewiring, oscillators with more outlying natural frequencies tend to have lower degree than some oscillators with natural frequencies closer to the mean of the distribution. This seems to occur, at least to some extent, across both values of the network density and for both frequency distributions. In order to think about this phenomena, we first remark that changes in the structure of the network (and thus the dynamics) must be due to which edges are \textit{broken} over time, since the location of a new edge is chosen at random. Second, oscillators with more disparate intrinsic frequencies will naturally want to rotate more rapidly than oscillators with more moderate intrinsic frequencies. Following this reasoning, we can then posit that the most outlying oscillators will have a greater chance of being most in phase with a focal node  -- which is selected at random --  than oscillators with moderate intrinsic frequencies whose phases will tend to change less quickly and thus reduce the chance of being nearby the focal node. Therefore, this thought process suggests that when the selected focal node assesses its phase difference with its neighbors, the oscillators with natural frequencies most different from the mean will have a greater probability of becoming disconnected, and this would account for the observed lower degree of these nodes at low coupling and at the start of the adaptation period. Preliminary results show that making $T$ smaller can sometimes slightly enhance or extend the region of couplings over which $C_{|\tilde{\omega}|, k}$ goes negative, which is consistent with the proposed logic. Another observation, though, is that the degree-frequency correlation can change sign over time (go from slightly negative to positive), which does not seem directly obvious to explain from the previously outlined arguments. In order to make wholly concrete statements and to understand in more detail how the adaptation process gives rise to various results -- such as the intricacies of the time-evolution of the network and also the dependence of that time-evolution on parameters like the global coupling, network density, and frequency distribution -- will require a much more in-depth investigation. While outside the main contributions of this paper, it would be useful to formulate and carry out a more formal theoretical analysis in future work.

\subsection{Spectral Analysis}
\label{s:spectral_analysis_uniform}

As an additional connection to recent literature on optimizing synchronization in heterogeneous oscillator populations, we also carry out a spectral analysis, following the work of \textcite{Skardal:2014a}. In a series of papers \cite{Skardal:2014a,Skardal:2015a,Skardal:2016a}, the authors derived conditions that describe the deviation from full phase locking, and also uncovered conditions for promoting synchronization. By considering a linearized form of the dynamics valid in the high-synchrony regime, one of the main analytical results is that the global order parameter, $R$ can in general be optimized by aligning the vector of intrinsic frequencies $\bm{\tilde{\omega}}$ with the dominant eigenvector $\bm{v}_{N}$ of the Laplacian matrix $\bm{L}$ (where the elements of the Laplacian are defined as $L_{ij} = \delta_{ij}k_{i} - A_{ij}$). Given the enhanced synchronization found here, it is interesting to ask whether the local rewiring rule that we study generates networks with similar spectral properties.  

\begin{figure}[h!]
\centering
\includegraphics[width=\columnwidth]{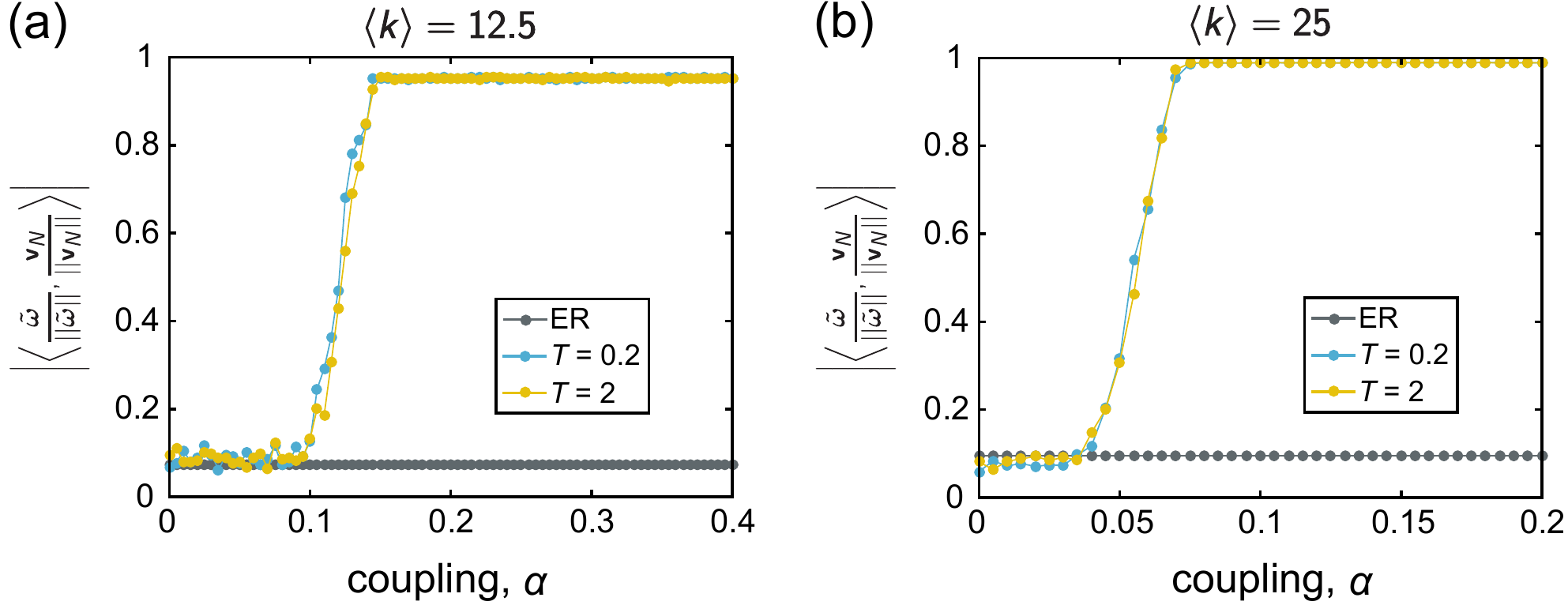}
\caption{Evolution of the overlap $| \langle \frac{\bm{\tilde{\omega}}}{||\bm{\tilde{\omega}}||} , \frac{\bm{v}_{N}}{||\bm{v}_{N}||} \rangle |$ between the intrinsic frequencies $\bm{\tilde{\omega}}$ and dominant Laplacian eigenvector $\bm{v}_{N}$ as a function of the coupling $\alpha$. In each panel, the gray data points correspond to the original, uncorrelated ER random graphs $\mathcal{G}_{o}$, and the blue and yellow curves correspond to the adapted networks $G_{\star}$ evolved under rewiring time scales of $T = 0.2$ and $T = 2$, respectively. The natural frequencies were drawn from the uniform distribution $\{\omega_{U}\}$, and the mean degree $\langle k \rangle$ of the networks are \emph{(a)} $\langle k \rangle = 12.5$ and \emph{(b)} $\langle k \rangle = 25$. All curves depict averages over 25 instantiations, and the lines between data points serve as guides for the eye.}
\label{f:spectral_analysis_uniform}
\end{figure}

To address this question, we examine the quantity $| \langle \frac{\bm{\tilde{\omega}}}{||\bm{\tilde{\omega}}||} , \frac{\bm{v}_{N}}{||\bm{v}_{N}||} \rangle |$, where $\frac{\bm{\tilde{\omega}}} {||\bm{\tilde{\omega}}||}$ is the (normalized) vector of natural frequency offsets, and $\frac{\bm{v}_{N}}{||\bm{v}_{N}||}$ is the (normalized) dominant Laplacian eigenvector. In order to understand the dependence of this frequency-eigenvector alignment on the coupling strength, at each value of $\alpha$ we compute $| \langle \frac{\bm{\tilde{\omega}}}{||\bm{\tilde{\omega}}||}, \frac{\bm{v}_{N}}{||\bm{v}_{N}||} \rangle |$ on the final, co-evolved networks $\mathcal{G}_{\star}$, as well as on the original ER networks $\mathcal{G}_{o}$, for comparison. Fig.~\ref{f:spectral_analysis_uniform} \emph{(a)} and \emph{(b)} depict the results of this analysis for the two different values of the mean degree (see Fig.~\ref{f:spectral_analysis_normal} for the corresponding analysis with normally distributed frequencies). At low coupling, $| \langle \frac{\bm{\tilde{\omega}}}{||\bm{\tilde{\omega}}||}, \frac{\bm{v}_{N}}{||\bm{v}_{N}||} \rangle |$ for the rewired networks remains low, near its initial value. Note that in this regime the order parameter is low as well (Fig.~\ref{f:R_vs_alpha_Uniform}). As $\alpha$ increases, however, the frequency-eigenvector overlap begins to grow, and rapidly increases until a leveling out with further increase in the coupling; the plateau occurs close to the maximal value of one for both values of the mean degree. As with the relationships between natural frequency and node degree, and between natural frequency and neighbor natural frequency considered in the previous section, the increasing projection of the intrinsic frequencies onto the dominant Laplacian eigenvector is accompanied by an increase in $R$. Thus, the adaptive reconfiguration can cause persistent changes to the organization of the network that are consistent with the conditions predicted by \textcite{Skardal:2014a} for optimizing synchronization, and the result becomes more prominent at higher coupling. The investigation for the case of normally-distributed frequencies yields similar conclusions (see Appendix~\ref{a:spectral_analysis_normal}, Fig.~\ref{f:spectral_analysis_normal}). This analysis provides a more analytical understanding of how the adaptive process is able to enhance the synchronization in the system.

\section{Discussion and Conclusions}
\label{s:discussion}

In this study, we examined co-evolution of network topology and Kuramoto phase-oscillator dynamics as a means to evolve initially unstructured networks towards organized architectures, and to simultaneously enhance synchronization in the system. In terms of the interplay between network topology and synchronization, it has been found that the presence of specific correlations between the structural layout of the network and the oscillator frequencies (which is a property of the dynamics) can greatly augment global synchronization. But these relationships usually arise through optimization strategies that utilize global information about the network or of node states as a function of time \cite{Brede:2008b, Brede:2009a,Fan:2009a, Buzna:2009a, Carareto:2009a,Kelly:2011a,Brede:2010a,Brede:2008a, Freitas:2015a,Skardal:2014a,Skardal:2015a,Skardal:2016a}. On the other hand, a different set of work has shown that adaptive strategies that suppress phase differences between Kuramoto oscillators \cite{Ren:2007a,Ren:2014a} can lead to heightened synchronization. But in the latter case, the interactions between the topology of the adaptive networks and the node frequencies has not been explored or examined in depth. An interesting line of investigation is to therefore understand whether an adaptive rule for updating the structure of the network --  based on local dynamical information -- can shape the topological patterns and correlations with dynamical properties whose emergence simultaneously enhances synchronization. To this end, we studied a type of disassortative mechanism in which the edge between a randomly selected node and its most instantaneously synchronized neighbor is stochastically rewired, while all other edges are maintained. Co-evolution of the dynamics and network connectivity occurs through the repetition of this feedback process, whereby an initially random network continually reconfigures in response to the states of locally connected oscillators. 

Through numerical simulation, we examined the time-evolution of this process and the dependence on the global coupling. We found that for a significant coupling range, the rewiring strategy was able to bring the system to a state of heightened collective behavior, as measured by the global order parameter. It is interesting to note that this eventual state of enhanced global coherency depended on the adaptive prevention of local synchronization, suggesting a trade-off between local and global dynamics. Other work on adaptive Kuramoto networks has shown that the opposite type of rule, i.e. a competitive strategy which strengthens connections between more in-phase oscillators at the expense of weakening connections elsewhere, can lead to modular organization and hence enhanced local rather than global synchrony \cite{Gutierrez:2011a, Assenza:2011a, Avalos-Gaytan:2012a}. Perhaps most importantly, the enhancement of synchronization indeed co-occurred with the emergence of correlations in the network that have been shown to arise through optimization of the global order parameter. In particular, when the oscillators exhibited more coherent dynamics, the evolved networks tended to exhibit (\emph{1}) positive correlations between node degrees and the magnitude of oscillators' difference in natural frequency from the mean of the population (i.e. magnitude of their frequency offset), and (\emph{2}) the preference of connections between oscillators that have natural frequency offsets of opposite sign. We found that the emergence of these relationships and how the adaptive scheme reorganized the network topology over time depended on the global coupling parameter and the intrinsic frequencies, and - to some extent - the density of the network. We note that in addition to enhancing synchronization \cite{Brede:2008a,Brede:2008b,Brede:2009a,Brede:2010a}, the purposeful placement of these types of correlations has also been associated with first order, or explosive, synchronization transitions \cite{Gomez-Gardenes:2011aa,Leyva:2013a,Leyva:2013b}. Far fewer studies, however, have considered how these structural patterns and relationships might arise in a network from local rearrangements or adaptation (though see \cite{Zhang:2015a} for one example).

It is important to state that the results found in this study are in line with previous work that has examined adaptive schemes in which weights grow or shrink as a function of phase differences, which also find that rules that actively suppress differences in state are able to improve synchronization. However, there are some important distinctions between the strategies studied in \cite{Ren:2007a} and the one studied here. In regards to the former, the network starts as completely disconnected, with no edges between any oscillators. The edge weights between \emph{all} pairs of nodes are then allowed to increase, up to some bound. In this way, the total density of the network is allowed to change, connections can theoretically occur between all pairs of oscillators at a given time, and the individual edge weights can fluctuate. Other work has considered a situation in which the \textit{topology} of the network is pre-defined and constant, while the weights can change \cite{Ren:2014a}. On the other hand, we wished to consider a situation in which the effect of topological organization alone could be isolated from other confounding features. We thus studied a case where the network begins connected, but in a random, disorganized arrangement, and then allowed the system to self-organize under the constraint of \emph{fixed} total density and also binary and undirected connectivity. Together, these conditions mean that only a fraction of the nodes can be directly connected at a given time, and the goal is to understand if simple \emph{rearrangements} in those connections, based on a local rule for determining the rewiring, can enhance synchronization. Thus, the path to a more coherent state is different here than in previous studies on co-evolutionary Kuramoto systems. 

Of course, there are still some methodological considerations to make note of, as well as possibilities for future work. For example, we studied an adaptive mechanism that utilized only local information of a given node, and that preserved binary connectivity and the total number of edges, so as to isolate the effects of rearrangements in network topology from other factors. However, incorporating these constraints required that there be a random component to the rewiring process, and there was not a natural or self-employed stopping condition for network reconfiguration. One could thus further explore how results are affected by the length of time the system is allowed to co-evolve, and also how this relates to changes in other parameters, such as the size of the network, the mean degree, and the spread in the intrinsic frequencies. Another parameter that may be important to examine more in depth is the time-scale of the adaptation in the network. In addition, while some stochasticity or noise is likely a realistic feature of natural systems, it would be interesting to incorporate that into a related adaptation model that allows for weighted rather than just binary connectivity between network units. It is also important to note that, while meaningful insights can be gained from the empirical and observational type of analysis carried out in this study, in forthcoming work it will be useful to try and understand the origins of various results from a more fundamental and theoretical standpoint. Finally, we point out that continued investigation into the role of network topology and adaptation in shaping dynamics and structure may lead to a better understanding of the development and function of real-world networks, such as neuronal assemblies or large-scale brain structure and activity patterns. Indeed, there are many computational models of these systems in which this can be studied \cite{Brunel:1999a, Brunel:2000a, Guardiola:2000a, Lago-Fernandez:2000a, Breakspear:2010a, Izhikevich:2006a, Jarman:2014a, Deco:2009a, Roxin:2004a, Rubinov:2009a, Rubinov:2011a}. In addition, for neural systems in particular, it is interesting to note that while synchronization is often a desired property, hyper-synchronization can also be detrimental, as is the case with epileptic seizures \cite{Stam:2005a}. Therefore, further examination of the tradeoff and transition between local and global synchrony in biologically motivated models \cite{Percha:2005a,mofakham2016interplay} -- and understanding how this might occur adaptively over time and influence the structure and function of a network -- continues to be an exciting line of study.

In conclusion, understanding the concurrent influence of network architecture on the emergence of collective dynamics \cite{Winfree:1980a,Strogatz:2003a, Boccaletti:2006a, Barthelemy:2008a, Vespignani:2012a}, and in turn, the effect of a dynamical process on reshaping or inducing network structure \cite{Gross:2009a,Gross:2008a,Sayama:2013a}, is currently an active area of research across a broad set of disciplines, including the physical, social, and biological sciences. Along these lines, we have studied a dynamical rewiring scheme for networks of Kuramoto oscillators. The adaptive rule for the network was specifically inspired by previous work on optimizing networks for synchronization of heterogeneous oscillators \cite{Brede:2008b, Brede:2009a,Fan:2009a, Buzna:2009a, Carareto:2009a,Kelly:2011a,Brede:2010a,Brede:2008a, Freitas:2015a,Skardal:2014a,Skardal:2015a,Skardal:2016a}, from which we wished to uncover how similar dynamics and organization could occur through a co-evolution process. We found that a restructuring of the network, in which the effect of incoherence is suppressed by maintaining edges between disparate oscillators, while randomly rewiring edges between the most locally and instantaneously in-phase oscillators, led to the emergence of distinct topological patterns and correlations that concurrently enhance the system's ability to synchronize. This study thus sheds light on a mechanism for how enhanced synchronization and network structure might arise in a system that evolves and reconfigures according to local information alone, without knowledge of global connectivity or node states. 

\section{Supplementary Material}

Please see the Supplementary Material (\emph{S.M.}) for an analysis of the time-evolution of locally synchronized clusters, and also for an examination of the robustness of some of the main results to simple variations in the network size, the initial network topology, and the frequency distribution. 

\section{Acknowledgments}
\label{s:acknowledge}

L.P. acknowledges support from the National Science Foundation Graduate Research Fellowship Program. J.K. acknowledges support from the National Science Foundation Graduate Research Fellowship Program and NIH T32-EB020087, PD: Felix W. Wehrli. D.S.B. also acknowledges support from the John D. and Catherine T. MacArthur Foundation, the Alfred P. Sloan Foundation, and the National Science Foundation (BCS-1441502, CAREER PHY-1554488, BCS-1631550, and CNS-1626008). We also thank two anonymous reviewers whose comments greatly improved the quality of this work. The content is solely the responsibility of the authors and does not necessarily represent the official views of any of the funding agencies. 

\clearpage
\appendix

\section{An additional measure to quantify relationships between network topology and oscillator frequencies}
\label{a:freq_totalFreq}

In Sec.~\ref{s:emerging_correlations}, we used two measures to quantify emerging relationships between the intrinsic frequencies of the oscillators and the network topology: the correlation $C_{|\tilde{\omega}|, k}$ between the magnitude of the oscillator frequency offset $|\tilde{\omega}_{i}|$ and the node degree $k_{i}$, and the mean fraction $f$ of an oscillator's neighbors that have natural frequency offsets of opposite sign compared to the frequency offset of the central oscillator (averaged across all nodes in the system) \cite{Brede:2008b, Brede:2009a}. Here we define a related measure to further understand the organization that arises in the adaptively rewired networks. In particular, for each oscillator $i$, we assess the association between the frequency offset $\tilde{\omega_{i}}$ and the \textit{sum} of oscillator $i's$ neighbors' frequency offsets $\sum_{j \in \mathcal{N}_{i}} \tilde{\omega}_{j}$. We quantify this relationship in the system by considering the correlation $C_{\tilde{\omega},\sum\tilde{\omega}}$ between $\tilde{\omega_{i}}$ and $\sum_{j \in \mathcal{N}_{i}} \tilde{\omega}_{j}$ across all nodes in the network.

\begin{figure}[h!] 
\centering
\includegraphics[width=\columnwidth]{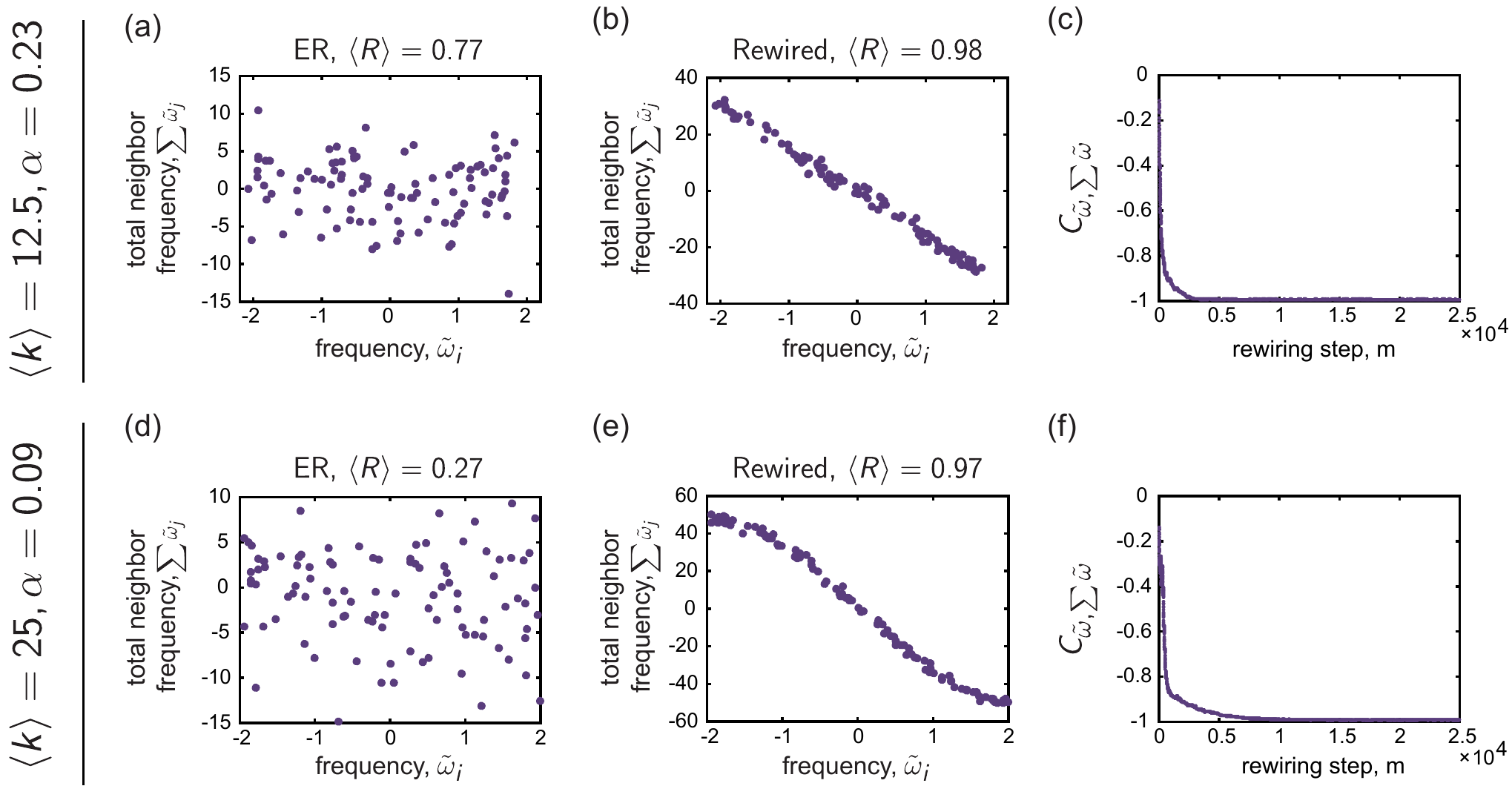}
\caption{The relationship between the oscillator frequency offset $\tilde{\omega}_{i}$ and the sum of neighbor frequency offsets $\sum_{j \in \mathcal{N}_{i}}\tilde{\omega}_{j}$ can be used to further quantify the interplay between the network structure and the intrinsic frequencies of the oscillators that arises due to the co-evolutionary process. The top row shows an example of this relationship for a network with $\langle k \rangle = 12.5$, and the bottom row shows examples for a network with $\langle k \rangle = 25$; in both cases, the frequencies were drawn from the uniform distribution $\{\omega_{U}\}$. For each network density, the first column corresponds to an ER random graph that exhibits only intermediate levels of synchrony at the displayed coupling $\alpha$ (as measured by $\langle R \rangle$), and the second column corresponds to the adapted network, which exhibits a higher level of synchrony. \emph{(a,b); (d,e)} The total neighbor frequency offset $\sum_{j \in \mathcal{N}_{i}}\tilde{\omega}_{j}$ \emph{vs.} frequency offset $\tilde{\omega}_{i}$. \emph{(c); (f)} The correlation $C_{\tilde{\omega},\sum\tilde{\omega}}$ between oscillator frequency offset $\tilde{\omega}_{i}$ and the sum of neighbor frequency offsets $\sum_{j \in \mathcal{N}_{i}} \tilde{\omega}_{j}$ \emph{vs.} the number of rewiring steps $m$.}
\label{f:freq_sumFreq_corr_uniform}
\end{figure}

Fig.~\ref{f:freq_sumFreq_corr_uniform} gives examples of this relationship for the same networks as those in Fig.~\ref{f:correlations_uniform} in the main text. The top and bottom rows correspond to networks with $\langle k \rangle = 12.5$ and $\langle k \rangle = 25$, respectively. Panels \emph{(a)} and \emph{(d)} show that in the ER networks, there is no clear relationship between frequency offset and the sum of neighbor frequency offsets, as expected. In the adaptively rewired networks -- panels \emph{(b)} and \emph{(e)} -- the order parameter has increased from its original value and there is a strong negative correlation between $\tilde{\omega}_{i}$ and $\sum_{j \in \mathcal{N}_{i}} \tilde{\omega}_{j}$. Not only do connections tend to form between oscillators that have frequency offsets of opposite sign (as measured by $f$), but in addition, edges become distributed in the network such that the sum of each oscillator's neighbor frequency offsets proportionately cancels out each oscillator's own difference from the mean frequency of the population. Fig.~\ref{f:freq_sumFreq_corr_uniform}\emph{c,f} shows how the strength of this relationship -- quantified by the correlation $C_{\tilde{\omega},\sum\tilde{\omega}}$ -- evolves as the network is rewired.

\begin{figure}[h!] 
\centering
\includegraphics[width=\columnwidth]{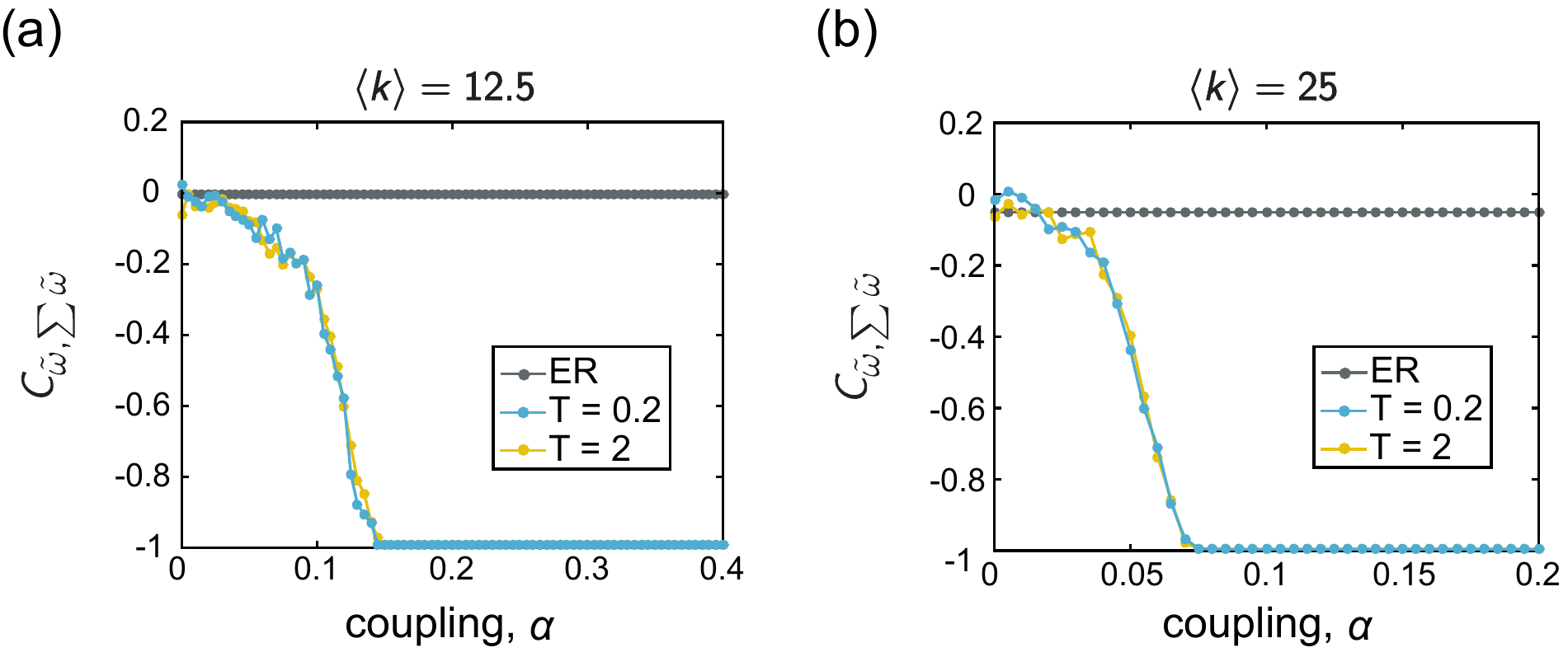}
\caption{These plots depict the correlation $C_{\tilde{\omega},\sum\tilde{\omega}}$ between oscillator frequency offset $\tilde{\omega}_{i}$ and the sum of neighbor frequency offsets $\sum_{j \in \mathcal{N}_{i}}\tilde{\omega}_{j}$ as a function of the coupling. In each panel, the gray data points correspond to the initial, uncorrelated ER random graphs $\mathcal{G}_{o}$, and the blue and yellow points correspond to the adapted networks $G_{\star}$ evolved under rewiring time scales of $T = 0.2$ and $T = 2$, respectively. The frequencies were drawn from the uniform distribution $\{\omega_{U}\}$, and the mean degree of the networks are \emph{(a)} $\langle k \rangle = 12.5$, and \emph{(b)} $\langle k \rangle = 25$. All curves depict averages over 25 instantiations, and the lines between data points serve as guides for the eye.}
\label{f:freq_sumFreq_vs_alpha_uniform}
\end{figure}

Fig.~\ref{f:freq_sumFreq_vs_alpha_uniform} shows $C_{\tilde{\omega},\sum\tilde{\omega}}$ as a function of the coupling $\alpha$. As $\alpha$ increases, $C_{\tilde{\omega},\sum\tilde{\omega}}$ decreases to a strong negative value, and then remains approximately constant for larger values of the coupling. (Compare to Fig.~\ref{f:degree_freq_corr_uniform} and Fig.~\ref{f:freq_neighbFreq_uniform}, which show $C_{|\tilde{\omega}|, k}$ \emph{vs.} $\alpha$ and $f$ \emph{vs.} $\alpha$, respectively). Finally, note that the strong decreases in $C_{\tilde{\omega},\sum\tilde{\omega}}$ occur in conjunction with increases in the order parameter (Fig.~\ref{f:R_vs_alpha_Uniform}).

\section{Analysis with normally distributed frequencies}
\label{a:normal_frequencies}

The analysis in the main text was carried out using natural frequencies $\{\omega_{U}\}$ drawn at random from the uniform distribution in the range $[-2,2]$. To demonstrate that the main results are not specific to a single choice of the frequency distribution, in this Appendix, we also consider the case of frequencies $\{\omega_{G}\}$ drawn from a normal distribution with zero mean and unit standard deviation. The findings shown here are largely consistent with those reported in the main text. 

\subsection{Dependence of the order parameter on time and global coupling}
\label{a:R_vs_time_alpha}

\begin{figure}
\centering
\includegraphics[width=\columnwidth]{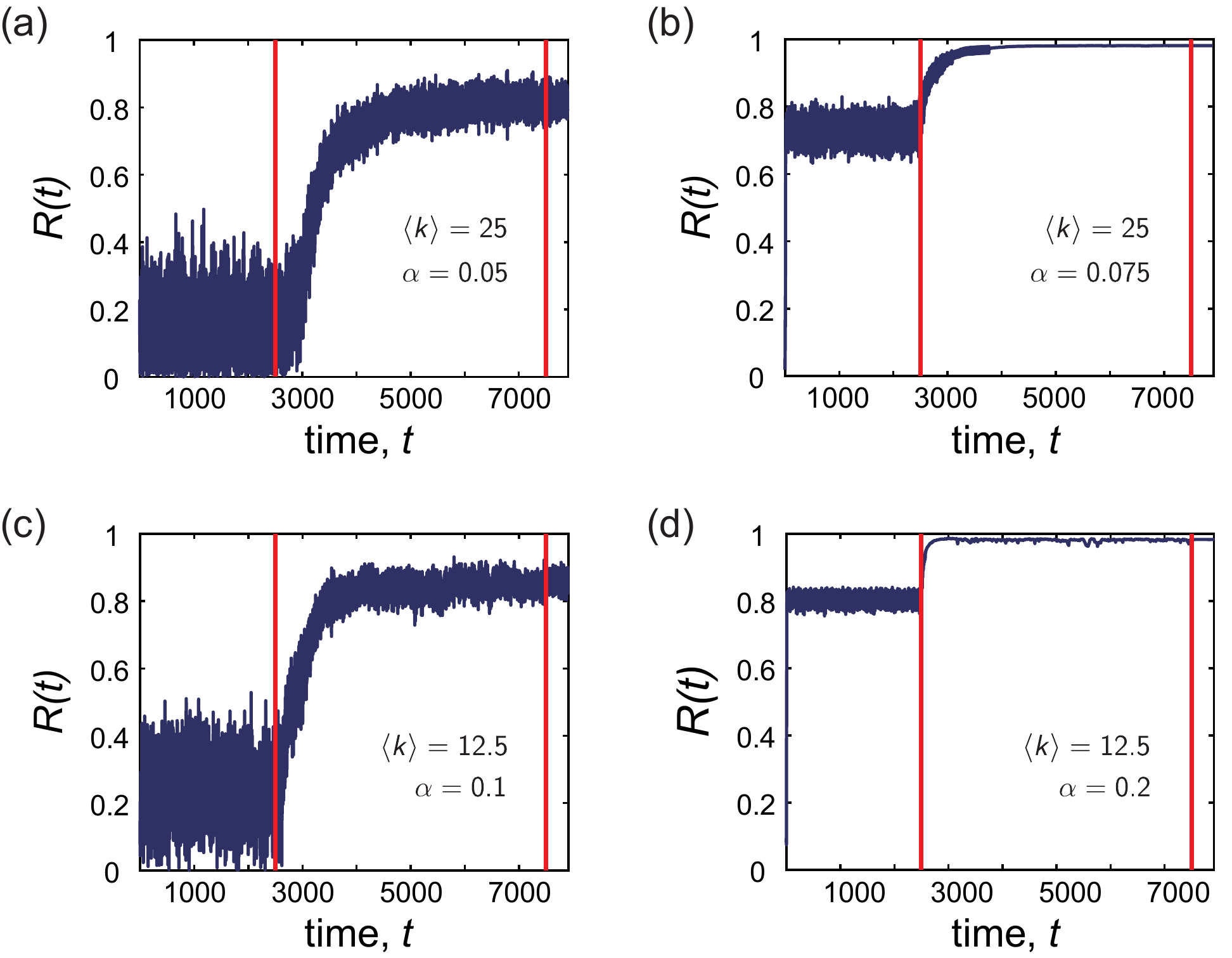}
\caption{Examples of the global order parameter $R(t)$ \emph{vs.} time $t$, for various representative couplings $\alpha$. In each case, the dynamics were first run atop an initially ER random graph with average degree $\langle k \rangle$, after which co-evolution of the network and dynamics took place between the two red lines. The natural frequencies were drawn from a normal distribution $\{ \omega_{G} \}$, and the mean degree $\langle k \rangle$ and coupling $\alpha$ used for each panel were \emph{(a)} $\langle k \rangle = 25$, $\alpha = 0.05$, \emph{(b)} $\langle k \rangle = 25$, $\alpha = 0.075$, \emph{(c)} $\langle k \rangle  = 12.5$, $\alpha = 0.1$, \emph{(d)} $\langle k \rangle = 12.5$, $\alpha = 0.2$. During the adaptation period, the network was continually rewired once every $T = 0.2$ time units. The co-evolving networks exhibit enhanced collective dynamics, as observed by increases in the global order parameter.}
\label{f:R_vs_T_Normal}
\end{figure}

Fig.~\ref{f:R_vs_T_Normal} shows examples of the order parameter $R(t)$ \emph{vs.} time $t$ at different values of the coupling and for the two different mean degrees $\langle k \rangle = 12.5$ and $\langle k \rangle = 25$. For these trials and values of the coupling, we observe that the order parameter increases due to the rewiring of the network, which takes place between the two red lines. (Fig.~\ref{f:R_vs_T_Uniform} from the main text shows examples for the case of uniformly distributed natural frequencies).

\begin{figure}[h]
\centering
\includegraphics[width=\columnwidth]{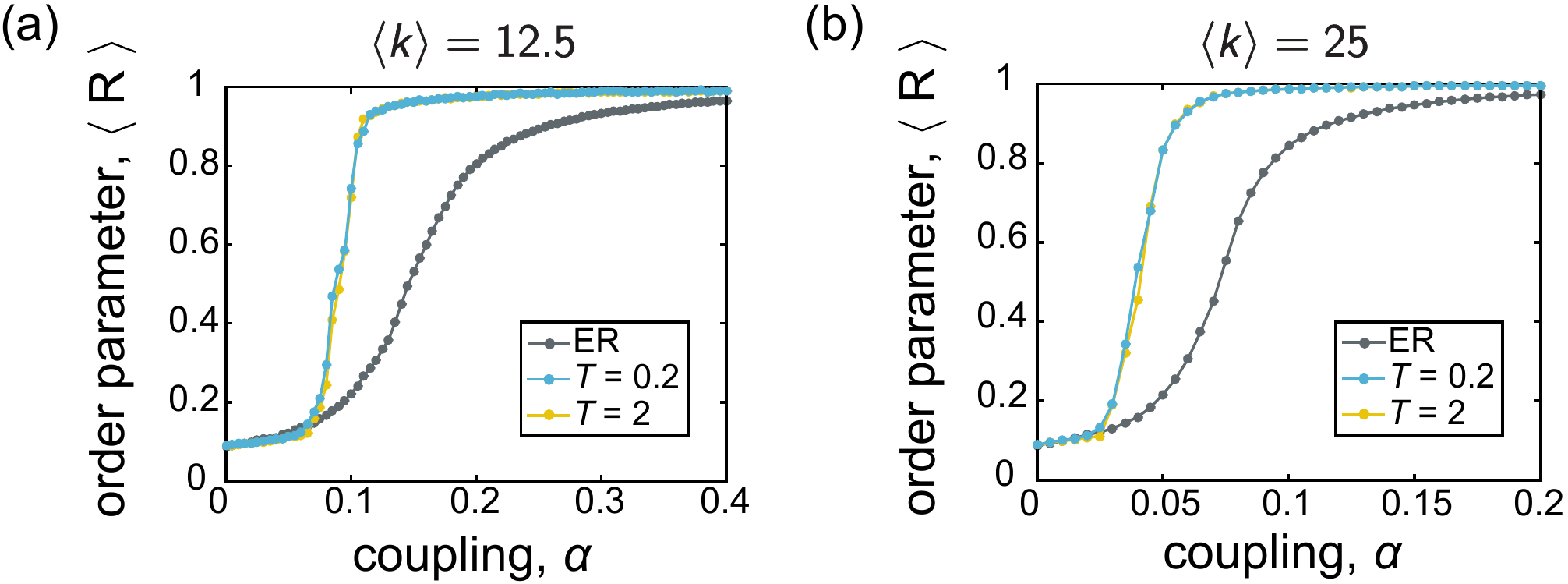}
\caption{The time-averaged order parameter $\langle R \rangle$ \emph{vs.} coupling $\alpha$. In each panel, the gray data points correspond to static ER random graphs $\mathcal{G}_{o}$, and the blue and yellow points correspond to the adapted networks $\mathcal{G}_{\star}$ evolved under rewiring time scales of $T = 0.2$ and $T = 2$, respectively.  The frequencies were drawn from the normal distribution $\{\omega_{G}\}$, and the mean degree of the networks were \emph{(a)} $\langle k \rangle = 12.5$, and \emph{(b)} $\langle k \rangle = 25$. All curves depict averages over 25 instantiations, and the lines between data points serve as guides for the eye.}
\label{f:R_vs_alpha_Normal}
\end{figure}

Fig.~\ref{f:R_vs_alpha_Normal} shows examples of the time-averaged order parameter $\langle R \rangle$ \emph{vs.} the coupling $\alpha$. As with the case of uniformly distributed frequencies (Sec.~\ref{s:enhanced_synchronization_uniform}, Fig.~\ref{f:R_vs_alpha_Uniform}), we find that the adapted networks exhibit heightened synchronization over a large coupling range.

\subsection{Correlations between network topology and the intrinsic frequencies}

\begin{figure*}
\centering
\includegraphics[width=0.75\textwidth]{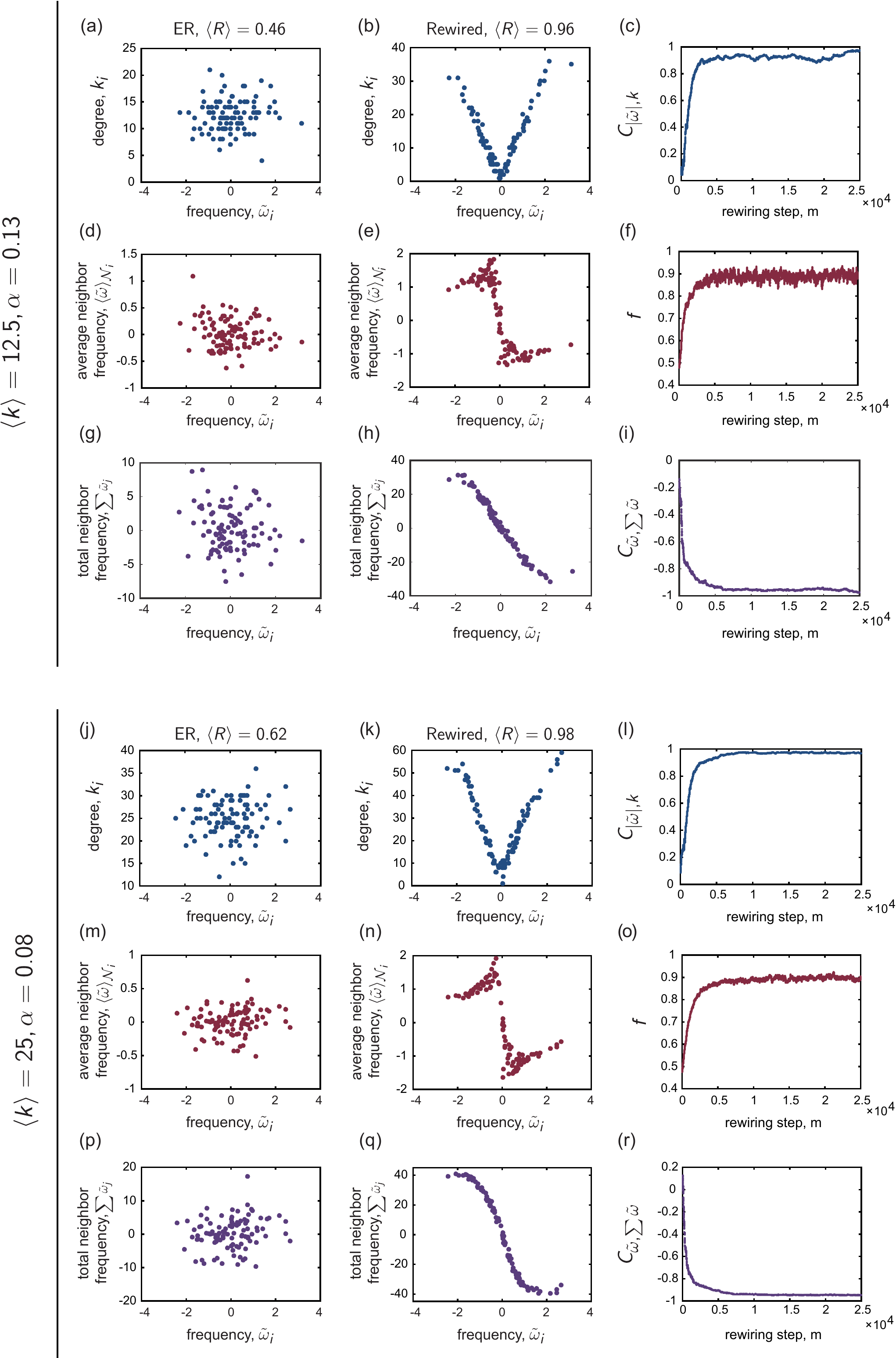}
\caption{Relationships between the network structure and the intrinsic frequencies of the oscillators.  The top three rows show examples for a network with $\langle k \rangle = 12.5$, and the bottom three rows show examples for a network with $\langle k \rangle = 25$; in both cases, the frequencies were drawn from the normal distribution $\{\omega_{G}\}$. For each network density, the first column corresponds to an ER random graph that exhibits only intermediate levels of synchrony at the displayed coupling $\alpha$ (as measured by $\langle R \rangle$), and the second column corresponds to the adapted network, which exhibits a higher level of synchrony. These plots highlight key relationships that emerge from the co-evolutionary network update rule. \emph{(a,b); (j,k)} Node degree $k_{i}$ \emph{vs.} frequency offset $\tilde{\omega}_{i}$. \emph{(d,e); (m,n)} Average neighbor frequency offset $\langle \tilde{\omega} \rangle_{\mathcal{N}_{i}}$ \emph{vs.} frequency offset $\tilde{\omega}_{i}$.  \emph{(g,h); (p,q)} Total neighbor frequency offset $\sum_{j \in \mathcal{N}_{i}}\tilde{\omega}_{j}$ \emph{vs.} frequency offset $\tilde{\omega}_{i}$. \emph{(c); (l)} The correlation $C_{|\tilde{\omega}|, k}$ between node degree $k_{i}$ and the magnitude of the frequency offset $|\tilde{\omega}_{i}|$ \emph{vs.} the number of rewiring steps $m$, and \emph{(f); (o)} The mean fraction $f$ (i.e. averaged over all nodes in the network) of an oscillator's neighbors that have frequency offsets of opposite sign compared to that of the central oscillator \emph{vs.} the number of rewiring steps $m$. \emph{(i); (r)} The correlation $C_{\tilde{\omega},\sum\tilde{\omega}}$ between oscillator frequency offset $\tilde{\omega}_{i}$ and the sum of neighbor frequency offsets $\sum_{j \in \mathcal{N}_{i}} \tilde{\omega}_{j}$ \emph{vs.} the number of rewiring steps $m$.}
\label{f:correlations_normal}
\end{figure*}

Fig.~\ref{f:correlations_normal} shows examples of node degree $k_{i}$ \emph{vs.} intrinsic frequency offset $\tilde{\omega}_{i}$, average neighbor frequency offset $\langle \tilde{\omega} \rangle_{\mathcal{N}_{i}}$ \emph{vs.} node frequency offset $\tilde{\omega}_{i}$, and total neighbor frequency offset $\sum_{j \in \mathcal{N}_{i}} \tilde{\omega}_{j}$ \emph{vs.} frequency offset $\tilde{\omega}_{i}$ for ER networks and the corresponding co-evolved networks (see Sec.~\ref{s:emerging_correlations} and Appendix~\ref{a:freq_totalFreq} for the definitions of these quantities). The top three and bottom three panels correspond to networks with $\langle k \rangle = 12.5$ and $\langle k \rangle = 25$, respectively, and the frequencies $\{\omega_{G}\}$ were normally distributed. At the values of coupling used for these examples, we see that as the system reconfigures, the network begins to exhibit distinct patterns in terms of the organization of oscillators with different intrinsic frequencies. Each of the three metrics considered -- $C_{|\tilde{\omega}|, k}$, $f$, and $C_{\tilde{\omega},\sum\tilde{\omega}}$  -- exhibit a progression that allows for the eventual heightened degree of synchrony in the rewired network. These findings are similar to those discussed for the uniform frequency distribution (see Fig.~\ref{f:correlations_uniform} and Fig.~\ref{f:freq_sumFreq_corr_uniform} and the corresponding text in Sec.~\ref{s:emerging_correlations}).

\begin{figure}[h!]
\centering
\includegraphics[width=\columnwidth]{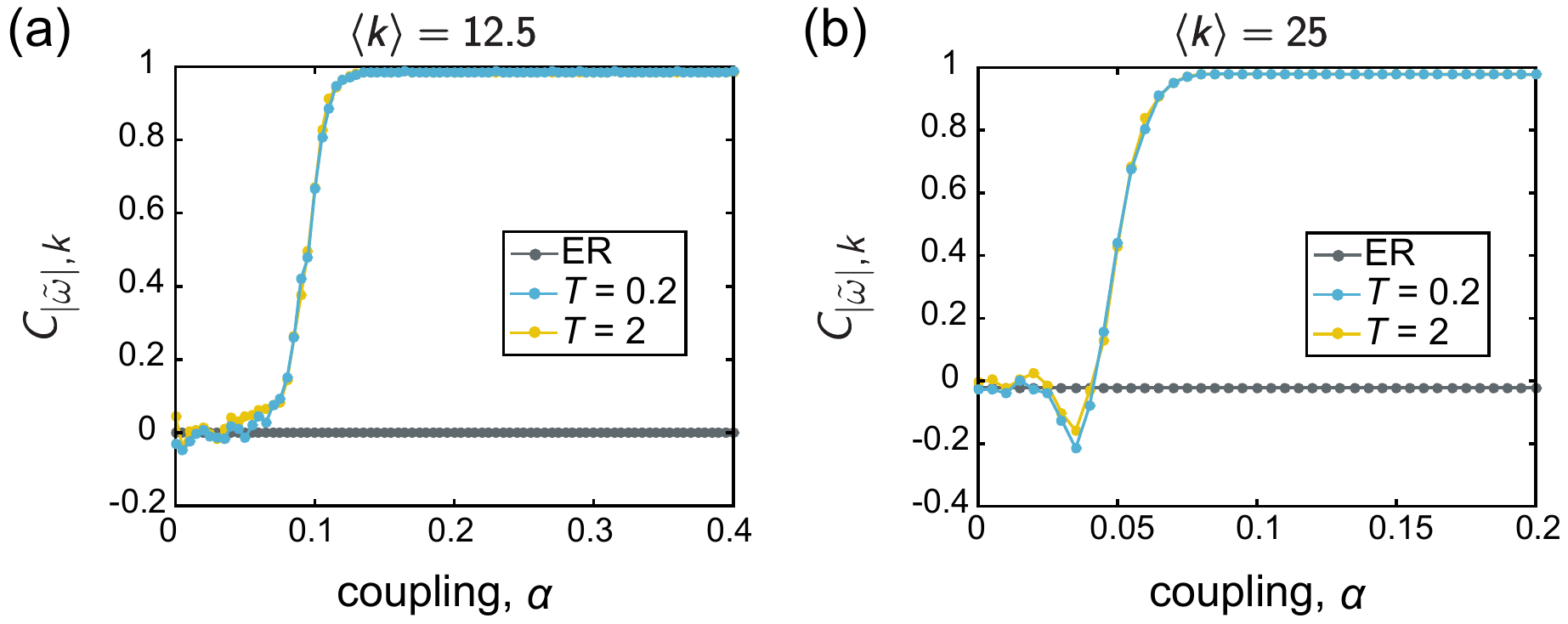}
\caption{These plots depict the correlation $C_{|\tilde{\omega}|, k}$ between node degree $k_{i}$ and the magnitude of the frequency offset $|\tilde{\omega}_{i}|$, as a function of the coupling. In each panel, the gray data points correspond to the initial, uncorrelated ER random graphs $\mathcal{G}_{o}$, and the blue and yellow points correspond to the adapted networks $G_{\star}$ evolved under rewiring time scales of $T = 0.2$ and $T = 2$, respectively. The frequencies were drawn from the normal distribution $\{\omega_{G}\}$, and the mean degree of the networks are \emph{(a)} $\langle k \rangle = 12.5$, and \emph{(b)} $\langle k \rangle = 25$. The dip observed in \emph{(b)} at $\alpha \approx 0.4$ is due to the localization of edges on a cluster of oscillators with natural frequencies near the mean of the distribution; this is examined further in Appendix~\ref{s:time_dependence_normal}. All curves depict averages over 25 instantiations, and the lines between data points serve as guides for the eye.}
\label{f:degree_freq_corr_normal}
\end{figure}

\begin{figure}[h!]
\centering
\includegraphics[width=\columnwidth]{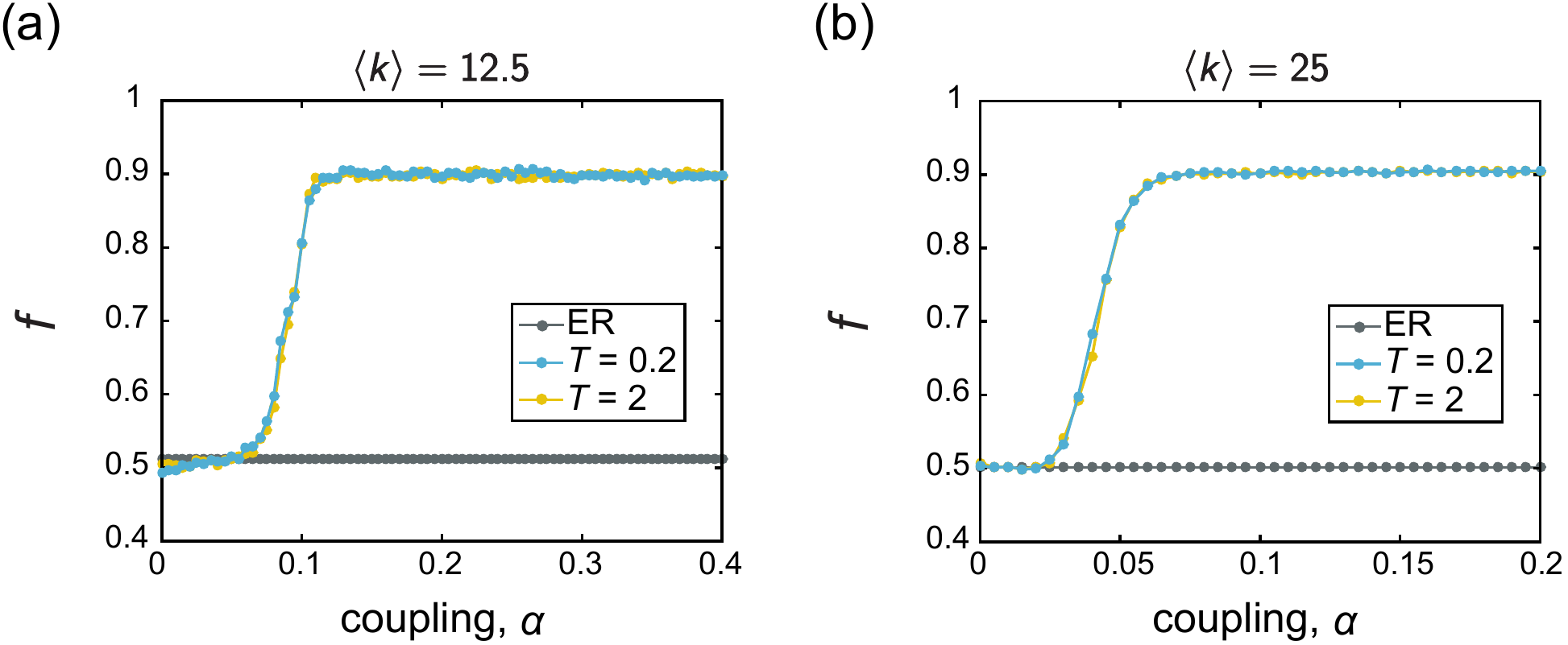}
\caption{These plots depict the mean fraction $f$ (i.e. averaged over all nodes in the network) of an oscillator's neighbors that have natural frequency offsets of opposite sign compared to that of the central oscillator, as a function of the coupling. In each panel, the gray data points correspond to the initial, uncorrelated ER random graphs $\mathcal{G}_{o}$, and the blue and yellow points correspond to the adapted networks $G_{\star}$ evolved under rewiring time scales of $T = 0.2$ and $T = 2$, respectively. The frequencies were drawn from the normal distribution $\{\omega_{G}\}$, and the mean degree of the networks are \emph{(a)} $\langle k \rangle = 12.5$, and \emph{(b)} $\langle k \rangle = 25$. All curves depict averages over 25 instantiations, and the lines between data points serve as guides for the eye.}
\label{f:freq_neighbFreq_normal}
\end{figure}

\begin{figure}[h!] 
\centering
\includegraphics[width=\columnwidth]{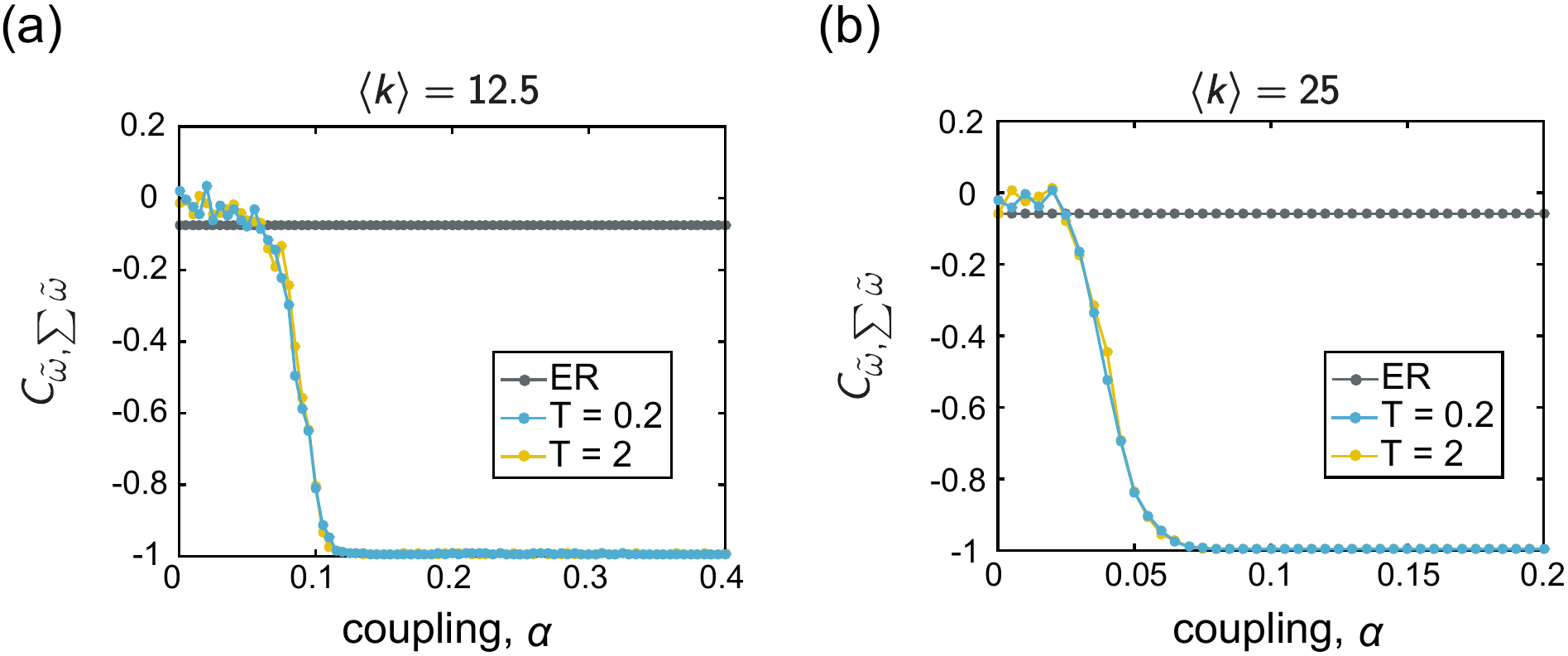}
\caption{These plots depict the correlation $C_{\tilde{\omega},\sum\tilde{\omega}}$ between oscillator frequency offset $\tilde{\omega}_{i}$ and the sum of neighbor frequency offsets $\sum_{j \in\ mathcal{N}(i)} \tilde{\omega}_{j}$, as a function of the coupling. In each panel, the gray data points correspond to the initial, uncorrelated ER random graphs $\mathcal{G}_{o}$, and the orange and yellow points correspond to the adapted networks $G_{\star}$ evolved under rewiring time scales of $T = 0.2$ and $T = 2$, respectively. The frequencies were drawn from the normal distribution $\{\omega_{G}\}$, and the mean degree of the networks are \emph{(a)} $\langle k \rangle = 12.5$, and \emph{(b)} $\langle k \rangle = 25$. All curves depict averages over 25 instantiations, and the lines between data points serve as guides for the eye.}
\label{f:freq_sumFreq_normal}
\end{figure}

We next show the evolution of the relationships between network topology and the intrinsic frequencies as a function of the coupling $\alpha$. Figs.~\ref{f:degree_freq_corr_normal},~\ref{f:freq_neighbFreq_normal}, and ~\ref{f:freq_sumFreq_normal} show $C_{|\tilde{\omega}|, k}$ \emph{vs.} $\alpha$, $f$ \emph{vs.} $\alpha$, and $C_{\tilde{\omega},\sum\tilde{\omega}}$ \emph{vs.} $\alpha$, respectively. The conclusions drawn for the case of normally distributed frequencies shown here are the same as those for the uniformly distributed frequencies examined in the main text. (See Figs.~\ref{f:degree_freq_corr_uniform},~\ref{f:freq_neighbFreq_uniform}, and ~\ref{f:freq_sumFreq_vs_alpha_uniform} and the corresponding discussions in Sec.~\ref{s:emerging_correlations} and Appendix~\ref{a:freq_totalFreq}). Briefly, each of the three measures exhibit transitions at similar values of the coupling, and the emergence of strong relationships between the network structure and oscillator frequencies arise near the coupling when the order parameter transitions from low to higher values.

\subsection{Time-dependence of the instantaneous frequencies and network structure}
\label{s:time_dependence_normal}

\begin{figure}
\centering
\includegraphics[width=\columnwidth]{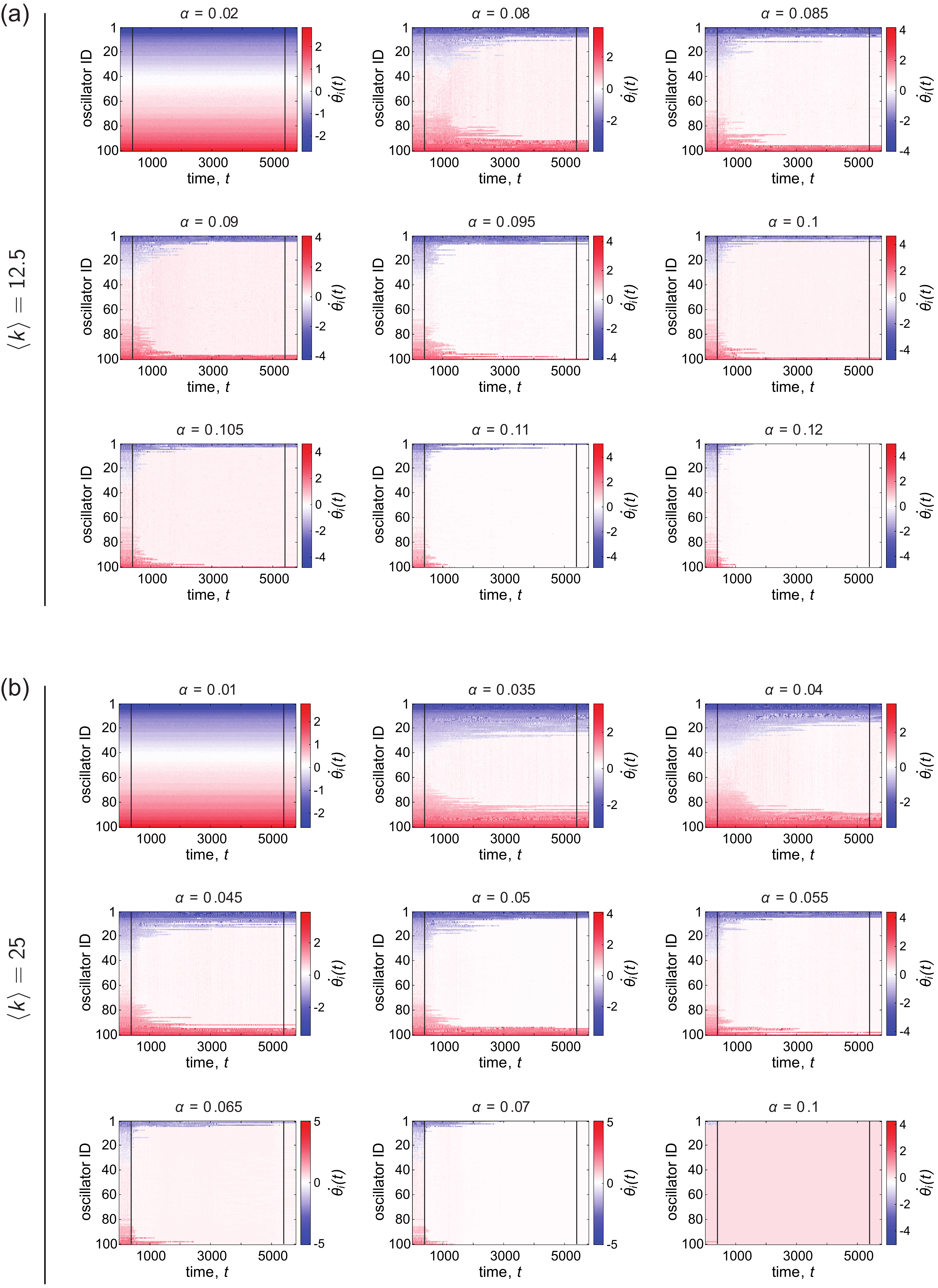}
\caption{Examples of the instantaneous frequencies $\dot{\theta}_{i}(t)$ \emph{vs.} time $t$, for various representative couplings $\alpha$. The mean degree of the networks are \emph{(a)} $\langle k \rangle = 12.5$ and \emph{(b)} $\langle k \rangle = 25$, and the natural frequencies $\{\omega_{G}\}$ were drawn from the normal distribution. In all panels, each row corresponds to one oscillator, and the rows are ordered by the quantity $\tilde{\omega}_{i} = \omega_{i} - \langle \omega \rangle$ (i.e. the offset from the mean intrinsic frequency of the population). For each coupling, the dynamics were first run atop an initially ER random graph, after which co-evolution of the network and dynamics took place between the two black lines. During the adaptation period, the network was continually rewired once every $T = 0.2$ time units.}
\label{f:freq_vs_time_allCouplingSubPlot_N100_T10_typeOmeganormal_spreadOmega1_kavg25_trial8_kavg125_trial8}
\end{figure}

Fig.~\ref{f:freq_vs_time_allCouplingSubPlot_N100_T10_typeOmeganormal_spreadOmega1_kavg25_trial8_kavg125_trial8} shows examples of $\dot{\theta_{i}}(t)$ \emph{vs.} $t$ for several values of the coupling $\alpha$ around the point in which the dynamics transition from an incoherent state to a synchronized state. The top set of panels \emph{(a)} are for a network with $\langle k \rangle = 12.5$, and the bottom set of panels \emph{(b)} are for a network with $\langle k \rangle = 25$; the frequencies $\{\omega_{G}\}$ were normally distributed and the same for both cases. Each row corresponds to one oscillator, and the rows are ordered by the quantity $\tilde{\omega}_{i} = \omega_{i} - \langle \omega \rangle$ (i.e. the offset from the mean intrinsic frequency of the population). Adaptation of the network takes place between the two black lines. The results are qualitatively consistent with those described in the main text for the uniformly distributed frequencies (see the discussion in Sec.~\ref{s:time_dependence_uniform} and Fig.~\ref{f:freq_vs_time_allCouplingSubPlot_N100_T10_typeOmegauniform_spreadOmega2_kavg25_trial1_kavg125_trial1} for comparison).

\begin{figure}
\centering
\includegraphics[width=\columnwidth]{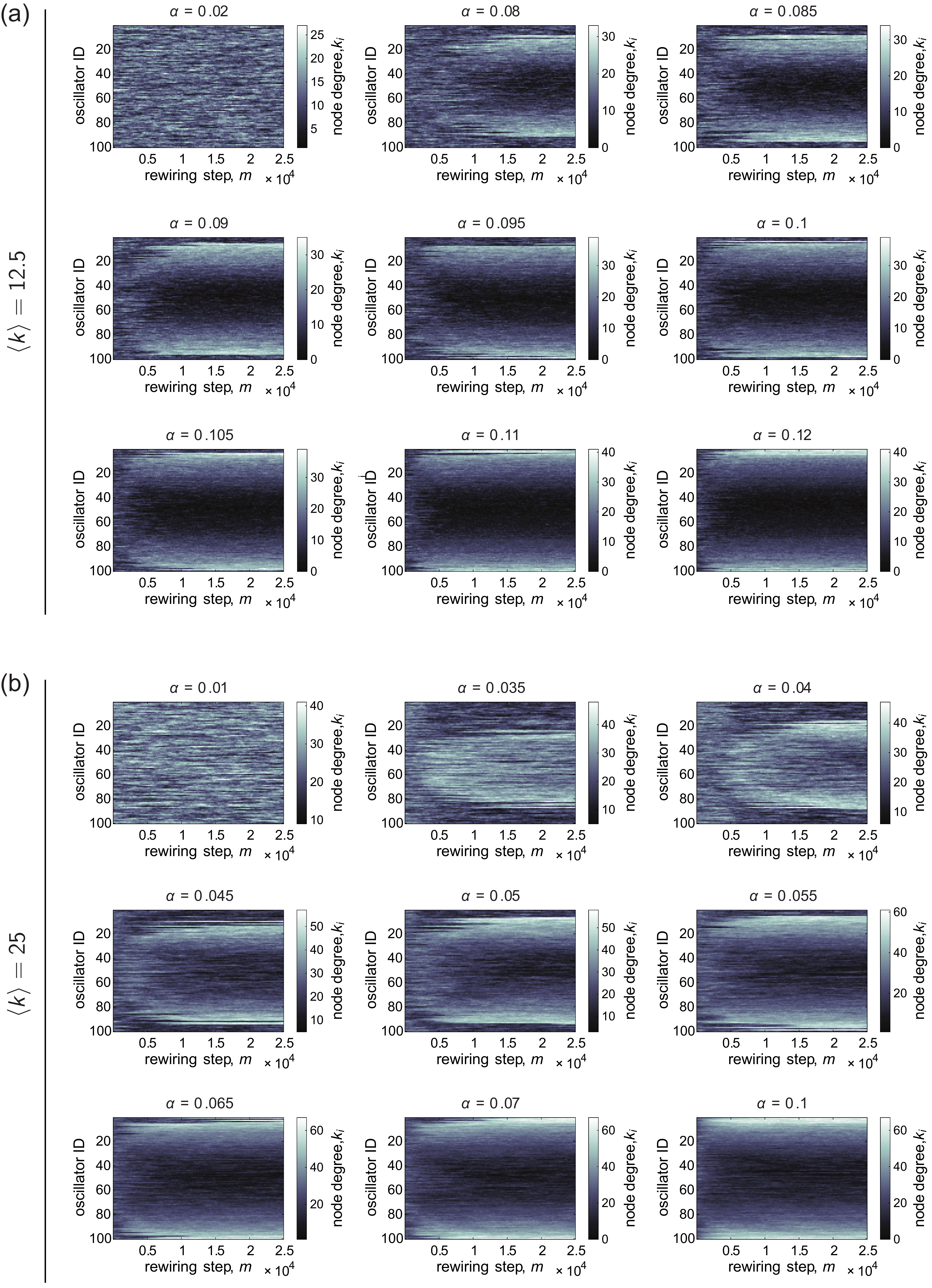}
\caption{Examples of the evolution of the node degree $k_{i}$ \emph{vs.} the rewiring step $m$, for various representative couplings $\alpha$. The mean degree of the networks are \emph{(a)} $\langle k \rangle = 12.5$ and \emph{(b)} $\langle k \rangle = 25$, and the natural frequencies $\{\omega_{G}\}$ were drawn from the normal distribution. In all panels, each row corresponds to one oscillator, and the rows are ordered by the quantity $\tilde{\omega}_{i} = \omega_{i} - \langle \omega \rangle$ (i.e. the offset from the mean intrinsic frequency of the population). The network was continually rewired once every $T = 0.2$ time units.}
\label{f:degree_vs_time_allCouplingSubPlot_N100_T10_typeOmeganormal_spreadOmega1_kavg25_trial8_kavg125_trial8}
\end{figure}

\begin{figure}
\centering
\includegraphics[width=\columnwidth]{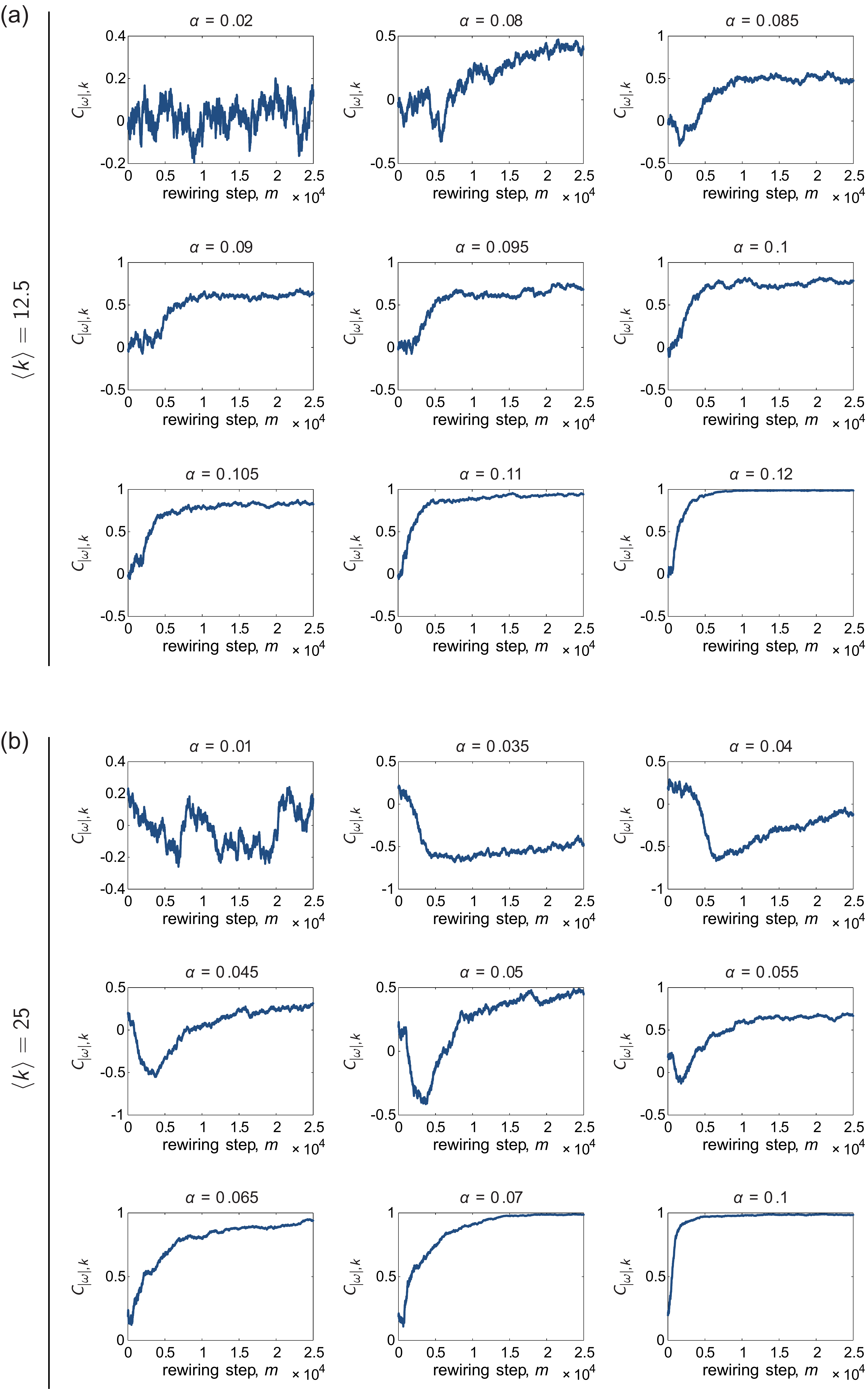}
\caption{Examples of the evolution of the correlation $C_{|\tilde{\omega}|, k}$ between the magnitude of the oscillator frequency offset $|\tilde{\omega}_{i}|$ and the node degree $k_{i}$ \emph{vs.} the rewiring step $m$, for various representative couplings $\alpha$. The mean degree of the networks are \emph{(a)} $\langle k \rangle = 12.5$ and \emph{(b)} $\langle k \rangle = 25$, and the natural frequencies $\{\omega_{G}\}$ were drawn from the normal distribution. In all panels, each row corresponds to one oscillator, and the rows are ordered by the quantity $\tilde{\omega}_{i} = \omega_{i} - \langle \omega \rangle$ (i.e. the offset from the mean intrinsic frequency of the population). The network was continually rewired once every $T = 0.2$ time units.}
\label{f:degree_freq_corr_vs_time_allCouplingSubPlot_N100_T10_typeOmeganormal_spreadOmega1_kavg25_trial8_kavg125_trial8}
\end{figure}

Fig.~\ref{f:degree_vs_time_allCouplingSubPlot_N100_T10_typeOmeganormal_spreadOmega1_kavg25_trial8_kavg125_trial8} shows the node degree $k_{i}$ \emph{vs.} the rewiring step $m$ and Fig.~\ref{f:degree_freq_corr_vs_time_allCouplingSubPlot_N100_T10_typeOmeganormal_spreadOmega1_kavg25_trial8_kavg125_trial8} shows the correlation $C_{|\tilde{\omega}|,k}$ between the absolute value of the frequency offset $|\tilde{\omega}_{i}|$ and the node degree $k_{i}$ \emph{vs.} the rewiring step $m$ for the same set of networks, natural frequencies, and coupling values as in Fig.~\ref{f:degree_vs_time_allCouplingSubPlot_N100_T10_typeOmeganormal_spreadOmega1_kavg25_trial8_kavg125_trial8}. In both figures, panel \emph{(a)} corresponds to a network with $\langle k \rangle = 12.5$, and panel \emph{(b)} corresponds to a network with $\langle k \rangle = 25$; the natural frequencies $\{\omega_{G}\}$ are normally distributed and are the same in both cases. We observe similar types of behavior and regimes in terms of the evolution of these quantities as we did for the situation of uniformly distributed frequencies in the main text (see the discussion in Sec.~\ref{s:time_dependence_uniform} and Figs.~\ref{f:degree_vs_time_allCouplingSubPlot_N100_T10_typeOmegauniform_spreadOmega2_kavg25_trial1_kavg125_trial1} and \ref{f:degree_freq_corr_vs_time_allCouplingSubPlot_N100_T10_typeOmegauniform_spreadOmega2_kavg25_trial1_kavg125_trial1} for comparison). One interesting result to point out for the examples using normally distributed frequencies shown here, is that the correlation $C_{|\tilde{\omega}|,k}$ plateaus at a value near 0.5 (for $\langle k \rangle = 12.5$) for several of the intermediate coupling values, before increasing further at higher coupling. As seen in Fig.~\ref{f:degree_vs_time_allCouplingSubPlot_N100_T10_typeOmeganormal_spreadOmega1_kavg25_trial8_kavg125_trial8}, this is due to the oscillators with the most outlying natural frequencies (which can sometimes be more extreme for the normal distribution than for the uniform distribution) remaining with low degree. Though the results are qualitatively similar between the two frequency distributions, further work is needed to quantify and understand what may be subtle and intricate differences.

\subsection{Spectral Analysis}
\label{a:spectral_analysis_normal}

\begin{figure}[h!]
\centering
\includegraphics[width=\columnwidth]{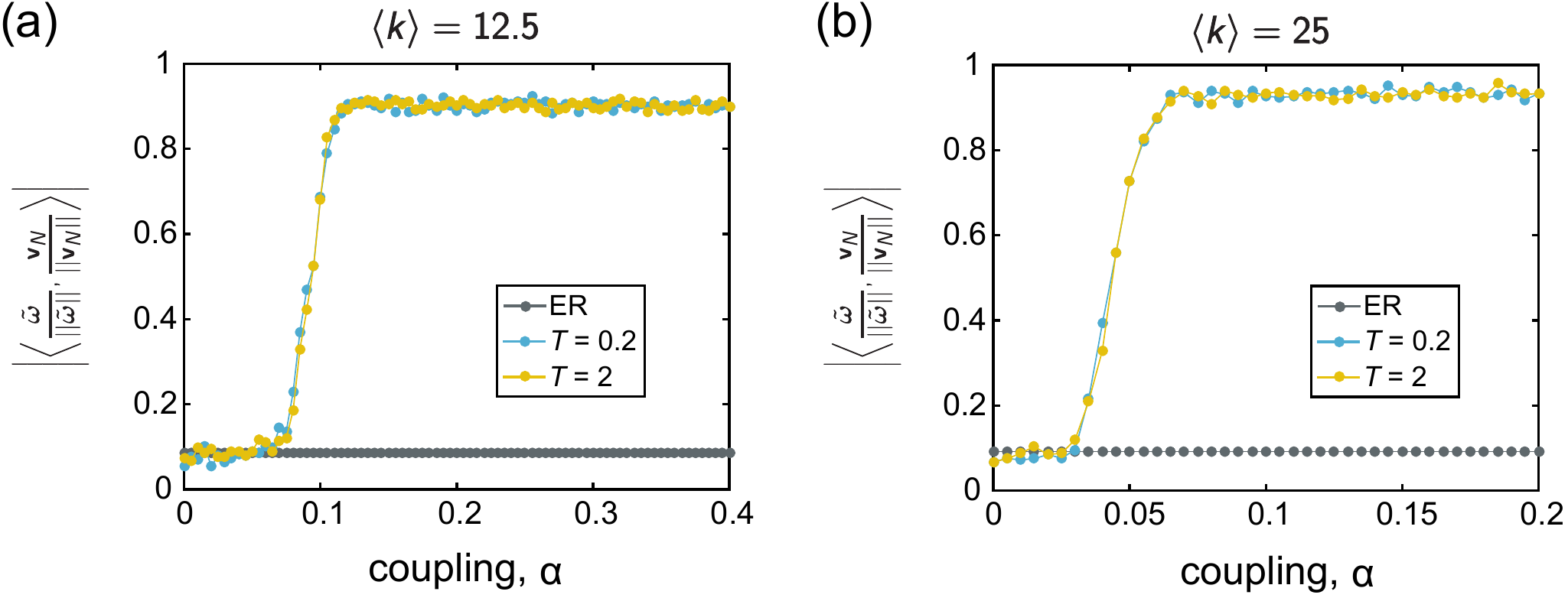}
\caption{Evolution of the overlap $| \langle \frac{\bm{\tilde{\omega}}}{||\bm{\tilde{\omega}}||} , \frac{\bm{v}_{N}}{||\bm{v}_{N}||} \rangle |$ between the intrinsic frequencies $\bm{\tilde{\omega}}$ and dominant Laplacian eigenvector $\bm{v}_{N}$ as a function of the coupling $\alpha$. In each panel, the gray data points correspond to the original, uncorrelated ER random graphs $\mathcal{G}_{o}$, and the blue and yellow curves correspond to the adapted networks $G_{\star}$ evolved under rewiring time scales of $T = 0.2$ and $T = 2$, respectively. The natural frequencies were drawn from the uniform distribution $\{\omega_{U}\}$, and the mean degree $\langle k \rangle$ of the networks are \emph{(a)} $\langle k \rangle = 12.5$ and \emph{(b)} $\langle k \rangle = 25$. All curves depict averages over 25 instantiations, and the lines between data points serve as guides for the eye.}
\label{f:spectral_analysis_normal}
\end{figure}

In Sec.~\ref{s:spectral_analysis_uniform} we carried out a spectral-based analysis of the co-evolved networks inspired by \cite{Skardal:2014a}. Fig.~\ref{f:spectral_analysis_normal} shows the results of this analysis for the case of normally distributed frequencies $\{\omega_{G}\}$, which are consistent with those previously discussed in Sec.~\ref{s:spectral_analysis_uniform} using uniformly distributed frequencies (compare to Fig.~\ref{f:spectral_analysis_uniform}). In short, for both values of the mean degree ($\langle k \rangle = 12.5$ and $\langle k \rangle = 25$) the overlap $| \langle \frac{\bm{\tilde{\omega}}}{||\bm{\tilde{\omega}}||} , \frac{\bm{v}_{N}}{||\bm{v}_{N}||} \rangle |$ between the normalized natural frequency offsets and the normalized dominant Laplacian eigenvector of the rewired networks increases and then plateaus at a high value as the coupling increases (Fig.~\ref{f:spectral_analysis_normal}\emph{a,b}). We refer the reader to Sec.~\ref{s:spectral_analysis_uniform} for a more detailed discussion of these findings.

\clearpage
\newpage

%


\end{document}


\title{Supplementary Material for ~\\~\\
``Development of structural correlations and synchronization from adaptive rewiring in networks of Kuramoto oscillators''}

\author{Lia Papadopoulos} 
\affiliation{Department of Physics and Astronomy, University of Pennsylvania, Pennsylvania, PA 19104, USA}
\author{Jason Kim}
\affiliation{Department of Bioengineering, University of Pennsylvania, Pennsylvania, PA 19104, USA}
\author{J{\"u}rgen Kurths}
\affiliation{Potsdam Institute for Climate Impact Research - Telegraphenberg A 31, 14473 Potsdam, Germany}
\affiliation{Department of Physics, Humboldt University Berlin - 12489 Berlin, Germany}
\affiliation{Institute for Complex Systems and Mathematical Biology, University of Aberdeen - Aberdeen AB24 3UE, UK}
\author{Danielle S. Bassett}
\affiliation{Department of Bioengineering, University of Pennsylvania, Pennsylvania, PA 19104, USA}
\affiliation{Department of Electrical \& Systems Engineering, University of Pennsylvania, Pennsylvania, PA 19104, USA}
\email{dsb@seas.upenn.edu}

\date{\today}

\maketitle

\section{Time-evolution of synchronized clusters}
\label{a:synchronized_clusters}

The order parameter $R$ is a global measure of phase synchronization, and when the coupling is high enough, we found that network adaptation could increase this quantity from a regime of low synchrony to a regime of intermediate to high synchrony (Sec. IV A, Figs. 2, 3, 13, 14). As a supplement to this measure, we can also quantify changes in the system during the adaptation process on a more local level by considering the time-evolution of synchronized clusters of oscillators. For this analysis, we follow \cite{Arenas:2006a}, who defined a local order parameter between pairs of oscillators as
\begin{equation}
\rho_{ij}(t) = \cos[\theta_{i}(t) - \theta{j}(t)],
\end{equation}
and then constructed a binarized phase-similarity matrix $\bm{\tilde{\rho}}(t)$ by setting entries less than or equal to some threshold $\Delta$ to zero and by setting entries greater than $\Delta$ to $1$. Specifically,
\begin{equation}
\begin{split}
\tilde{\rho}_{ij}(t) = \left\{ \begin{array}{ll}
        0  & \mbox{if $\rho_{ij}(t)\leq \Delta$} \\
        1  & \mbox{if $\rho_{ij}(t) > \Delta$}. \end{array} \right. 
 \end{split}
\label{eq:sync_matrix_threshold}
\end{equation}
We use a value of $\Delta = 0.999$. The binarized matrix $\bm{\tilde{\rho}}(t)$ contains information about which pairs of oscillators are strongly synchronized at a given time, and synchronized \textit{clusters} or groups of oscillators can be extracted by finding the connected components of $\bm{\tilde{\rho}}(t)$. Finally, we can track the number of synchronized components $n_{sc}(t)$ as a function of time $t$, and observe how the behavior in this quantity changes when the network begins to co-evolve with the dynamics. For low values of the coupling - where the adaptive strategy does not have a large influence on the dynamics - we expect $n_{sc}(t)$ to remain high as a function of time. However when the coupling is increased and the rewiring causes consistent changes in the network and the dynamics, we expect a decrease in the number of synchronized components as the network evolves from a configuration that does not support collective dynamics to one that does. An interesting question is whether $n_{sc}(t)$ decreases in abrupt, sudden bursts, or smoothly over time, and how this behavior depends on the magnitude of the overall coupling $\alpha$.

\begin{figure}
\centering
\includegraphics[width=\columnwidth]{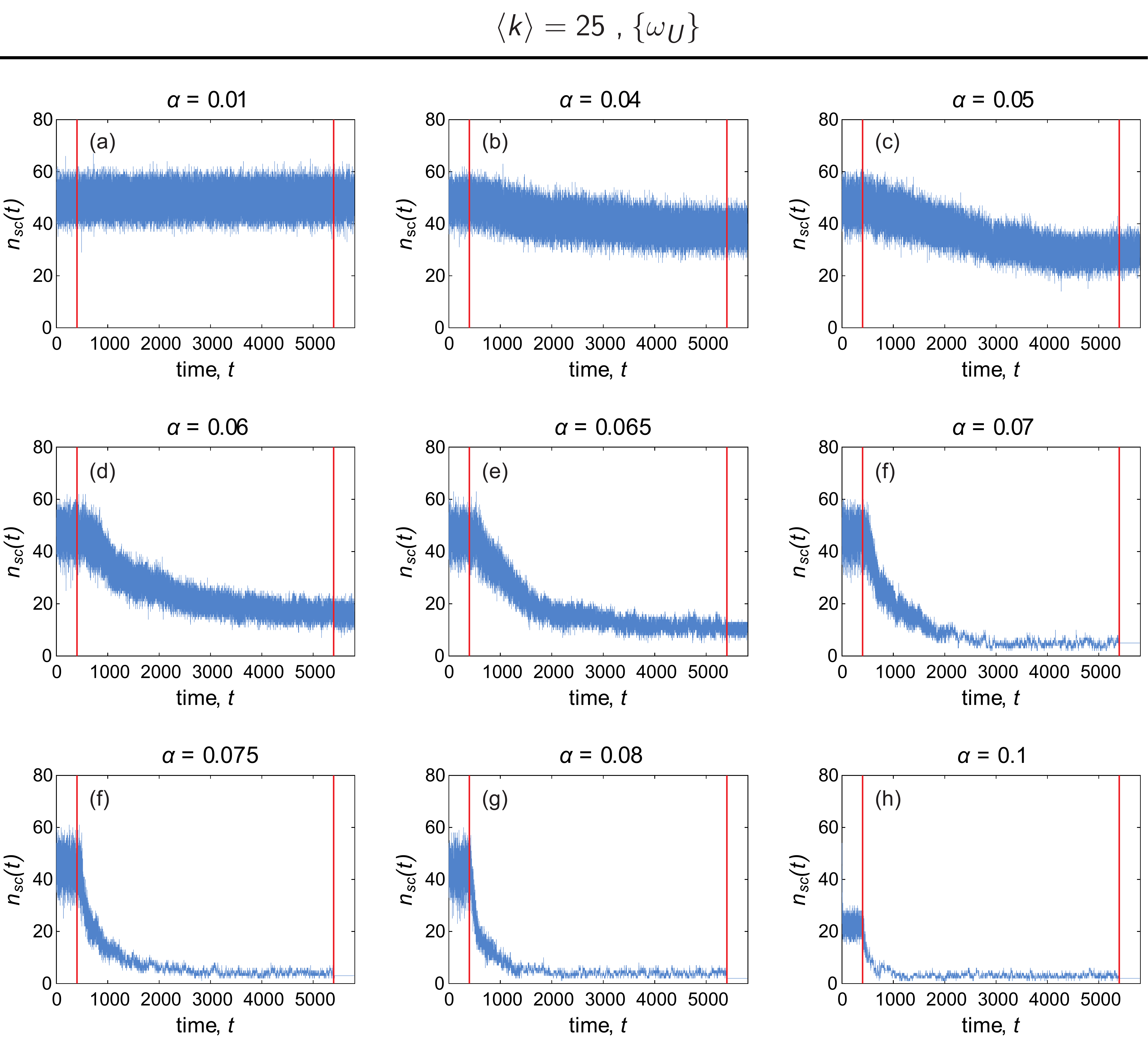}
\caption{Examples of the number of synchronized clusters $n_{sc}(t)$ \emph{vs.} time $t$, for various representative couplings $\alpha$. In each case, the dynamics were first run atop an ER random graph, after which co-evolution of the network and dynamics took place between the two red lines. For these plots, the natural frequencies were drawn from the uniform distribution $\{ \omega_{U} \}$ and the mean degree of the network was $\langle k \rangle = 25$. During the adaptation period, the network was continually rewired once every $T = 0.2$ time units. For high enough coupling, co-evolution results in a decrease in the number of synchronized clusters over time.}
\label{f:Nsc_vs_time_allCouplingSubPlot_syncThresh0999_N100_avgDegree25_T10_typeOmegauniform_spreadOmega2}
\end{figure}

\begin{figure}
\centering
\includegraphics[width=\columnwidth]{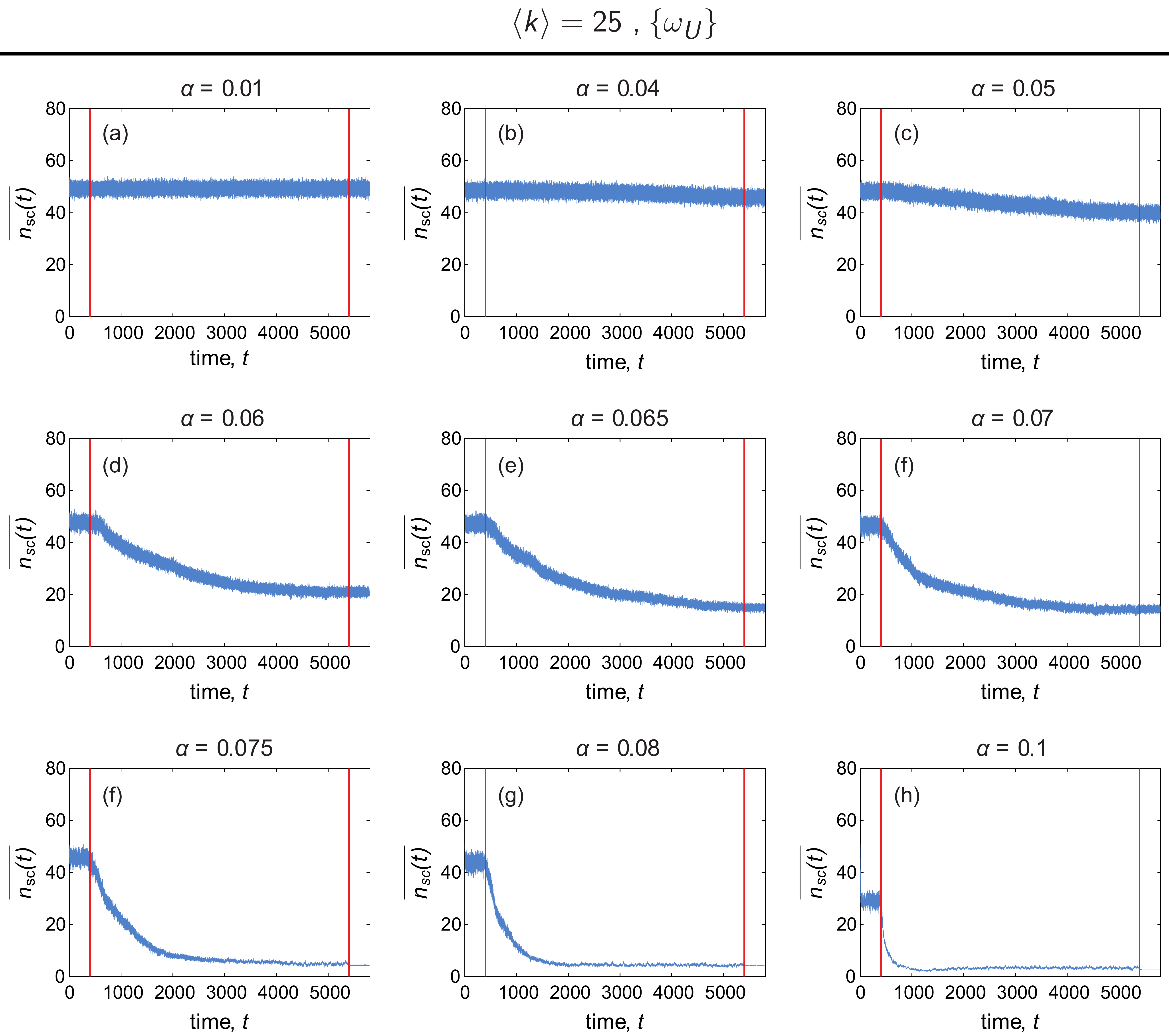}
\caption{Ensemble averages of the number of synchronized clusters $\overline{n_{sc}(t)}$ \emph{vs.} time $t$, for various representative couplings $\alpha$. In each case, the dynamics were first run atop an ER random graph, after which co-evolution of the network and dynamics took place between the two red lines. For these plots, the natural frequencies were drawn from the uniform distribution $\{ \omega_{U} \}$, the mean degree of the network was $\langle k \rangle = 25$, and each curve corresponds to an average over 10 different simulations. During the adaptation period, the network was continually rewired once every $T = 0.2$ time units. For high enough coupling, co-evolution results in a decrease in the number of synchronized clusters over time.}
\label{f:avgNsc_vs_time_allCouplingSubPlot_syncThresh0999_N100_avgDegree25_T10_typeOmegauniform_spreadOmega2}
\end{figure}

Fig.~S\ref{f:Nsc_vs_time_allCouplingSubPlot_syncThresh0999_N100_avgDegree25_T10_typeOmegauniform_spreadOmega2} shows examples of $n_{sc}(t)$ \emph{vs.} $t$ for a single simulation of the dynamics, and for various values of the coupling $\alpha$ (the mean degree is $\langle k \rangle = 25$ and the natural frequencies $\{ \omega_{U} \}$ are uniformly distributed). We see that the number of synchronized clusters is initially large, since the oscillators' phases are distributed at random at $t = 0$. Before the red line, the system evolves on a static ER network, and depending on the coupling strength, the number of components either fluctuates around a high value or - after a short initial transient - rapidly decreases to a lower value, denoting partial synchronization. Network adaptation begins at the time denoted by the red line, after which the behavior of $n_{sc}(t)$ is again determined by the global coupling. For very low $\alpha$, the co-evolution process does not significantly affect the number of synchronized clusters. However, as the coupling increases, the network rewiring begins to have an impact on the dynamical clustering. At intermediate values of $\alpha$, the number of synchronized components slowly decreases as a function of time. For higher coupling values, this transition takes place more rapidly. However, it appears that no matter the coupling regime, the transition from high to lower $n_{sc}$ does not occur abruptly at a single instance of time, but instead occurs over several time steps through the merging of clusters. In addition to the curves depicting individual examples of the number of synchronized clusters \emph{vs.} time, Fig.~S\ref{f:avgNsc_vs_time_allCouplingSubPlot_syncThresh0999_N100_avgDegree25_T10_typeOmegauniform_spreadOmega2} also shows averages of $n_{sc}(t)$ \emph{vs.} $t$ over 10 simulations using different instantiations of the initial conditions.

\begin{figure}
\centering
\includegraphics[width=\columnwidth]{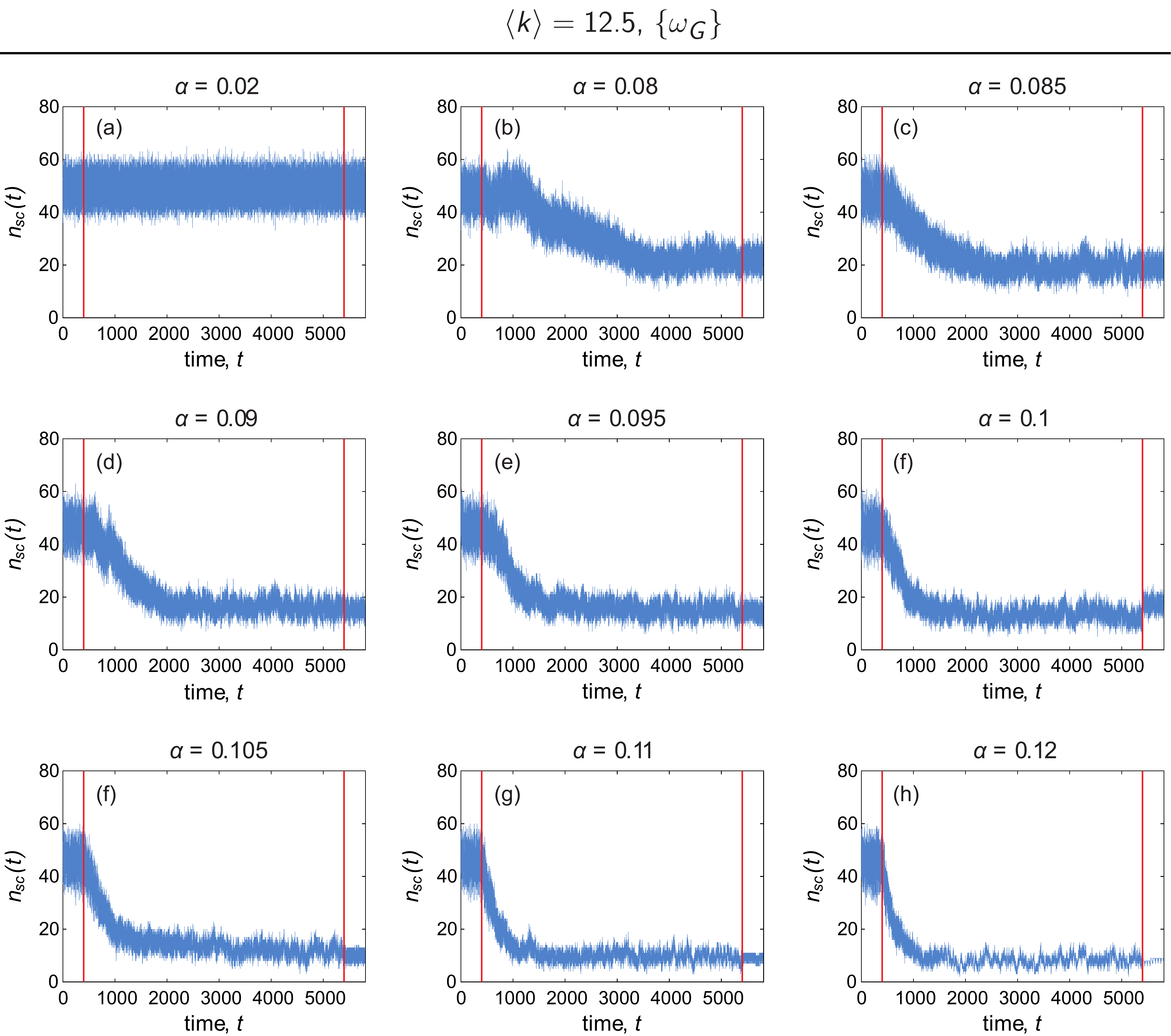}
\caption{Examples of the number of synchronized clusters $n_{sc}(t)$ \emph{vs.} time $t$, for various representative couplings $\alpha$. In each case, the dynamics were first run atop an ER random graph, after which co-evolution of the network and dynamics took place between the two red lines. For these plots, the natural frequencies were drawn from the normal distribution $\{ \omega_{G} \}$ and the mean degree of the network was $\langle k \rangle = 12.5$. During the adaptation period, the network was continually rewired once every $T = 0.2$ time units. For high enough coupling, co-evolution results in a decrease in the number of synchronized clusters over time.}
\label{f:Nsc_vs_time_allCouplingSubPlot_syncThresh0999_N100_avgDegree125_T10_typeOmeganormal_spreadOmega1}
\end{figure}

\begin{figure}
\centering
\includegraphics[width=\columnwidth]{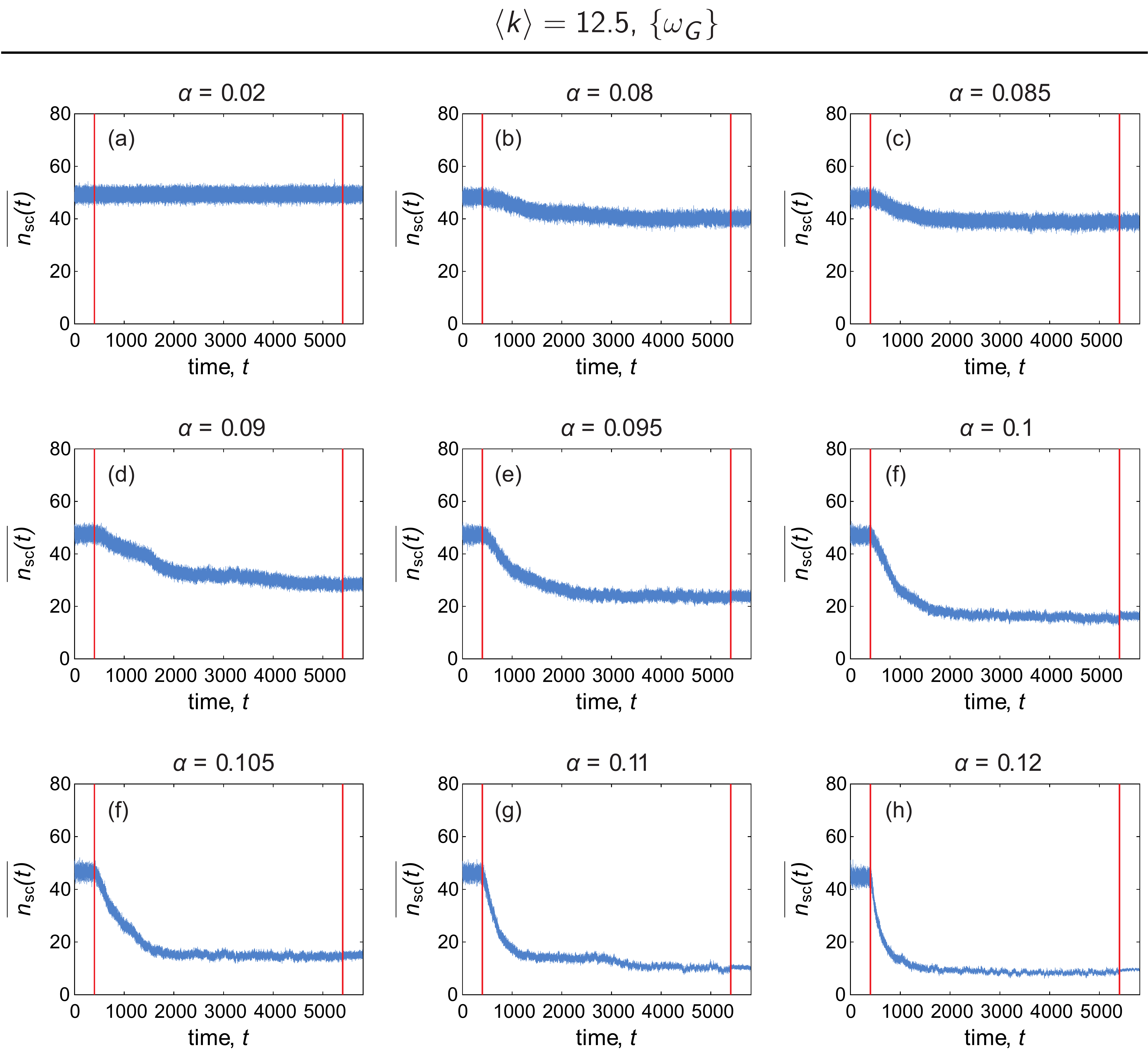}
\caption{Ensemble averages of the number of synchronized clusters $\overline{n_{sc}(t)}$ \emph{vs.} time $t$, for various representative couplings $\alpha$. In each case, the dynamics were first run atop an ER random graph, after which co-evolution of the network and dynamics took place between the two red lines. For these plots, the natural frequencies were drawn from the normal distribution $\{ \omega_{G} \}$, the mean degree of the network was $\langle k \rangle = 12.5$, and each curve corresponds to an average over 10 different simulations. During the adaptation period, the network was continually rewired once every $T = 0.2$ time units. For high enough coupling, co-evolution results in a decrease in the number of synchronized clusters over time.}
\label{f:avgNsc_vs_time_allCouplingSubPlot_syncThresh0999_N100_avgDegree125_T10_typeOmeganormal_spreadOmega1}
\end{figure}

Fig.~S\ref{f:Nsc_vs_time_allCouplingSubPlot_syncThresh0999_N100_avgDegree125_T10_typeOmeganormal_spreadOmega1} and Fig.~S\ref{f:avgNsc_vs_time_allCouplingSubPlot_syncThresh0999_N100_avgDegree125_T10_typeOmeganormal_spreadOmega1} show $n_{sc}(t)$ \emph{vs.} time $t$ and the averages $\overline{n_{sc}(t)}$ \emph{vs.} $t$, respectively, for the case of mean degree $\langle k \rangle = 12.5$ and normally distributed natural frequencies $\{ \omega_{G} \}$. We find similar results to those for the denser networks. At low values of the coupling, network adaptation does not cause a consistent change in $n_{sc}$. As the coupling increases, however, the number of components begins to decrease from its initial value, signifying the merging or growth of synchronized clusters due to rearrangements in the network. This decrease becomes more rapid at larger couplings, though does not occur discontinuously as a function of time. 

\section{Variation of network size, initial network topology, and frequency distribution}
\label{s:robustness_check}

In the main text (and Appendix), we fixed the system size at $N = 100$ and initialized the coupling between oscillators to be ER random graphs with mean degree $\langle k \rangle = 25$ or $\langle k \rangle = 12.5$. We then examined two symmetric distributions (uniform $\{\omega_{U}\}$ and normal $\{\omega_{G}\}$) for the intrinsic frequencies of the oscillators. In this section, we explore the robustness of some of the main results to simple variations in the network size, the initial network topology, and the frequency distribution.

\subsection{Larger networks}
\label{s:larger_networks}

\begin{figure}[h]
\centering
\includegraphics[width=\columnwidth]{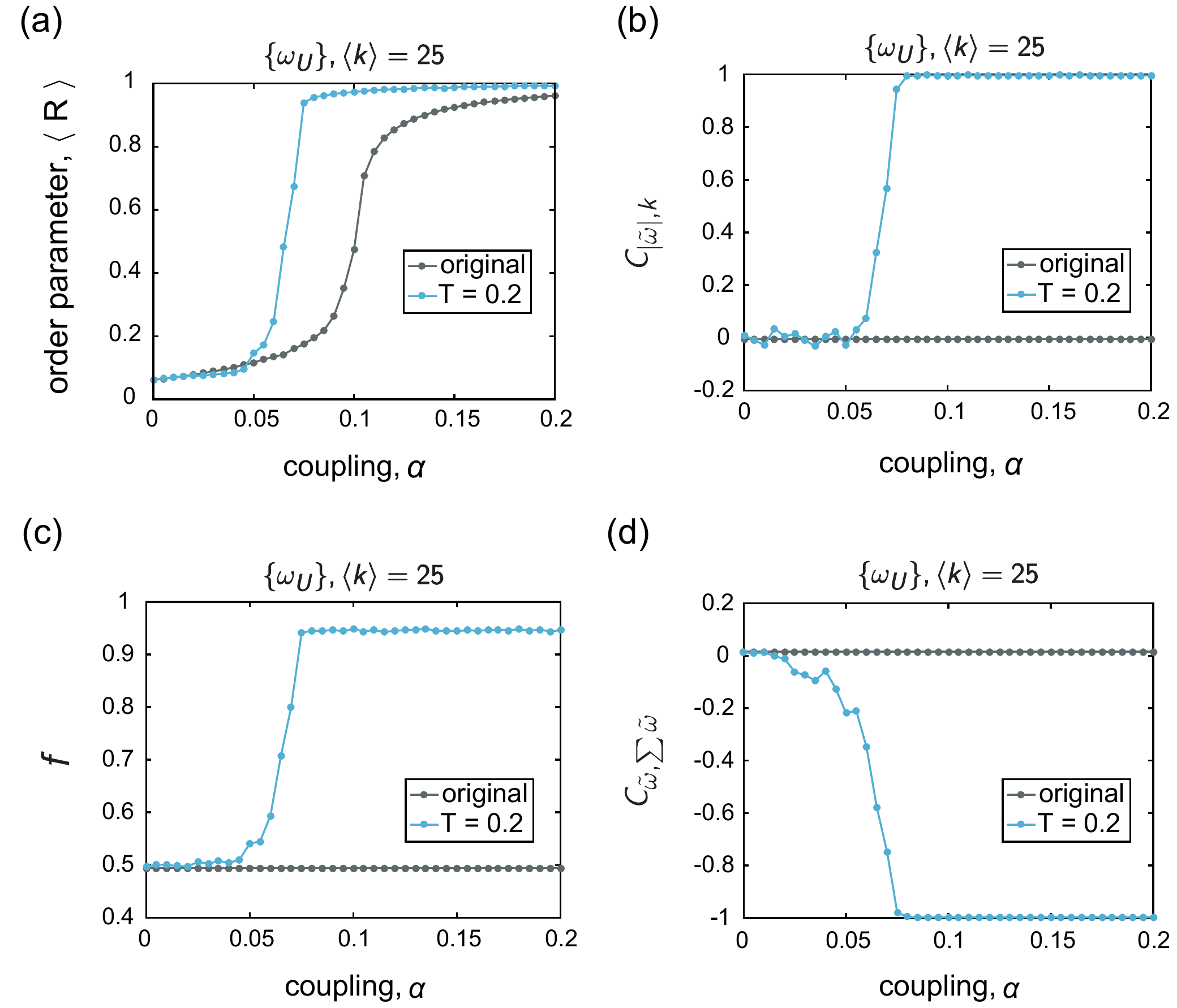}
\caption{Analysis of the synchronization and various correlations between the network topology and the intrinsic frequencies for networks of size $N = 200$. In each panel, the gray data points correspond to the initial ER networks $\mathcal{G}_{o}$ with mean degree $\langle k \rangle = 25$, and the blue points correspond to the adapted networks $G_{\star}$ evolved under a rewiring time scale of $T = 0.2$ time units. The natural frequencies were drawn from the uniform distribution $\{ \omega_{U} \}$. \emph{(a)} The time-averaged order parameter \emph{vs.} coupling. \emph{(b)} The correlation $C_{|\tilde{\omega}|, k}$ between node degree $k_{i}$ and the magnitude of the frequency offset $|\tilde{\omega}_{i}|$, as a function of the coupling. \emph{(c)} The mean fraction $f$ of an oscillator's neighbors that have frequency offsets of opposite sign compared to the central oscillator, as a function of the coupling. \emph{(d)} The correlation $C_{\tilde{\omega},\sum\tilde{\omega}}$ between oscillator frequency offset $\tilde{\omega}_{i}$ and the sum of neighbor frequency offsets $\sum\tilde{\omega}_{j}$, as a function of the coupling. All curves depict averages over 10 instantiations, and the lines between data points serve as guides for the eye.}
\label{f:N200_kavg25_randNetwork_freqUniform}
\end{figure}

We first investigate the effect of the co-evolutionary mechanism on the degree of synchronization in the system when the network is doubled in size to $N = 200$ (while also proportionately doubling the number of rewirings allowed for the adaptation process). Fig.~S\ref{f:N200_kavg25_randNetwork_freqUniform}\emph{a} shows the order parameter $\langle R \rangle$ \emph{vs.} coupling $\alpha$ for the case of uniformly distributed frequencies $\{ \omega_{U} \}$, mean degree $\langle k \rangle = 25$, and a rewiring time scale of $T = 0.2$ time units. We find qualitatively similar results to that of $N = 100$ (Fig. 3\emph{b} of the main text), in that the curve corresponding to the adapted networks (blue) is shifted to the left compared to the curve corresponding to a set of static, random ER networks (gray). Thus, over a large range of global coupling, the co-evolved networks exhibit a heightened degree of collective dynamics.

In addition to the order parameter, we also assessed the co-evolved networks $\mathcal{G}_{\star}$ for the emergence of correlations between their topology and the intrinsic frequencies of the oscillators (as done in Sec. IV B and Appendix A of the main text). The results of this analysis are shown in Fig.~S\ref{f:N200_kavg25_randNetwork_freqUniform}\emph{(b - d)}. As found previously for the case of $N=100$ (Fig. 5\emph{b},  6\emph{b}, and 12\emph{b}), we find that the same set of relationships appear when the system size is doubled. Following the analysis described in the main text, for each node $i$, we computed its offset from the mean frequency $\tilde{\omega}_{i} = \omega_{i} - \langle \omega \rangle$, where $\langle \omega \rangle$ is the mean natural frequency of the population. We then considered the correlations between $|\tilde{\omega}_{i}|$ and node degree $k_{i}$ as a function of the coupling (Fig.~S\ref{f:N200_kavg25_randNetwork_freqUniform}\emph{b}), and between $\tilde{\omega}_{i}$ and the total neighbor frequency offset $\sum{\tilde{\omega}_{j}}$ ($\forall j \in \mathcal{N}(i)$ as a function of the coupling (Fig.~S\ref{f:N200_kavg25_randNetwork_freqUniform}\emph{d}). In addition, for each oscillator $i$, we computed the fraction $f_{i}$ of $i's$ neighbors that have a value of $\tilde{\omega}$ of opposite sign to that of $\tilde{\omega}_{i}$, and then computed the average $f$ over all nodes in the network (Fig.~S\ref{f:N200_kavg25_randNetwork_freqUniform}\emph{c}). As the coupling increases, there is an increasingly positive correlation $C_{|\tilde{\omega}|, k}$ between node degree and the magnitude of the intrinsic frequency offset, and an increasingly negative correlation $C_{\tilde{\omega},\sum\tilde{\omega}}$ between the intrinsic frequency offset of each oscillator and the sum of its neighbors' frequency offsets. Furthermore, oscillators tend to become connected to other oscillators that have intrinsic frequency offsets on the opposite side of the mean. Although an in-depth examination into the effect of system size is beyond the scope of this study, it is important that the results are consistent for at least somewhat larger networks. We refer the reader to \cite{Brede:2010a} for a detailed finite-size study and analysis of the critical coupling and critical exponents in oscillator populations with purposefully constructed correlations.

\subsection{Smaller mean degree}

\begin{figure}[h]
\centering
\includegraphics[width=\columnwidth]{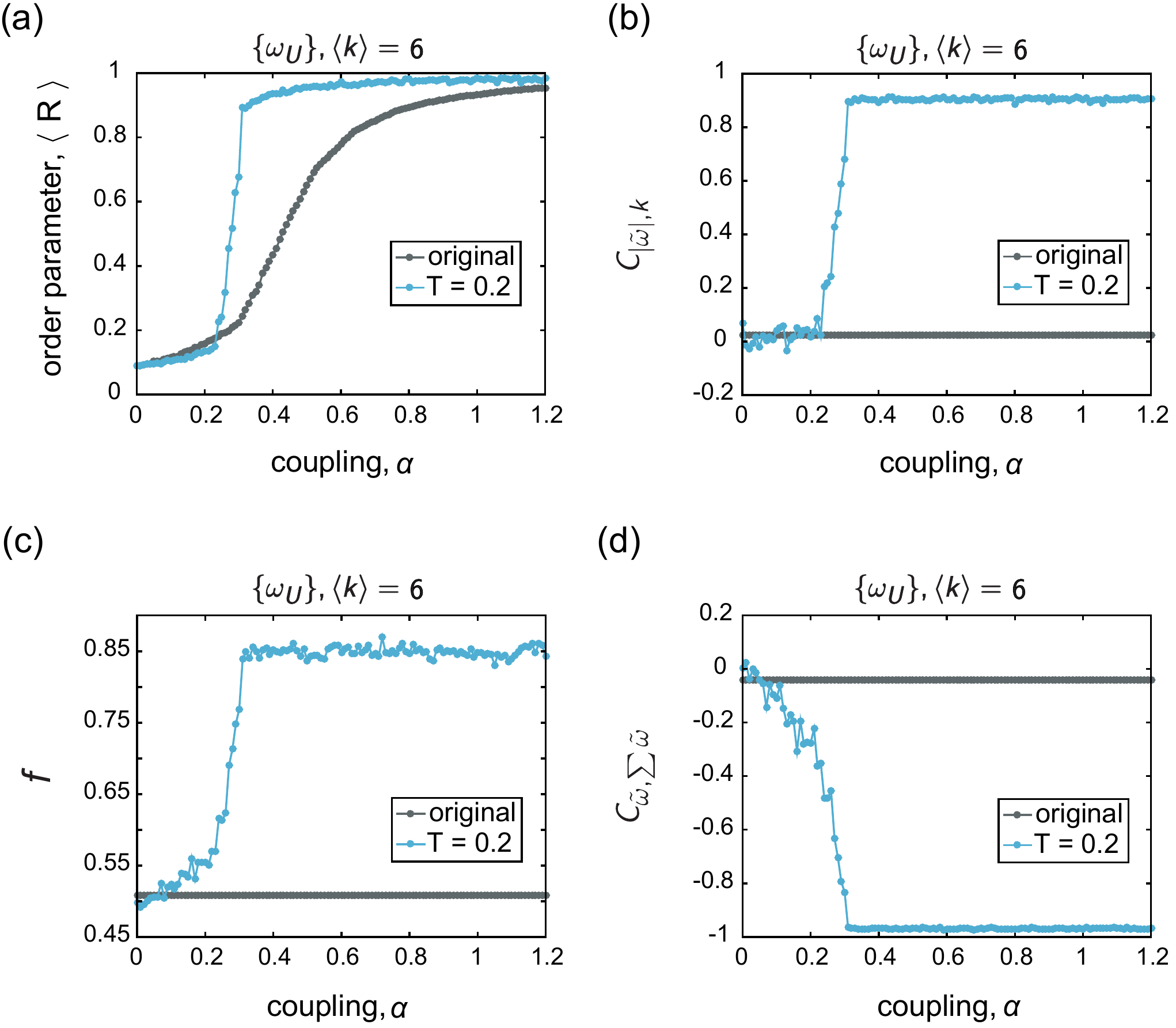}
\caption{Analysis of the synchronization and various correlations between the network topology and the intrinsic frequencies for networks with sparser connectivity. In each panel, the gray data points correspond to the initial ER networks $\mathcal{G}_{o}$ with mean degree $\langle k \rangle = 6$, and the blue points correspond to the adapted networks $G_{\star}$ evolved under a rewiring time scale of $T = 0.2$ time units. The natural frequencies were drawn from the uniform distribution $\{ \omega_{U} \}$. \emph{(a)} The time-averaged order parameter \emph{vs.} coupling. \emph{(b)} The correlation $C_{|\tilde{\omega}|, k}$ between node degree $k_{i}$ and the magnitude of the frequency offset $|\tilde{\omega}_{i}|$, as a function of the coupling. \emph{(c)} The mean fraction $f$ of an oscillator's neighbors that have frequency offsets of opposite sign compared to the central oscillator, as a function of the coupling. \emph{(d)} The correlation $C_{\tilde{\omega},\sum\tilde{\omega}}$ between oscillator frequency offset $\tilde{\omega}_{i}$ and the sum of neighbor frequency offsets $\sum\tilde{\omega}_{j}$, as a function of the coupling. All curves depict averages over 10 instantiations, and the lines between data points serve as guides for the eye.}
\label{f:N100_kavg6_randNetwork_freqUniform}
\end{figure}

This study has focused on networks with relatively large mean degree (yielding density values similar to that of large-scale anatomical brain networks, for example \cite{bassett2011conserved}, but many other real-world systems - such as social networks \cite{Newman:2003aa} - are more sparsely connected. Here, we thus briefly examine whether various results hold for networks with a lower mean degree of $\langle k \rangle = 6$. Fig.~S\ref{f:N100_kavg6_randNetwork_freqUniform}\emph{a} shows the time-averaged order parameter $\langle R \rangle$ \emph{vs.} coupling $\alpha$ for the static ER networks (gray curve) and the adapted networks (blue curve). The frequencies $\{ \omega_{U} \}$ were uniformly distributed the rewiring time scale was $T = 0.2$ time units. As for the denser networks considered in the main text (Fig. 3), here we also observe a significant enhancement of the order parameter in the co-evolved systems.

We also examined the locally adapted systems for the previously described relationships between the natural frequencies and the topological organization of the network. Fig.~S\ref{f:N100_kavg6_randNetwork_freqUniform} panels \emph{(b-d)} show $C_{|\tilde{\omega}|, k}$, $f$, and $C_{\tilde{\omega},\sum\tilde{\omega}}$, respectively, as a function of the coupling. Again we find that as the coupling increases, there is an increasingly positive correlation between degree and the magnitude of intrinsic frequency offset, and an increasingly negative correlation between the intrinsic frequency offset of an oscillator and the sum of its neighbors' frequency offsets. Furthermore, oscillators tend to become connected to other oscillators with frequency offsets on the opposite side of the mean. Although the curves shown here for the case of $\langle k \rangle = 6$ exhibit somewhat more fluctuations compared to the case of $\langle k \rangle = 25$, the findings are in general robust and consistent with those found for networks with higher mean degree (compare to Figs. 5, 6, and 12 of the main text).

\subsection{Initial network topology}

\begin{figure}[h]
\centering
\includegraphics[width=0.75\columnwidth]{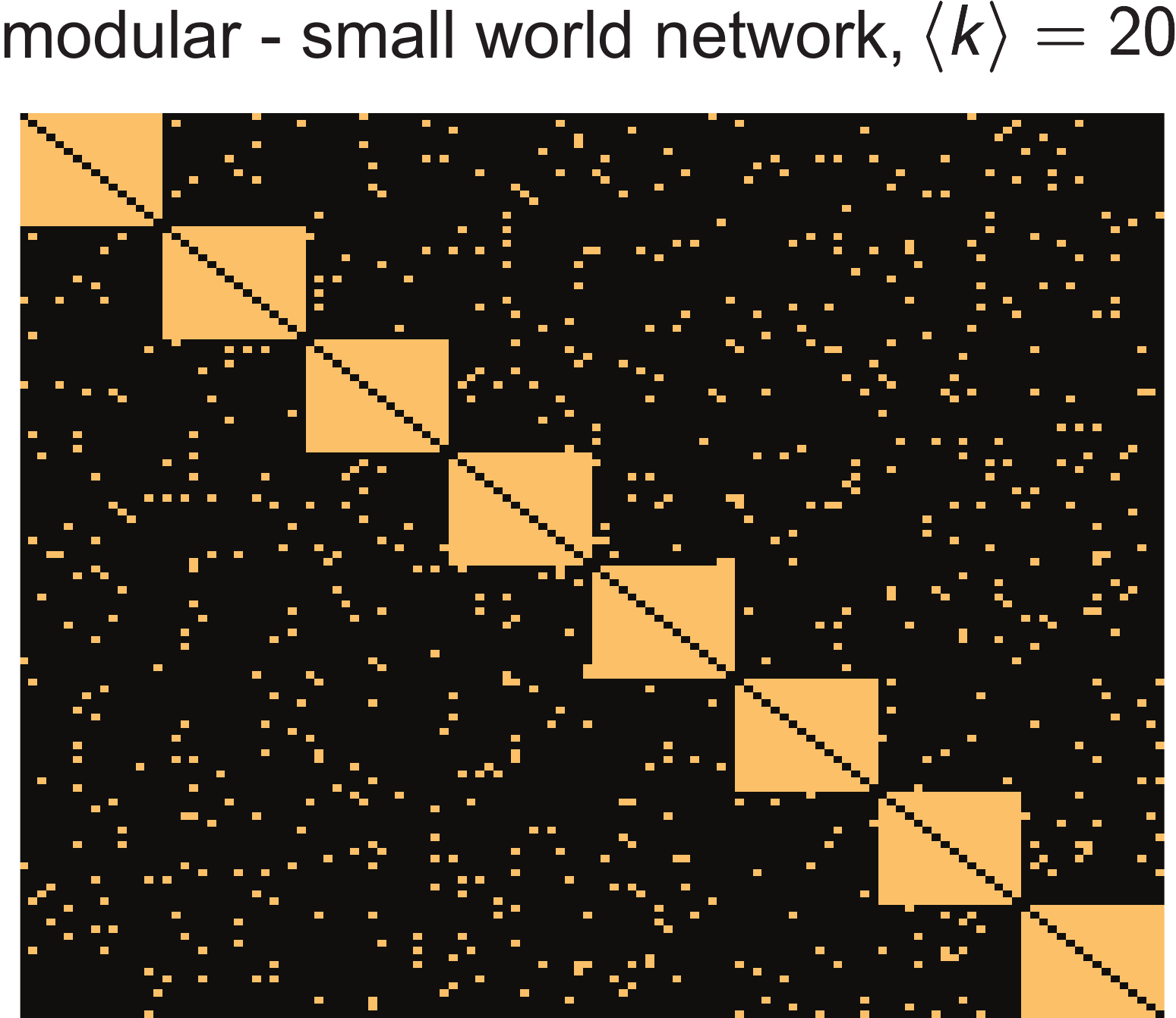}
\caption{An example connection pattern of a modular - small world (MSW) network of size $N = 128$, with 8 communities and a mean degree $\langle k \rangle = 20$.}
\label{f:modsw_network_example}
\end{figure}

The main interest and goal of this paper has been to define and explore a simple, local mechanism that evolves some initial network of coupled oscillators to a structured configuration that in turn supports enhanced collective dynamics. Given this objective, ER random graphs - which are considered to be relatively ``disordered" - are a natural family of networks on which to begin the co-evolution and also to compare the adapted systems against. However, it may still be useful to examine if the results hold for a different initial network topology. To test this, we constructed a set of modular - small world (MSW) networks, which consist of a specified number of fully connected communities that are then linked together by placing edges at random between the modules. In contrast to the ER random graph model, these networks contain a significant amount of planted structure. For this analysis, we used networks of size $N = 128$, with $N_{c} = 8$ modules, and a total number of edges such that the average degree was $\langle k \rangle = 20$. Fig.~S\ref{f:modsw_network_example} shows an example connection pattern.

\begin{figure}[h]
\centering
\includegraphics[width=\columnwidth]{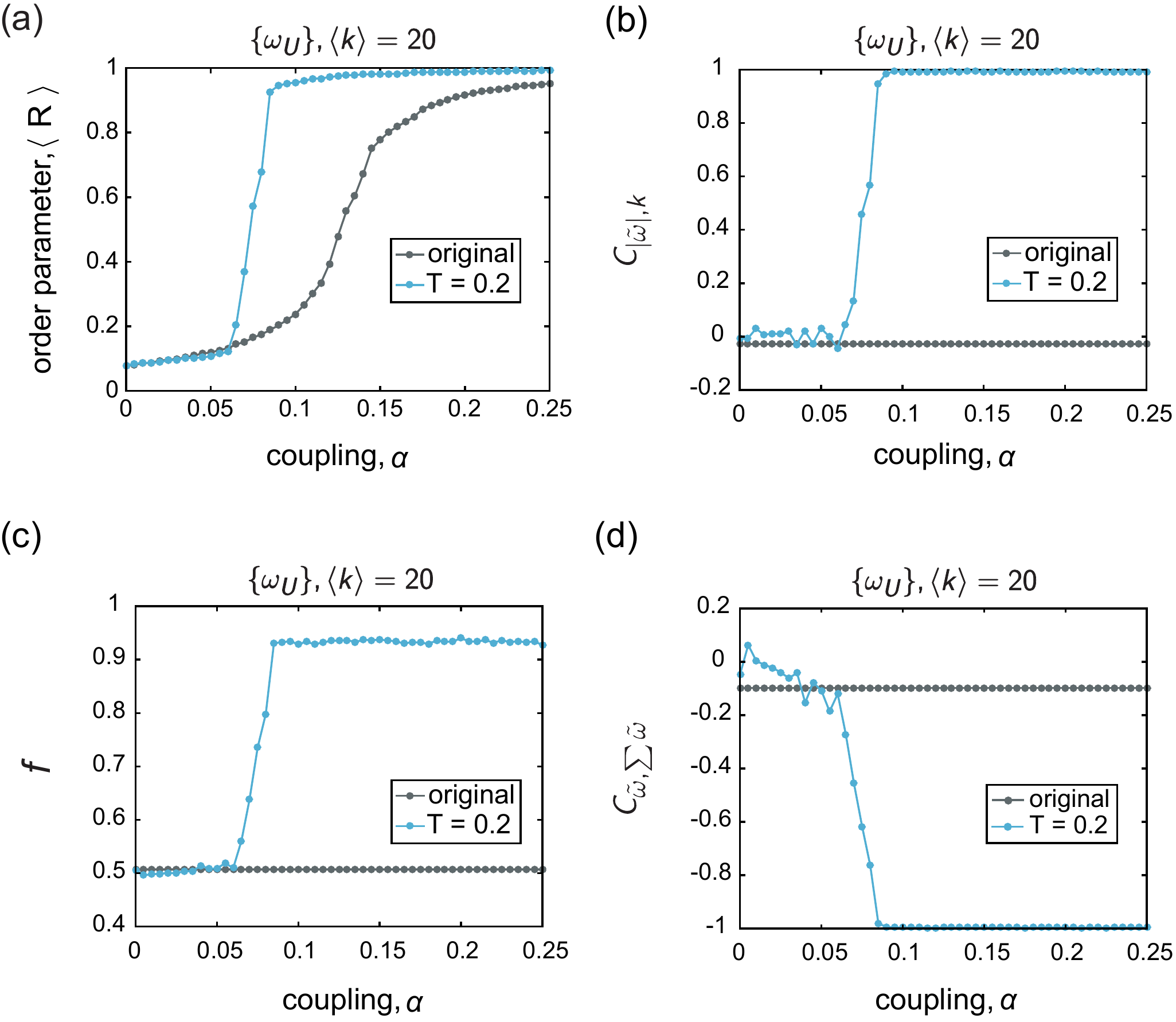}
\caption{Analysis of the synchronization and various correlations between the network topology and the intrinsic frequencies when the initial network is modular-small world, rather than a random graph. In each panel, the gray data points correspond to the initial MSW networks $\mathcal{G}_{o}$, and the blue points correspond to the adapted networks $G_{\star}$ evolved under a rewiring time scale of $T = 0.2$ time units. The natural frequencies were drawn from the uniform distribution $\{ \omega_{U} \}$. \emph{(a)} The time-averaged order parameter \emph{vs.} coupling. \emph{(b)} The correlation $C_{|\tilde{\omega}|, k}$ between node degree $k_{i}$ and the magnitude of the frequency offset $|\tilde{\omega}_{i}|$, as a function of the coupling. \emph{(c)} The mean fraction $f$ of an oscillator's neighbors that have frequency offsets of opposite sign compared to the central oscillator, as a function of the coupling. \emph{(d)} The correlation $C_{\tilde{\omega},\sum\tilde{\omega}}$ between oscillator frequency offset $\tilde{\omega}_{i}$ and the sum of neighbor frequency offsets $\sum\tilde{\omega}_{j}$, as a function of the coupling. All curves depict averages over 10 instantiations, and the lines between data points serve as guides for the eye.}
\label{f:N128_nc8_kavg20_modsw_freqUniform}
\end{figure}

Due to the nature of the adaptive mechanism, we expect that it will undo the seeded community structure and evolve this class of networks towards topologies with the same set of features we have discussed previously. To test this expectation, we first simulated the Kuramoto dynamics on non-adaptive MSW networks, and then used the same set of MSW networks as the initial network configurations for the adaptive mechanism. Fig.~S\ref{f:N128_nc8_kavg20_modsw_freqUniform}\emph{a} shows the time-averaged order parameter $\langle R \rangle$ \emph{vs.} coupling $\alpha$ for the case of uniformly distributed frequencies $\{ \omega_{U} \}$, and a rewiring time scale of $T = 0.2$ time units. The adaptively rewired networks (blue curve) show heightened global synchronization across a large range of coupling values compared to the static MSW networks (gray curve). Turning next to the correlations between the natural frequencies and the network structure, we find results consistent with those described in the main text. In particular, strong degree - frequency correlations and frequency - neighbor frequency mixing emerge in the self-organized systems (see Fig.~S\ref{f:N128_nc8_kavg20_modsw_freqUniform}\emph{b - d} for plots of $C_{|\tilde{\omega}|, k}$, $f$, and $C_{\tilde{\omega},\sum\tilde{\omega}}$ \emph{vs.} the coupling $\alpha$). This investigation provides some evidence that the adaptive mechanism is agnostic to the initial network configuration from which it begins.

\subsection{Natural frequency distribution}

One can also explore in more depth how the distribution of natural frequencies affects certain results. In this study, we considered two standard distributions - uniform and normal - that are often used in studies on the Kuramoto model. Here, we also show some brief results for the case of natural frequencies $\{\omega_{P}\}$ drawn from a power-law distribution in which $P({\omega}) \propto \omega^{-\gamma}$, with $\gamma = 3$. The dynamics and network co-evolution start from ER random graphs with $N = 200$ nodes and mean degree $\langle k \rangle = 25$. We find qualitatively similar results to those in the main text when using this skewed distribution for the intrinsic frequencies.

\begin{figure}[h]
\centering
\includegraphics[width=\columnwidth]{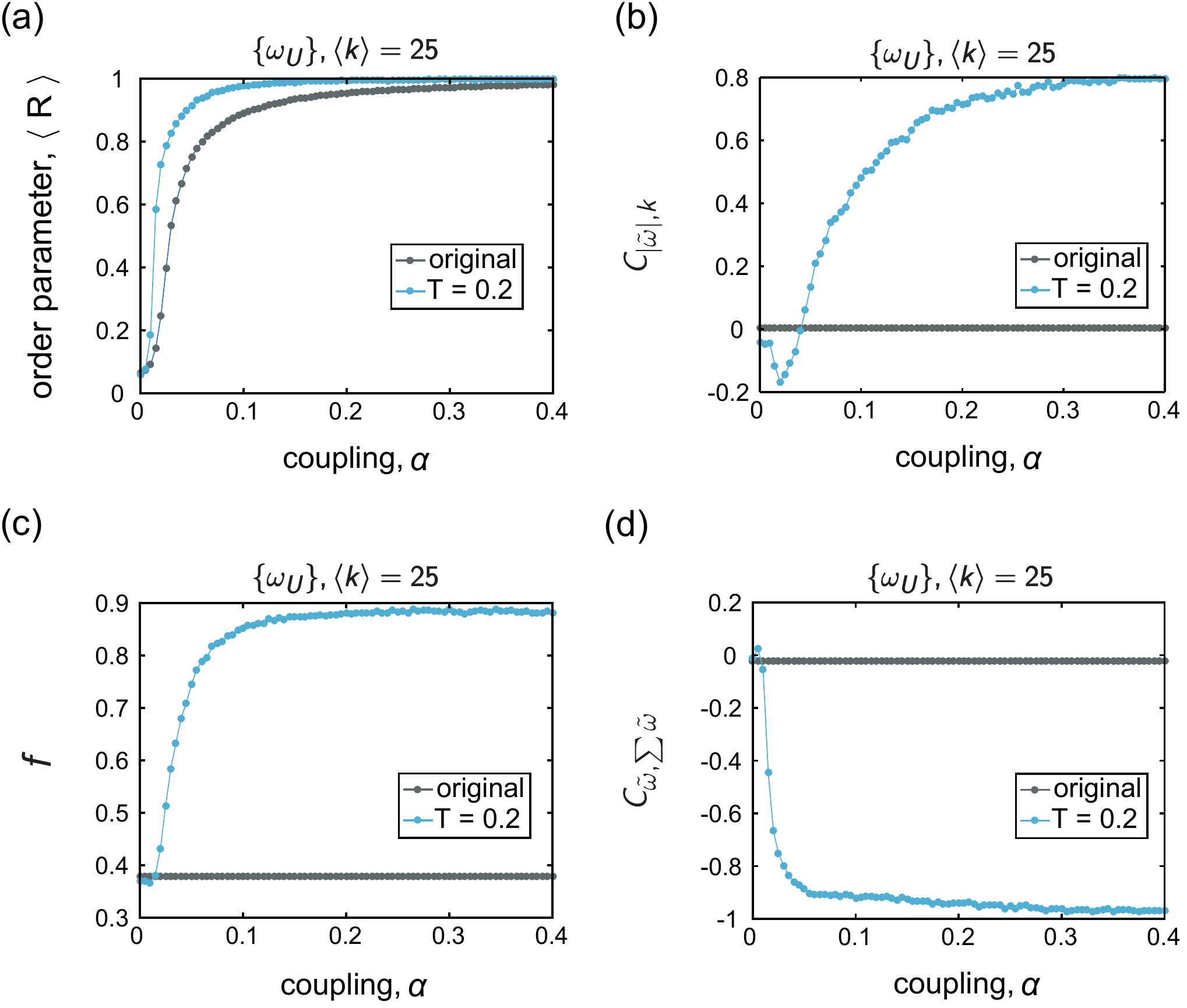}
\caption{Analysis of the synchronization and various correlations between the network topology and the intrinsic frequencies when the frequencies are drawn from a power law distribution with exponent $\gamma = 3$. In each panel, the gray data points correspond to the initial ER networks $\mathcal{G}_{o}$, and the blue points correspond to the adapted networks $G_{\star}$ evolved under a rewiring time scale of $T = 0.2$ time units. The network size is $N=200$ and their mean degree is $\langle k \rangle = 25$. \emph{(a)} The time-averaged order parameter \emph{vs.} coupling. \emph{(b)} The correlation $C_{|\tilde{\omega}|, k}$ between node degree $k_{i}$ and the magnitude of the frequency offset $|\tilde{\omega}_{i}|$. \emph{(c)} The mean fraction $f$ of an oscillator's neighbors that have frequency offsets of opposite sign compared to the central oscillator, as a function of the coupling. \emph{(d)} The correlation $C_{\tilde{\omega},\sum\tilde{\omega}}$ between oscillator frequency offset $\tilde{\omega}_{i}$ and the sum of neighbor frequency offsets $\sum\tilde{\omega}_{j}$, as a function of the coupling. All curves depict averages over 10 instantiations, and the lines between data points serve as guides for the eye.}
\label{f:N200_kavg25_randNetwork_freqPowerlaw}
\end{figure}

Fig.~S\ref{f:N200_kavg25_randNetwork_freqPowerlaw} shows the time-averaged order parameter $\langle R \rangle$ \emph{vs.} coupling $\alpha$ for the static ER networks (gray curve) and adaptively rewired networks (blue curve). The rewiring time scale was $T = 0.2$ time units. As with the two symmetric frequency distributions, we observe an increase in the order parameter for the networks evolved according to the local mechanism. We also examined the co-evolved networks $G_{\star}$ for the correlations between the natural frequencies of the oscillators and the topological organization of the network.  Fig.~S\ref{f:N200_kavg25_randNetwork_freqPowerlaw}\emph{b - d} show plots of $C_{|\tilde{\omega}|, k}$, $f$, and $C_{\tilde{\omega},\sum\tilde{\omega}}$ \emph{vs.} the coupling $\alpha$. Consistent with the findings in the main text, we observe the emergence of strong degree-frequency correlations and frequency-neighbor frequency mixing patterns.

\clearpage
\newpage

%
